\documentclass[11pt]{article}
\pdfoutput=1

\usepackage{jheppub}
\usepackage{url,comment}
\usepackage{times}
\usepackage{latexsym}
\usepackage[dvipsnames]{xcolor}
\usepackage{graphicx, graphics, amsmath, amssymb, slashed, bbm,bm,amsthm, array}
\usepackage{subfigure}
\usepackage{listings} 
\usepackage{trimclip}
\usepackage{xspace}
\usepackage{rotating}
\usepackage{afterpage}

\usepackage{makecell}
 
\usepackage{multirow}

\usepackage{hyperref}	
\hypersetup{  colorlinks=true,     linkcolor=blue,     filecolor=magenta,          urlcolor=blue}

\newcommand{\nc}{\newcommand}

\nc{\beq}{\begin{equation}}
\nc{\eeq}{\end{equation}}
\nc{\barray}{\begin{eqnarray}}
\nc{\earray}{\end{eqnarray}}
\nc{\barrayn}{\begin{eqnarray*}}
\nc{\earrayn}{\end{eqnarray*}}
\nc{\bcenter}{\begin{center}}
\nc{\ecenter}{\end{center}}
\nc{\mc}{\mathcal}
\nc{\er}[1]{(\ref{eq:#1})}
\nc{\onehalf}{\frac{1}{2}} 
\nc{\partialbar}{\bar{\partial}}
\nc{\psit}{\widetilde{\psi}}
\nc{\Tr}{\mbox{Tr}}
\nc{\hc}{\mbox{H.c.}}
\nc{\ev}{\;\mathrm{eV}}
\nc{\mev}{\;\mathrm{MeV}}
\nc{\gev}{\;\mathrm{GeV}}
\nc{\kev}{\;\mathrm{keV}}
\nc{\tev}{\;\mathrm{TeV}}
\nc{\pev}{\;\mathrm{PeV}}
\nc{\eev}{\;\mathrm{EeV}}

\def\chii0{\chi_i^0}
\def\chij0{\chi_j^0}

\newcommand{\gsim}{\lower.7ex\hbox{$\;\stackrel{\textstyle>}{\sim}\;$}}
\newcommand{\lsim}{\lower.7ex\hbox{$\;\stackrel{\textstyle<}{\sim}\;$}}
\nc{\ttbar}{t\bar t}

\newcommand{\fref}[1]{Fig.~\ref{#1}}

\newcommand{\sref}[1]{Section~\ref{#1}}

\newcommand{\cref}[1]{Chapter~\ref{#1}}

\graphicspath{{plots/}}

\newcommand{\glossarycolor}{black}

\newcommand{\tower}{\textcolor{\glossarycolor}{tower module}\xspace}
\newcommand{\towers}{\textcolor{\glossarycolor}{tower modules}\xspace}

\newcommand{\Tower}{\textcolor{\glossarycolor}{Tower module}\xspace}
\newcommand{\Towers}{\textcolor{\glossarycolor}{Tower modules}\xspace}

\newcommand{\SiPM}{\textcolor{\glossarycolor}{SiPM}\xspace} 
\newcommand{\SiPMs}{\textcolor{\glossarycolor}{SiPMs}\xspace} 

\newcommand{\WLSF}{\textcolor{\glossarycolor}{WLSF}\xspace}

\newcommand{\fiber}{\textcolor{\glossarycolor}{fiber}\xspace} 

\newcommand{\mbar}{\textcolor{\glossarycolor}{bar}\xspace} 
\newcommand{\mBar}{\textcolor{\glossarycolor}{Bar}\xspace}
\newcommand{\mbars}{\textcolor{\glossarycolor}{bars}\xspace} 
\newcommand{\mBars}{\textcolor{\glossarycolor}{Bars}\xspace}

\newcommand{\barass}{\textcolor{\glossarycolor}{bar assembly}\xspace} 
\newcommand{\Barass}{\textcolor{\glossarycolor}{Bar assembly}\xspace}
\newcommand{\barasss}{\textcolor{\glossarycolor}{bar assemblies}\xspace}

\newcommand{\sublayer}{\textcolor{\glossarycolor}{sublayer}\xspace} 
\newcommand{\Sublayer}{\textcolor{\glossarycolor}{Sublayer}\xspace}
\newcommand{\sublayers}{\textcolor{\glossarycolor}{sublayers}\xspace}

\newcommand{\layer}{\textcolor{\glossarycolor}{tracking layer}\xspace} 
\newcommand{\layers}{\textcolor{\glossarycolor}{tracking layers}\xspace}
\newcommand{\Layer}{\textcolor{\glossarycolor}{Tracking layer}\xspace}
\newcommand{\Layers}{\textcolor{\glossarycolor}{Tracking layers}\xspace}

\newcommand{\plane}{\textcolor{\glossarycolor}{tracking plane}\xspace} 
\newcommand{\planes}{\textcolor{\glossarycolor}{tracking planes}\xspace}
\newcommand{\Plane}{\textcolor{\glossarycolor}{Tracking plane}\xspace}

\newcommand{\trackingmodule}{\textcolor{\glossarycolor}{tracking module}\xspace} 
\newcommand{\trackingmodules}{\textcolor{\glossarycolor}{tracking modules}\xspace} 
\newcommand{\Trackingmodule}{\textcolor{\glossarycolor}{Tracking module}\xspace}

\newcommand{\ceilingtrackingmodule}{\textcolor{\glossarycolor}{ceiling tracking module}\xspace} 
\newcommand{\ceilingtrackingmodules}{\textcolor{\glossarycolor}{ceiling tracking modules}\xspace} 
\newcommand{\Ceilingtrackingmodule}{\textcolor{\glossarycolor}{Ceiling tracking module}\xspace} 
\newcommand{\Ceilingtrackingmodules}{\textcolor{\glossarycolor}{Ceiling tracking modules}\xspace} 

\newcommand{\walltrackingmodule}{\textcolor{\glossarycolor}{wall tracking module}\xspace} 
\newcommand{\walltrackingmodules}{\textcolor{\glossarycolor}{wall tracking modules}\xspace} 
\newcommand{\Walltrackingmodule}{\textcolor{\glossarycolor}{Wall tracking module}\xspace} 
\newcommand{\Walltrackingmodules}{\textcolor{\glossarycolor}{Wall tracking modules}\xspace}

\newcommand{\wallvetosublayer}{\textcolor{\glossarycolor}{front wall veto \sublayer}\xspace} 
\newcommand{\wallvetosublayers}{\textcolor{\glossarycolor}{front wall veto \sublayers}\xspace} 
\newcommand{\Wallvetosublayer}{\textcolor{\glossarycolor}{Front wall veto \sublayer}\xspace}

\newcommand{\wallvetolayer}{\textcolor{\glossarycolor}{front wall  veto layer}\xspace} 
\newcommand{\wallvetolayers}{\textcolor{\glossarycolor}{front wall  veto layers}\xspace} 
\newcommand{\Wallvetolayer}{\textcolor{\glossarycolor}{Front wall veto layer}\xspace} 
\newcommand{\Wallvetolayers}{\textcolor{\glossarycolor}{Front wall veto layers}\xspace}

\newcommand{\wallveto}{\textcolor{\glossarycolor}{front wall veto detector}\xspace} 
 
\newcommand{\Wallveto}{\textcolor{\glossarycolor}{Front wall veto detector}\xspace}

\newcommand{\columndetector}{\textcolor{\glossarycolor}{column detector}\xspace} 
\newcommand{\columndetectors}{\textcolor{\glossarycolor}{column detectors}\xspace} 
\newcommand{\Columndetector}{\textcolor{\glossarycolor}{Column detector}\xspace}

\newcommand{\floorvetolayer}{\textcolor{\glossarycolor}{floor veto layer}\xspace} 
\newcommand{\floorvetolayers}{\textcolor{\glossarycolor}{floor veto layers}\xspace} 
\newcommand{\Floorvetolayer}{\textcolor{\glossarycolor}{Floor veto layer}\xspace}

\newcommand{\floorvetostrip}{\textcolor{\glossarycolor}{floor veto strip}\xspace} 
\newcommand{\floorvetostrips}{\textcolor{\glossarycolor}{floor veto strips}\xspace} 
\newcommand{\Floorvetostrip}{\textcolor{\glossarycolor}{Floor veto strip}\xspace} 
\newcommand{\Floorvetostrips}{\textcolor{\glossarycolor}{Floor veto strips}\xspace}

\newcommand{\floorveto}{\textcolor{\glossarycolor}{floor veto detector}\xspace} 
 
\newcommand{\Floorveto}{\textcolor{\glossarycolor}{Floor veto detector}\xspace}

\newcommand{\veto}{\textcolor{\glossarycolor}{veto detector}\xspace} 
 
\newcommand{\Veto}{\textcolor{\glossarycolor}{Veto detector}\xspace}

\newcommand{\barlength}{\textcolor{\glossarycolor}{2.35~m}\xspace}
\newcommand{\barwidth}{\textcolor{\glossarycolor}{3.5~cm}\xspace}
\newcommand{\barthickness}{\textcolor{\glossarycolor}{1~cm}\xspace}
\newcommand{\moduleside}{\textcolor{\glossarycolor}{9~m}\xspace}

\title{
Conceptual Design Report for the MATHUSLA Long-Lived Particle Detector near CMS
}

\author{
\includegraphics[width=4cm]{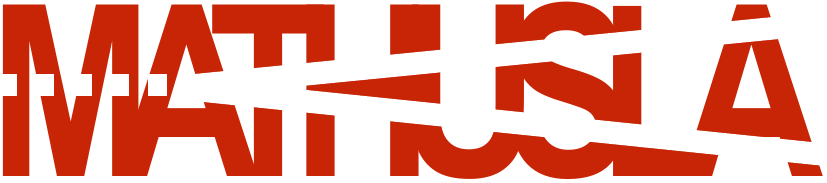}
\\
{\normalfont 
\href{https:mathusla-experiment.web.cern.ch}{\texttt{mathusla-experiment.web.cern.ch}}
\\
\vspace{5mm}}}

\author[1]{\hspace*{-1mm}Branden   Aitken,}
\author[2]{Cristiano  Alpigiani,}
\author[3]{Juan Carlos   Arteaga-Vel\'azquez,}
\author[4]{Mitchel   Baker,}
\author[5]{Kincso   Balazs,}
\author[6]{Jared   Barron,}
\author[7]{Brian   Batell,}
\author[8]{Austin   Batz,}
\author[9]{Yan   Benhammou,}
\author[5]{Tamara Alice  Bud,}
\author[10]{Karen Salom\'e   Caballero-Mora,}
\author[11]{John Paul   Chou,}
\author[12]{David   Curtin,}
\author[5]{Albert   de Roeck,}
\author[12]{Miriam   Diamond,}
\author[13]{Mariia   Didenko,}
\author[14, 15]{Keith R.    Dienes,}
\author[16]{William  Dougherty,}
\author[5]{Liam Andrew   Dougherty,}
\author[17]{Marco   Drewes,}
\author[11]{Sameer   Erramilli,}
\author[9]{Erez   Etzion,}
\author[18]{Arturo   Fern\'andez T\'ellez,}
\author[11]{Grace  Finlayson,}
\author[19]{Oliver   Fischer,}
\author[20]{Jim   Freeman,}
\author[5]{Jonathan   Gall,}
\author[2]{Ali   Garabaglu,}
\author[21]{Bhawna   Gomber,}
\author[11]{Stephen Elliott   Greenberg,}
\author[12, 22]{Jaipratap Singh  Grewal,}
\author[1]{Zoe   Hallman,}
\author[11]{Bahgat   Hassan,}
\author[23]{Yuekun   Heng,}
\author[12]{Keegan   Humphrey,}
\author[1]{Trystan   Humphrey,}
\author[8]{Graham  D.  Kribs,}
\author[12]{Alex   Lau,}
\author[12]{Jiahao   Liao,}
\author[15]{Zhen   Liu,}
\author[24]{Giovanni   Marsella,}
\author[5]{Matthew   McCullough,}
\author[25]{David   McKeen,}
\author[6]{Patrick   Meade,}
\author[1]{Caleb  Miller,}
\author[9]{Gilad   Mizrachi,}
\author[10]{O. G.   Morales-Olivares,}
\author[25]{David   Morrissey,}
\author[26]{Abdulrahman Ahmed   Morsy,}
\author[5]{John   Osborn,}
\author[12]{Gabriel  Owh,}
\author[27]{Michalis  Panagiotou,}
\author[2]{Mason   Proffitt,}
\author[12]{Runze  Ren,}
\author[4]{Steven H.   Robertson,}
\author[18]{Mario   Rodr\'iguez-Cahuantzi,}
\author[1]{Heather   Russell,}
\author[13]{Victoria   S\'anchez,}
\author[27]{Halil  Saka,}
\author[1]{Mamoksh   Samra,}
\author[28]{Rodney   Schnarr,}
\author[29]{Jessie   Shelton,}
\author[9]{Yiftah   Silver,}
\author[30]{Daniel   Stolarski,}
\author[31]{Martin A.   Subieta Vasquez,}
\author[32]{Sanjay Kumar   Swain,}
\author[11]{Steffie Ann   Thayil,}
\author[33]{Brooks   Thomas,}
\author[13]{Emma   Torro,}
\author[34]{Yuhsin   Tsai,}
\author[1]{Bennett   Winnicky-Lewis,}
\author[9]{Igor  Zolkin,}
\author[13]{Jose   Zurita}

\affiliation[1]{University of Victoria, Canada}
\affiliation[2]{University of Washington, Seattle, USA}
\affiliation[3]{Universidad Michoacana de San Nicol\'as de Hidalgo, Mexico (UMSNH)}
\affiliation[4]{University of Alberta, Canada}
\affiliation[5]{CERN, Switzerland}
\affiliation[6]{YITP Stony Brook, USA}
\affiliation[7]{University of Pittsburgh, USA}
\affiliation[8]{University of Oregon, USA}
\affiliation[9]{Tel Aviv University, Israel}
\affiliation[10]{Universidad Aut\'onoma de Chiapas, Mexico (UNACH)}
\affiliation[11]{Rutgers, the State University of New Jersey, USA}
\affiliation[12]{University of Toronto, Canada}
\affiliation[13]{Instituto de F\'isica Corpuscular (CSIC-UV), Valencia, Spain}
\affiliation[14]{University of Arizona, USA}
\affiliation[15]{University of Maryland, USA}
\affiliation[16]{Kenmore, Washington, USA}
\affiliation[17]{Universit\'{e} catholique de Louvain, France}
\affiliation[18]{Benem\'erita Universidad Aut\'onoma de Puebla, Mexico (BUAP)}
\affiliation[19]{Liverpool U., UK}
\affiliation[20]{Fermi National Accelerator Laboratory (FNAL), USA}
\affiliation[21]{Hyderabad University, India}
\affiliation[22]{University of California San Diego, USA}
\affiliation[23]{Institute of High Energy Physics, Beijing}
\affiliation[24]{Universit\`a degli di Studi di Palermo, Palermo, Italy}
\affiliation[25]{TRIUMF, Canada}
\affiliation[26]{Ain Shams University, Cairo, Egypt}
\affiliation[27]{University of Cyprus, Cyprus}
\affiliation[28]{Carleton University, Canada}
\affiliation[29]{University of Illinois Urbana-Champaign, USA}
\affiliation[30]{Carleton Unversity, Canada}
\affiliation[31]{Instituto de Investigaciones F\'isicas (IIF), Observatorio de F\'isica C\'osmica de \^a Chacaltaya\^a, Universidad Mayor de San Andr\'es (UMSA)}
\affiliation[32]{National Institute of Science Education and Research, HBNI, Bhubaneswar, India}
\affiliation[33]{Lafayette College, USA}
\affiliation[34]{University of Notre Dame, USA}

\emailAdd{mathusla.experiment@cern.ch}

\abstract{
We present the Conceptual Design Report (CDR) for the 
MATHUSLA (MAssive Timing Hodoscope for Ultra-Stable neutraL pArticles) long-lived particle detector at the HL-LHC,
covering the design, fabrication and installation at CERN Point 5.
MATHUSLA is a 40~m-scale  detector with an air-filled decay volume that is instrumented with scintillator tracking detectors, to be located near CMS.
Its large size, close proximity to the CMS interaction point and about 100~m of rock shielding from HL-LHC backgrounds allows it to detect LLP production rates and lifetimes that are  one to two orders of magnitude beyond the  ultimate sensitivity of the HL-LHC main detectors 
for many highly motivated LLP signals.
Data taking is projected to commence with the start of HL-LHC operations.
We present a new 40m design for the detector: its individual scintillator \mbars and wavelength-shifting fibers,  their organization into \layers, \trackingmodules, \towers and the \veto; define a high-level design for the supporting electronics, DAQ and trigger system, including supplying a hardware trigger signal to CMS to record the LLP production event; outline computing systems, civil engineering and safety considerations; and present preliminary cost estimates and timelines for the project.
We also conduct detailed simulation studies of the important cosmic ray and HL-LHC muon backgrounds, implementing full track/vertex reconstruction and background rejection, to ultimately demonstrate high signal efficiency and $\ll 1$ background event in realistic LLP searches for the main physics targets at MATHUSLA. 
This sensitivity is robust with respect to detector design or background simulation details.
Appendices provide various supplemental information. 
}

\usepackage{lineno}
\begin{document}

\begin{flushright}
%\phantom{\small{.}}
%CERN-LHCC-2023-XXX\\
%LHCC-X-XXX-XXX
\end{flushright}
%%%%%%%%%%%%%%%%%
\maketitle
\newpage
%%%%%%%%%%%%%%%%%

%%%%%%%%%%%%%%%%%
\section*{Foreword}
%%%%%%%%%%%%%%%%%

This document represents the evolution of the MATHUSLA experimental proposal to the conceptual design stage. Previous versions of the proposal advocated for a 100m detector, but in response to recent changes in the funding landscape, the proposal has been resized to a 40m decay volume without sacrificing the ability to improve upon the LLP reach of the main detectors by orders of magnitude for the target LLP physics scenarios. The modular and scalable nature of the MATHUSLA detector means that this change essentially amounts to building 16 instead of 100 \towers, and reducing their vertical height to reduce civil engineering demands. 
Detailed simulations have now been conducted to demonstrate that this design allows the MATHUSLA detector to discover LLPs with zero or very low background, while R\&D conducted in two test stand laboratories at the University of Toronto and the University of Victoria, building on previous DAQ and trigger design studies
and earlier design work at participating institutions, allow the collaboration to present this Conceptual Design Report (CDR) for the MATHUSLA detector.

\newpage
%%%%%%%%%%%%%%%%%

%%%%%%%%%%%%%%%%%
\section{Introduction}
\label{s.introduction}
%%%%%%%%%%%%%%%%%

We present the MATHUSLA (MAssive Timing Hodoscope for Ultra-Stable neutraL pArticles) detector Conceptual Design Report.
The motivation for auxiliary long-lived particle (LLP) detectors located above the LHC main detectors has been covered in many published papers, most comprehensively in the MATHUSLA physics case~\cite{Curtin:2018mvb}, as well as the Letters of Intent~\cite{MATHUSLA:2018bqv, MATHUSLA:2020uve} and the recent U.S. Snowmass study~\cite{MATHUSLA:2022sze}. 
The rest of this section briefly summarizes the physics objectives of MATHUSLA, as well as an overview of the current benchmark design for for the detector. 
Section~\ref{s.detectoratcms} discusses LLP searches at the MATHUSLA detector above CMS: 
experimental conditions at the surface of LHC Point 5, the resulting backgrounds to LLP searches, and detailed simulation studies that demonstrate how MATHUSLA can reconstruct physics target LLP signals and reject all significant backgrounds. This informs updated reach projections  of this 40m MATHUSLA design for BSM benchmark models.
In Section~\ref{s.detectordesign}, detector design considerations are discussed in detail, including physics requirements, different detector components, assembly, environmental conditions, and calibration. 
Electronics, DAQ, and trigger are discussed in Section~\ref{s.electronicsdaqtrigger}.
MATHUSLA's computing requirements are discussed in ~\ref{s.computing}. 
An engineering concept for the detector structure -- a modular design with \barasss as its basic building blocks -- is presented this section with further details supplied in ~\ref{s.civeng}, along with a civil engineering concept for the experimental hall that has to be constructed on CERN-owned land near CMS. 
Safety considerations for construction and operations are discussed in Section~\ref{s.safety}.
Finally, timeline and cost estimates for the experiment are presented in \sref{s.costschedule}.
Appendices include summaries of past and ongoing Detector R\&D efforts (Appendix~\ref{s.detectorRD}), as well as a list of supporting documents released by the MATHUSLA collaboration (Appendix~\ref{s.supportingdocuments}), a Glossary of frequently used terms (Appendix~\ref{s.glossary}), and an outline of the MATHUSLA collaboration organizational structure (Appendix~\ref{s.organization}).

%%%%%%%%%%%%%%%%%
\subsection{General Overview}
\label{s.overview}
%%%%%%%%%%%%%%%%%

\begin{figure}
\begin{center}
\includegraphics[width=0.8\textwidth]{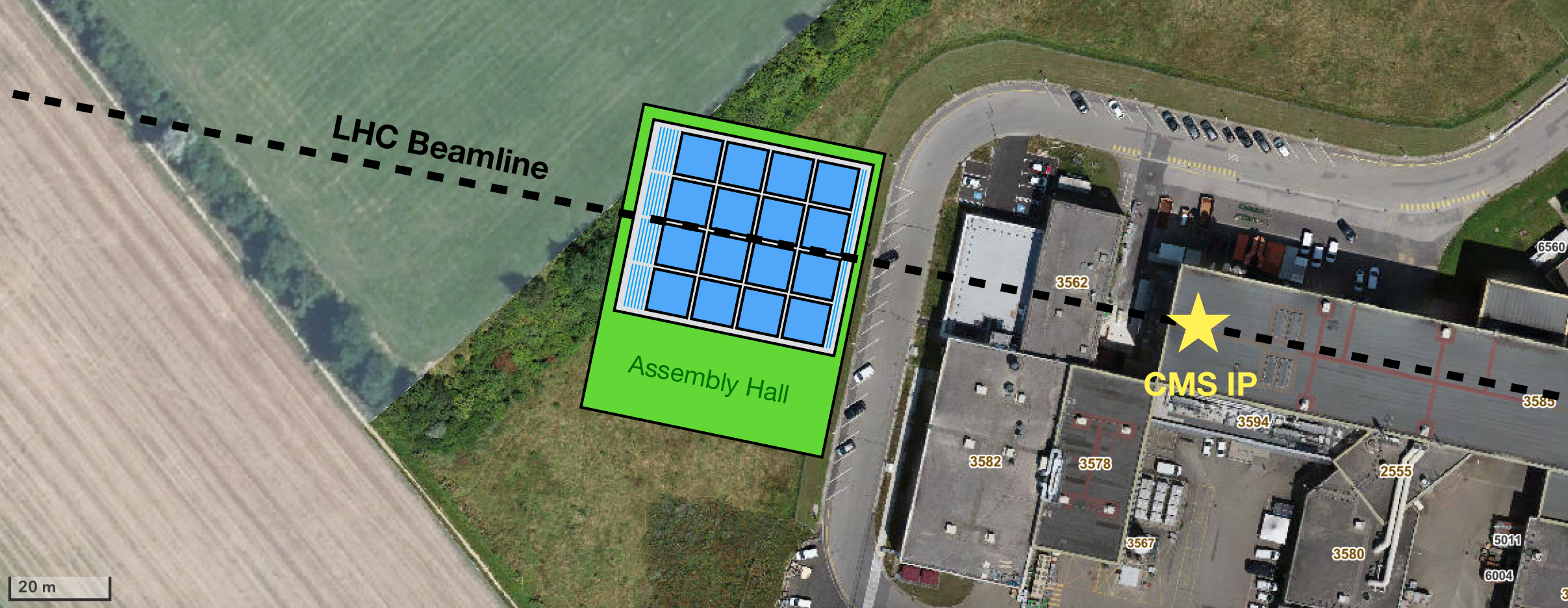}
\end{center}
\caption{Location of proposed MATHUSLA detector at the CMS site. 
}
\label{fig:layout_P5}
\end{figure}

\begin{figure}
\begin{center}
\includegraphics[width=1.0\textwidth]{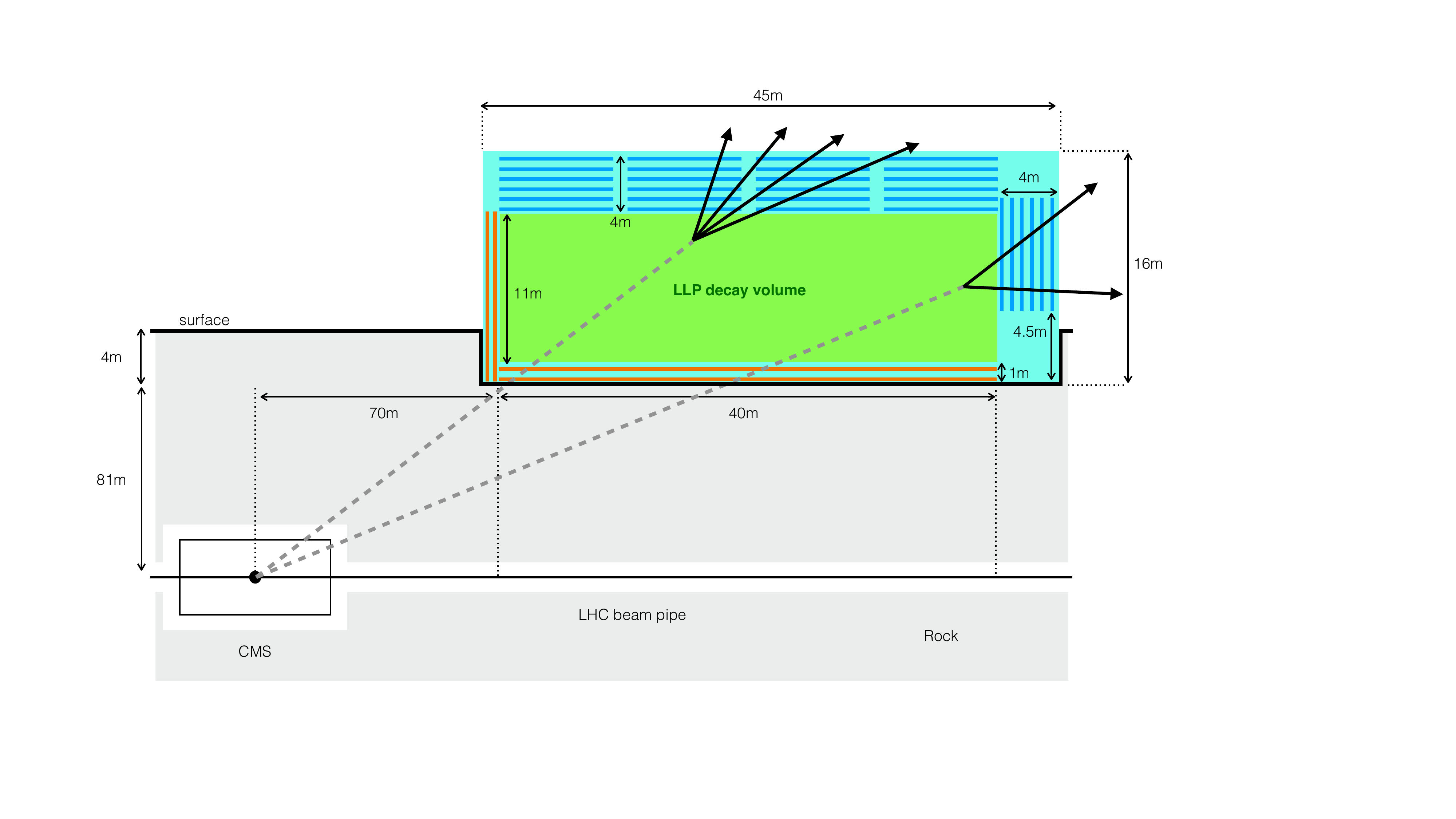}
\end{center}
\caption{
Schematic MATHUSLA geometry relative to the CMS collision point. 
LLPs (gray dashed) can decay into SM charged states (black arrows) in the 
$\sim 40~\mathrm{m} \times 40~\mathrm{m} \times 11~\mathrm{m}$ \textbf{LLP decay volume (green)}
and be reconstructed as displaced vertices by the \textbf{\trackingmodules (blue)}.
Both wall and ceiling \trackingmodules consist of six $(9~\mathrm{m})^2$ \layers separated by 80cm for a total thickness of 4m, arranged in a $4 \times 4$ grid on the ceiling and a row of 4 on the rear wall, with neighboring \trackingmodules separated by $\sim 1~\mathrm{m}$. 
Rejection of LHC muon and cosmic ray backgrounds is aided by the  double-layer \textbf{\veto} (orange), comprising the
\textbf{\wallveto} and \textbf{\floorveto}, which is close to hermetic for LHC muons and downward traveling cosmics. (The realistic structure of the \floorveto has been simplified for this illustration.) For consistency, we show the modest amount of surface excavation (at most 4m below grade) that may be required to fit the MATHUSLA detector into an experimental hall that conforms to local building height restrictions, see Section~\ref{s.civeng} for more details.}
\label{fig:mathuslacms}
\end{figure}

MATHUSLA is a proposed large-volume detector for the HL-LHC~\cite{Chou:2016lxi, MATHUSLA:2018bqv, MATHUSLA:2020uve} that is dedicated to finding exotic long-lived particles (LLPs) produced in $pp$ collisions in the CMS detector at Point 5.
Its approximately 45 m $\times$ 50 m footprint\footnote{Compared to previous MATHUSLA publications, this is a smaller detector geometry to bring the scale of the project more in line with realistic future funding envelopes at CERN, Europe and North America.} will be situated on CERN-owned, available land adjacent to the CMS complex (\fref{fig:layout_P5}). The basic detector design is simple in principle, consisting of an empty LLP decay volume  that is instrumented with \layers (\fref{fig:mathuslacms}). LLP decays into charged, Standard Model (SM) particles can be reconstructed as displaced vertices (DVs) and distinguished from cosmic ray and other backgrounds using both their direction of travel and a variety of other timing and geometric criteria.

MATHUSLA will add a crucial capability to the LHC physics program, beyond the current capabilities of the LHC main detectors.  As discussed in the MATHUSLA physics case white paper~\cite{Curtin:2018mvb},  LLP signals are broadly motivated and ubiquitous in BSM scenarios. Their discovery and subsequent characterization could resolve many fundamental mysteries of high energy physics, including the Hierarchy Problem, the nature of Dark Matter, the origin of the Universe's matter-antimatter asymmetry,  neutrino masses and the strong CP problem.
The LLP search program at the LHC has undergone dramatic development in recent years~\cite{Alimena:2019zri}, but the main detectors suffer from trigger limitations~\cite{Alimena:2021mdu} and complex backgrounds. This severely curtails their sensitivity to many classes of LLPs, in particular for lifetimes larger than the detector size, where the rate of observed decays is necessarily low. 
MATHUSLA has similar geometric acceptance for LLPs with lifetimes $c \tau \gtrsim 100~\mathrm{m}$ as ATLAS or CMS, but its location on the surface, shielded from the high-energy LHC environment, frees it entirely from their trigger and background limitations. This results in MATHUSLA being orders of magnitude more sensitive to the production of many important types of LLPs~\cite{Chou:2016lxi,Curtin:2018mvb,Beacham:2019nyx}.

%%%%%%%%%%%%%%%%%
\subsection{Physics Objectives}
\label{s.physicsobjectives}
%%%%%%%%%%%%%%%%%

The main physics goal of MATHUSLA is the search for electromagnetically neutral LLPs produced at the HL-LHC. The primary physics target for MATHUSLA 
is informed by the blind-spots of the HL-LHC main detectors, and closing those blind-spots informs much of the detector design we present in the rest of this document. 

The HL-LHC main detectors face certain limitations in searching for LLPs.\footnote{We focus on neutral LLPs, but some of these limitations also apply to charged LLPs.} First, full track reconstruction, let alone vertex reconstruction, is not available at hardware trigger level~\cite{Alimena:2021mdu} (though displaced tracks can sometimes be used in the high level trigger selection).
As a result, triggering has to rely on other features of the LLP decay or associated object produced alongside the LLP, such as high mass/energy final states, leptons, or photons. (A notable exception is the ATLAS Muon System, which can act as a L1 trigger.)
Second, while the displaced geometric nature of LLP decays is highly distinctive and can be used to severely reduce backgrounds, the busy, pile-up-rich environment of the HL-LHC presents huge combinatorial challenges to the reconstruction of DVs, and includes many complicated processes like material interactions, beam halo, and cavern radiation, all of which are hard to simulate and/or veto~\cite{Alimena:2019zri}. 
This is the case even when considering the capability of the ATLAS~\cite{ATLAS:2012av, ATLAS:2018tup,ATLAS:2019jcm,ATLAS:2022gbw} muon system, which is unique amongst the sub-detectors, to act as a hardware trigger for LLP decays. CMS is developing similar capability~\cite{Alimena:2021mdu}.  On the other hand, reducing these backgrounds by imposing additional signal requirements (for example, requiring a lepton from associated production when searching for LLPs produced in  exotic Higgs boson decays) can significantly reduce the effective signal cross section and hence LLP discovery potential.
All of this severely limits the sensitivity of many LLP searches, especially for \emph{medium-mass} LLPs (10 to few 100 GeV) that decay hadronically, or \emph{low-mass} LLPs ($\lesssim$ few GeV) of any decay mode. 
While the latter will be well-covered by the recently approved SHiP experiment~\cite{Beacham:2019nyx}, the former blind spot requires the full collision energy and luminosity of the HL-LHC as well as a background-free environment to resolve. Maximizing the discovery potential of the HL-LHC in this fashion is MATHUSLA's mission.

\textbf{The main physics target of MATHUSLA is therefore hadronically decaying LLPs in the 10 to few 100 GeV mass range. This is a highly motivated new physics scenario that specifically occurs in many theories, including the neutral naturalness solutions to the hierarchy problem~\cite{Chacko:2005pe, Craig:2015pha, Curtin:2015fna, Batz:2023zef}) and, for example, exotic Higgs boson decays to LLPs in general~\cite{Curtin:2013fra}. This also makes MATHUSLA a crucial component of the precision Higgs physics program at the HL-LHC.}  

LLPs of this very general type are the most glaring coverage gap of the LHC main detectors. The absence of hard leptons or photons in the decay and relatively modest mass scale makes triggering challenging, and QCD backgrounds are very high. This is effectively illustrated by HL-LHC sensitivity projections of the ATLAS muon-system single-DV search~\cite{Coccaro:2016lnz, Chou:2016lxi, Curtin:2018mvb} and even the lower-background CMS muon system search~\cite{CMS:2023arc}. The absence of backgrounds and trigger limitations allows MATHUSLA to probe LLP production rates \emph{1-2 orders of magnitude} smaller than the main detector at long lifetimes.

For lifetimes larger than $\sim 100$~m, MATHUSLA has a geometric acceptance for LLP decays of similar order as CMS or ATLAS, as the greater longitudinal depth of MATHUSLA compensates for the smaller solid angle coverage. 
Therefore, MATHUSLA's main physics objectives are neutral LLP scenarios where the main detector sensitivity is curtailed by trigger or background limitations.
For those scenarios, MATHUSLA effectively realizes something closer to the ``idealized geometric acceptance'',  and hence new physics sensitivity, of the main detectors, provided the LLP decays to a high multiplicity of final states.

Note that hadronically decaying LLPs with even higher masses in the TeV-range can of course be equally well discovered at MATHUSLA. However, in those scenarios the main detectors are likely to have excellent coverage, since the high energy released in the decay grants them high acceptance to e.g. $H_T$ based triggers, and reduces background to very low levels. The exception are high-mass LLPs that only release $\lesssim \mathcal{O}(100 \gev)$ of hadronic energy when they decay due to a modest mass gap: such a scenario behaves like a medium-mass LLP and is a primary MATHUSLA target.

A large number of previous studies (see~\cite{Chou:2016lxi, MATHUSLA:2018bqv, Curtin:2018mvb, MATHUSLA:2020uve} and references therein) have already established that HL-LHC production rates for LLPs and MATHUSLA's geometric acceptance for their \emph{decays} inside its detector volume are sufficient to fulfill these physics objectives.
In Section~\ref{s.detectoratcms} we study realistic track and vertex reconstruction within MATHUSLA, and conduct extensive signal and background simulations to establish the necessary veto strategies for background-free LLP searches. This yields expected acceptances for LLP decays in realistic searches, which are used to study the reach of the present MATHUSLA design across the parameter space of our physics target LLP benchmark models. We can conclude that MATHUSLA has excellent sensitivity to its LLP physics targets.

Crucial to MATHUSLA's success is the fact that the extremely dominant cosmic ray backgrounds can be carefully studied \emph{in situ} during the 50\% beam-off duty cycle of the HL-LHC, ensuring that the complete vetoing of backgrounds can be verified without any contamination by potential LLP signals. \textbf{In the event of a positive LLP detection, MATHUSLA can therefore robustly claim discovery of a new fundamental particle.}

%%%%%%%%%%%%%%%%%
\subsection{Detector Overview}
\label{s.detectorintro}
%%%%%%%%%%%%%%%%%

\begin{figure}[]
\begin{center}
\includegraphics[width=0.95\textwidth]{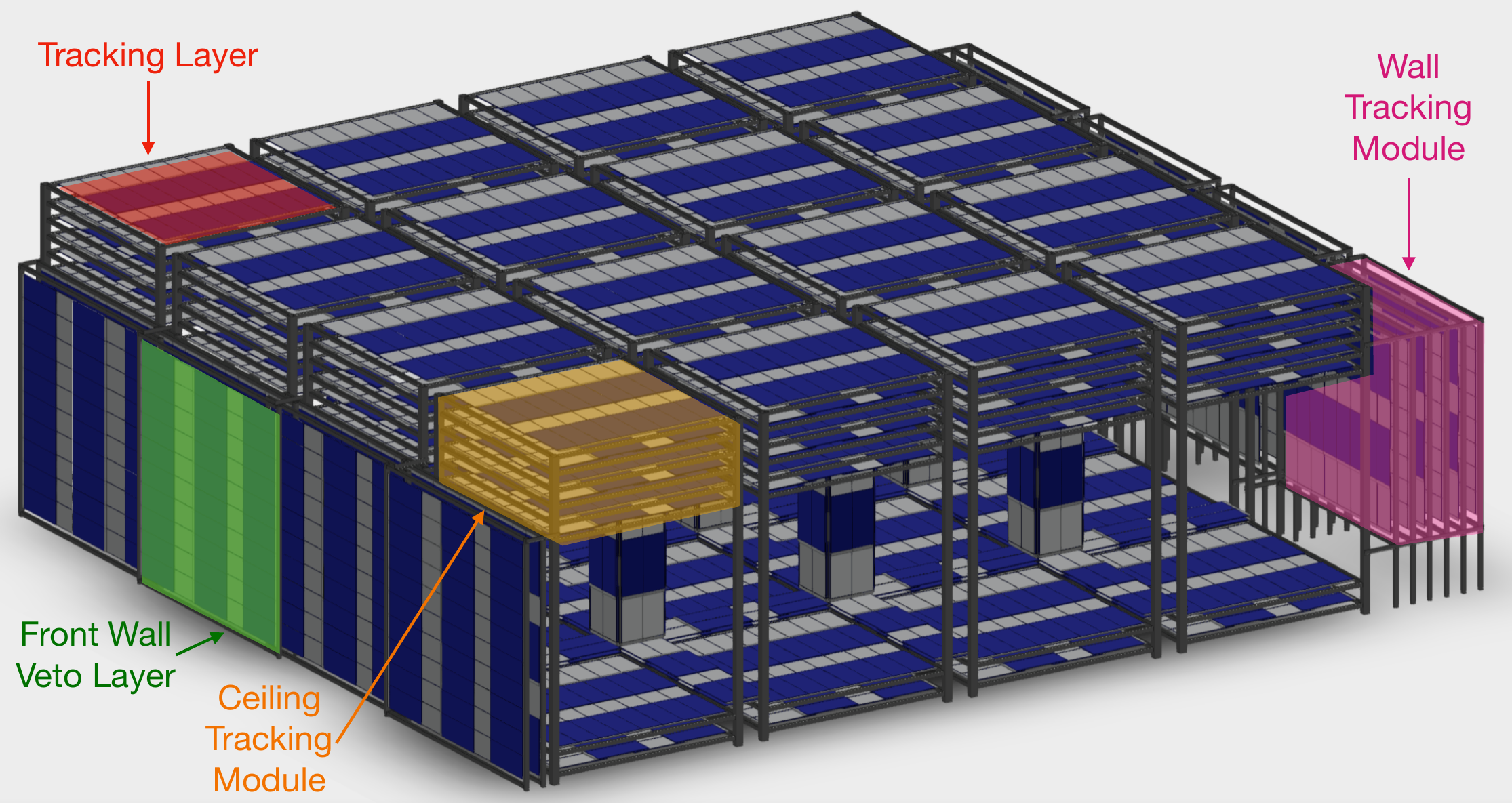}
\\ \vspace{3mm}
\includegraphics[width=0.95\textwidth]{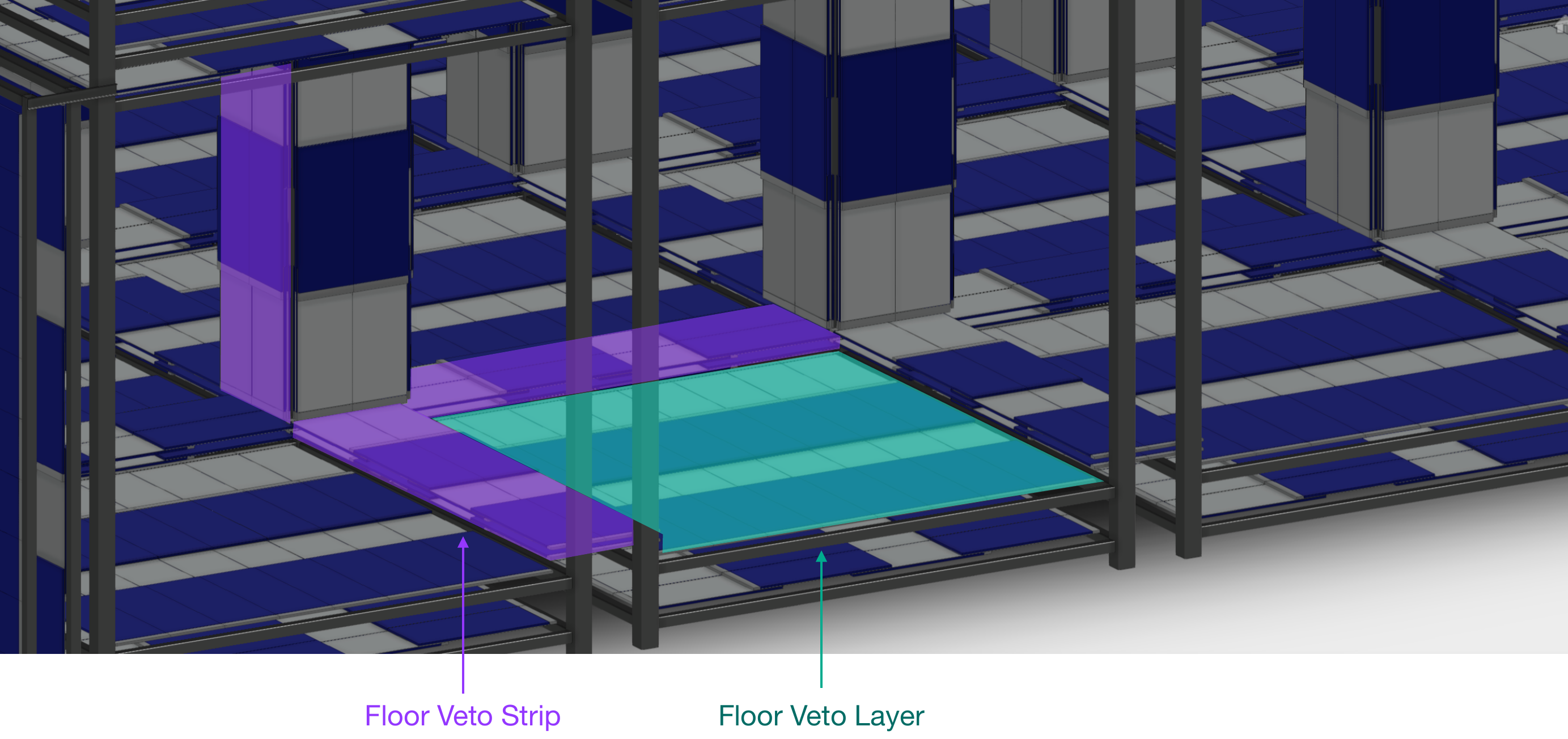}
\end{center}
\vspace*{-5mm}
\caption{
\emph{Top:}
Engineering concept for a realistic MATHUSLA detector structure. 
16 \Ceilingtrackingmodules, and 4 \walltrackingmodules in the rear, are each comprised of 6 \layers with 80cm separation. The \ceilingtrackingmodules are mounted in \towers which are arranged in a $4 \times 4$ grid, with 1m separation for maintenance access. 
The \wallveto is comprised of $(11.2~\mathrm{m})^2$ \wallvetolayers, 8 of which are arranged in 2 rows of 4 with overlap to provide hermetic coverage. 
\emph{Bottom:}
The \floorveto comprises \floorvetolayers, identical to those in the \ceilingtrackingmodules and mounted 0.5m and 1m above the floor in the \towers, 
as well as $\sim$ 2.3m $\times$ 9m \floorvetostrips, which cover the gaps between \towers. Four vertical \floorvetostrips also each constitute a single \columndetector, instrumenting the vertical support columns. 
In combination, this \veto system is hermetic for LHC muons and downward cosmics.
}
\label{fig:MATHUSLAstructure}
\end{figure}

\begin{table}[h!]
    \centering
    \caption{Summary of several attributes of the MATHUSLA detector benchmark design presented in this report. The third column references figure or section with more details. 
    }
    \hspace*{-3mm}
    \begin{tabular}{|m{0.3 \textwidth}|m{0.55\textwidth}|m{0.13\textwidth}|}
    \hline
    Distance from CMS IP
    &
    82-93m vertical, 70-110m horizontal along beam axis
    %: $\sim$ 110-130m total
    & Figure~\ref{fig:mathuslacms}
    \\ \hline
    Detector volume & $\sim$ $40~\mathrm{m} \times 45~\mathrm{m} \times 16~\mathrm{m} $ & Figure~\ref{fig:mathuslacms}
    \\ \hline
    Decay volume
    &
    $\sim$ $40~\mathrm{m} \times 40~\mathrm{m} \times 11~\mathrm{m}$ & Figures~\ref{fig:mathuslacms},~\ref{fig:MATHUSLAstructure}
    \\ \hline
    Number of tracking modules
    & 
    20 total: a grid of $4 \times 4$ \towers each has a \ceilingtrackingmodule, and 4 \walltrackingmodules are mounted on the rear wall.
    & Figures~\ref{fig:mathuslacms},~\ref{fig:MATHUSLAstructure}, Sec.~\ref{s.detectordesign}
    \\ \hline
    Tracking module Dimensions
    &
    9~m $\times$ 9~m, height $\sim$ 4~m
    & Figures~\ref{fig:mathuslacms},~\ref{fig:modular} 
    \\ \hline
    \Layers
    & 
    6 in ceiling (top 4m, 0.8m apart) and 6 in rear wall (starting $\sim$ 4.5m above the floor, also 0.8m apart).
    & Figure~\ref{fig:modular} \\ \hline
    %Vertical positions
    %& & \\ \hline
    Hermetic wall detector
    & 
    Double layer in wall facing IP to detect LHC muons.
    & Figure~\ref{fig:modular}  \\ \hline
    Hermetic floor detector
    & 
    2 \floorvetolayers at heights 0.5m and 1m in each of the 16 \towers, 24 $(9~\mathrm{m} \times 2.8~\mathrm{m})$ \floorvetostrips to cover gaps between \towers, and 9 \columndetectors each utilizing 4 vertical \floorvetostrips to cover the vertical support columns. 
    & Figure~\ref{fig:modular}  \\ \hline
    Detector technology
    & Extruded plastic scintillator \mbars, \barwidth wide, \barthickness thick, \barlength  long, arranged in alternating orientations with each vertical \layer. \mBars are threaded with wavelength-shifting fibers connected to \SiPMs. & Section~\ref{s.detectordesign}\\ \hline
    Number of \barasss
    &
    $6224$, 32 channels each
    &
    Section~\ref{s.detectordesign}
    \\ \hline
    Number of Channels
    &
    $\sim 2\times 10^5$ \SiPMs
    &
    Section~\ref{s.detectordesign}
    \\ \hline
    Tracking resolution
    & 
    $\sim$~1~ns timing resolution;
    $\sim$ 1~cm (15~cm) along transverse (longitudinal) direction of scintillator bar.
    & Section~\ref{s.detectordesign} \\ \hline
    Trigger
    & 
    $3 \times 3$ groups of \trackingmodules perform simplified tracking/vertexing to trigger on upwards-traveling tracks and vertices. 
    Corresponding time stamps flag regions of MATHUSLA datastream for full reconstruction and permanent storage. 
    MATHUSLA can also send hardware trigger signal to CMS to record LLP production event.
    &  Section~\ref{s.electronicsdaqtrigger} \\ \hline
    Data rate
    & 
    Each \trackingmodule and section of \floorveto detector associated with each \tower produces $\lesssim$~0.6 TB/day. (The \wallveto data rate is a small addition.)  Less than 0.1\% of full detector data will be selected for permanent storage using a trigger system, corresponding to about 8~TB/year. 
    & Section~\ref{s.DAQ}, \ref{s.computing} \\ \hline
    \end{tabular}
    \label{t.summary}
\end{table}

An overview of the engineering concept for a realistic implementation of the proposed MATHUSLA detector is shown in Fig.~\ref{fig:MATHUSLAstructure}.
Several key attributes of the benchmark design are summarized in Table~\ref{t.summary}.

In order to improve on the LLP sensitivity of the LHC main detectors by orders of magnitude, while also complying with maximum building height regulations near LHC Point 5 and minimizing potentially expensive excavation, MATHUSLA's  \textbf{LLP decay volume} has a footprint of $\sim (40~\mathrm{m})^2$ and height of $\sim$ 11~m, with about a meter on the bottom being taken up by the \textbf{\floorveto} and 4~m on top occupied by the \textbf{\ceilingtrackingmodules}. Additional \textbf{\walltrackingmodules} on the back wall relative to the LHC IP greatly enhance signal reconstruction efficiency for LLPs decaying in the rear of the detector, while a \textbf{\wallveto} enhances rejection of LHC muon backgrounds.

Each \trackingmodule, to be installed in the ceiling or rear wall, is comprised of 6 \layers with an area of $\sim (9~\mathrm{m})^2$, separated by 80~cm for a total height of 4~m. 
Each \ceilingtrackingmodule is mounted in a \tower with four vertical supports, which also support the components of the \floorveto. 
\Towers are arranged in a $4 \times 4$ grid to cover the entire decay volume, and are separated by 1~m-wide gaps.
The gaps have little impact on signal efficiency but are crucial for allowing maintenance access. 
The rear \walltrackingmodules are similarly arranged in a row with 1~m gaps between them. 

The 6 \layers in each tracking module are comprised of \textbf{scintillating \mbars} of \barwidth width that are arranged with alternating transverse orientation to their neighboring \layers. 
This configuration provides position and timing coordinates of charged particles resulting from the decay of LLPs in the \mbox{MATHUSLA} detector decay volume with $\sim 1~\mathrm{ns}$ timing and $\sim$~cm transverse spatial resolution. 
The \layers in each tracking module on the ceiling or wall are also used as {\bf trigger layers} in addition to providing tracking information.
Having six \layers ensures that an ideal track has 3 hits with high spatial resolution along each coordinate. The number of layers was also optimized with full simulations to ensure the primary physics target LLP signals can be searched for with effectively zero background, even if a single tracking layer fails in parts or the entirety of the detector. 
All the \layers at a given height or horizintal distance from the rear wall constitute one \plane.

Each of the two layers of the \wallveto  is implemented by a slightly staggered arrangement of four $(11.2~\mathrm{m})^2$ {\wallvetolayers} to provide hermetic coverage for the full 40m width of the front wall. The bars in each layer are oriented perpendicular to each other to provide $\sim$ cm spatial resolution for particles that hit both layers.
The \floorveto is comprised of \floorvetolayers and \floorvetostrips. 
The \floorvetolayers are identical to \layers in the \trackingmodules, and are mounted at heights of 0.5m and 1m above the floor in the tower modules to cover the majority of the floor area.
\Floorvetostrips have physical dimensions of 9m $\times$ 2.8m and each provide about 9m $\times$ 2.3m of double-layer sensor coverage. They are are mounted horizontally above the \floorvetolayers to cover the gaps between \towers, and are also mounted vertically around the support columns to constitute \columndetectors that enclose the space at the corners of the \towers. 
In addition to making the floor detector hermetic with respect to cosmics, these \columndetectors also provide explicit material veto capabilities for inelastic cosmic ray interactions in the support column.

The dominant background source by rate is downward traveling charged CRs, which are easily rejected by timing from the well separated detector \layers. Other backgrounds can be rejected using geometric and timing cuts, as discussed in Section~\ref{s.detectoratcms}.
It is noteworthy that MATHUSLA's spatial resolution is not a significant limiting factor for physics reach. 
If we ignore deflection due to material interactions of charged particles with the scintillator, the maximum boost for an LLP decay that can be resolved at MATHUSLA is roughly $b^\mathrm{max} \sim 1000 \cdot (1\mathrm{cm})/\Delta x$, where $\Delta x$ is the smallest separation between two hits to resolve them separately~\cite{Curtin:2018mvb}. All physics targets involve LLPs with boosts below $\mathcal{O}(10)$. Therefore, as long as the tracker can distinguish hits that are few to 10~cm apart, the signal efficiency will not depend overly sensitively on the resolution.
Furthermore, non-negligible scattering of charged particles within the scintillator \layers means that spatial resolutions below the cm-scale are unlikely to yield significant improvements in the spatial uncertainty of reconstructed vertices.

The modular structure of the  detector makes a staged installation of the tracking modules possible. As each  \layer is completed, it will be installed in a \ceilingtrackingmodule in a \tower or, towards the end of the installation, the \walltrackingmodules. Once each \tower is complete, it will begin collecting CR and calibration data. 
All 16 \towers, 4 \walltrackingmodules and the entire \veto are foreseen to be installed in the 2.5~years prior to HL-LHC $pp$ collisions.

The proposed location of MATHUSLA near CMS is shown in Figure \ref{fig:layout_P5}. The enclosing building would include some assembly area adjacent to the detector, in addition to the $\sim$ 45m $\times$ 50m total physical footprint of the detector itself (including front veto wall, rear tracker modules, and access stairs on the side of the detector structure, see Section~\ref{s.safety}). 
The structure would be located on the surface near the CMS Interaction point (IP), fitting entirely on CERN-owned land and allowing MATHUSLA to be centered on the LHC beamline. The site allows for the detector to be as close as 68~m horizontally from the IP, which is marked with a yellow star in Figure \ref{fig:layout_P5}, underlying our benchmark assumption that the actual decay volume has a horizontal distance of 70~m from the IP.

\begin{figure}[]
\begin{center}
\includegraphics[height=3.3cm]{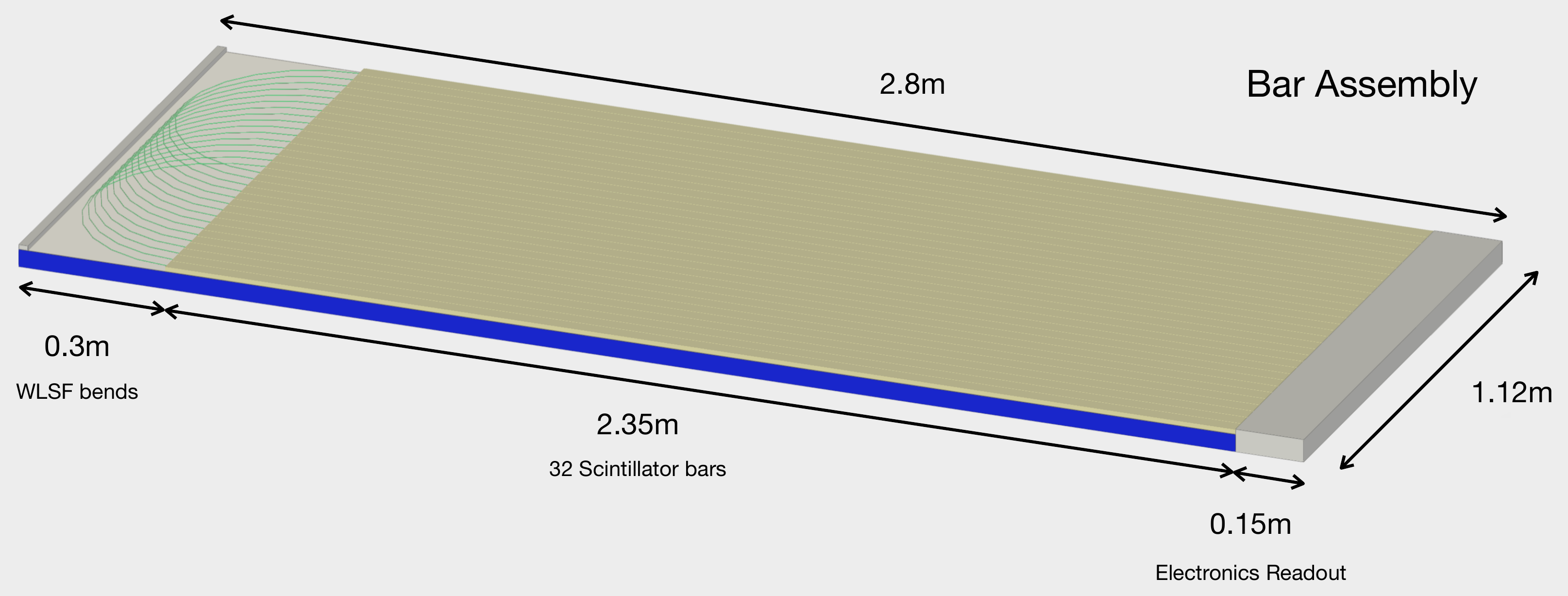}
\includegraphics[height=3.3cm]{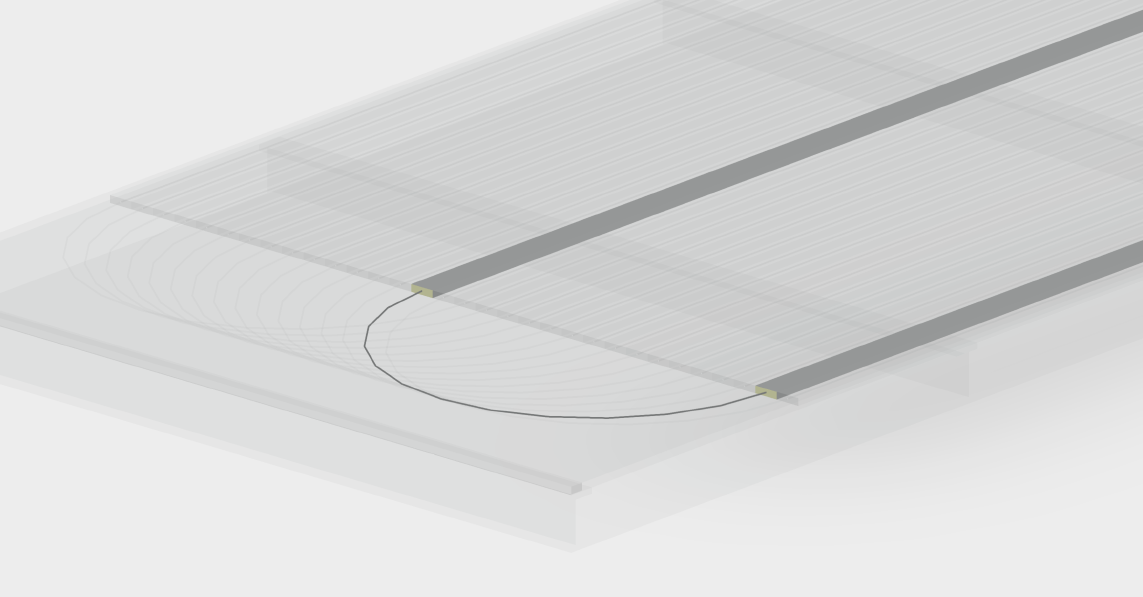}
\end{center}
\caption{Details of the \barass made of 32 scintillator \mbars. 
Left: overview of the \barass. The electronics readout box contains the Silicon Photomultipliers (\SiPMs) and electronics board. At the other end, the \WLSF "bends" as the fibers go down one bar and back through another bar. In this way all \SiPMs for the \barass are on the same end. Right: details of the \WLSF bend region.}
\label{fig:FNAL_extrusions}
\end{figure}

\begin{figure}
    \centering
    \begin{tabular}{m{0.5\textwidth}m{0.5\textwidth}}

    \begin{tabular}{c}
    \includegraphics[width=0.8\linewidth]{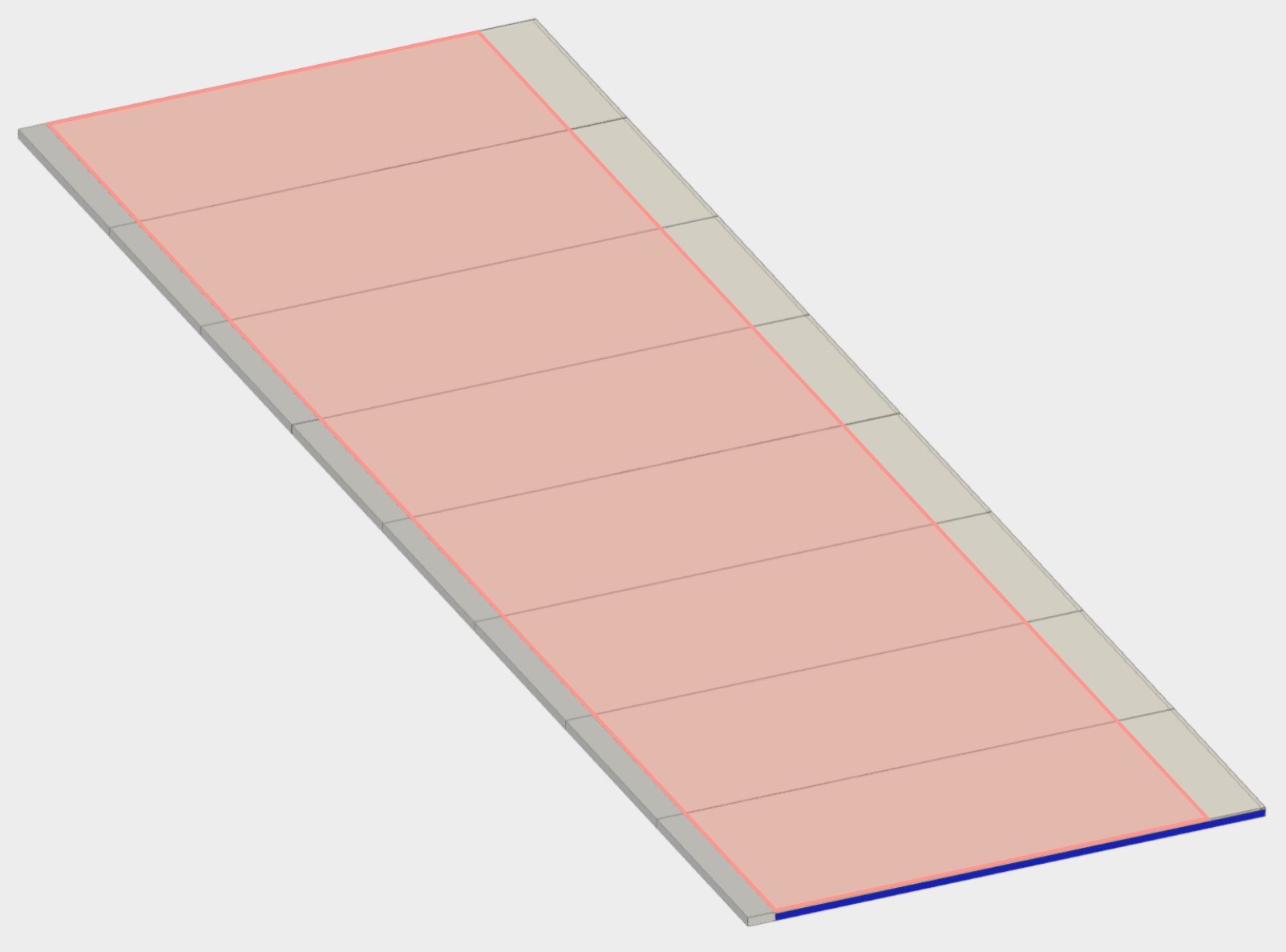}
    \\

    (a)

    \\
   \includegraphics[width=1\linewidth]{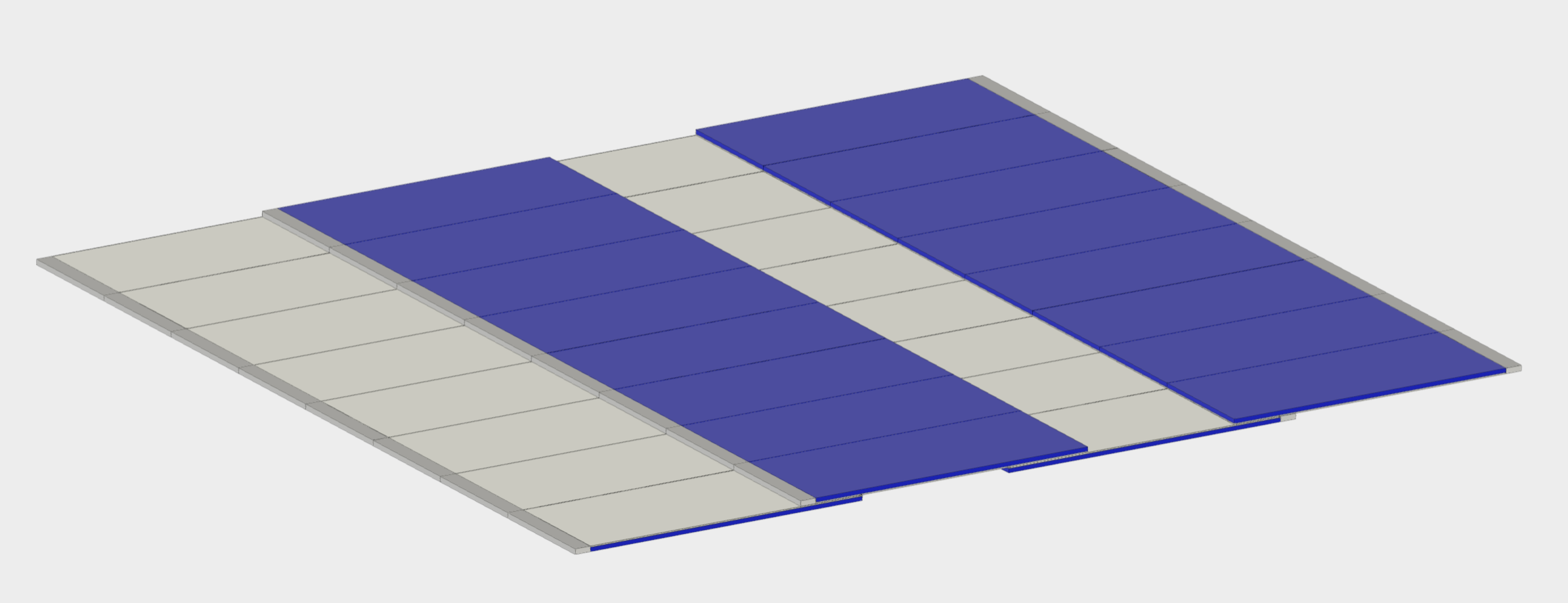}
    \\

    (b)

    \\

    \includegraphics[width=1\linewidth]{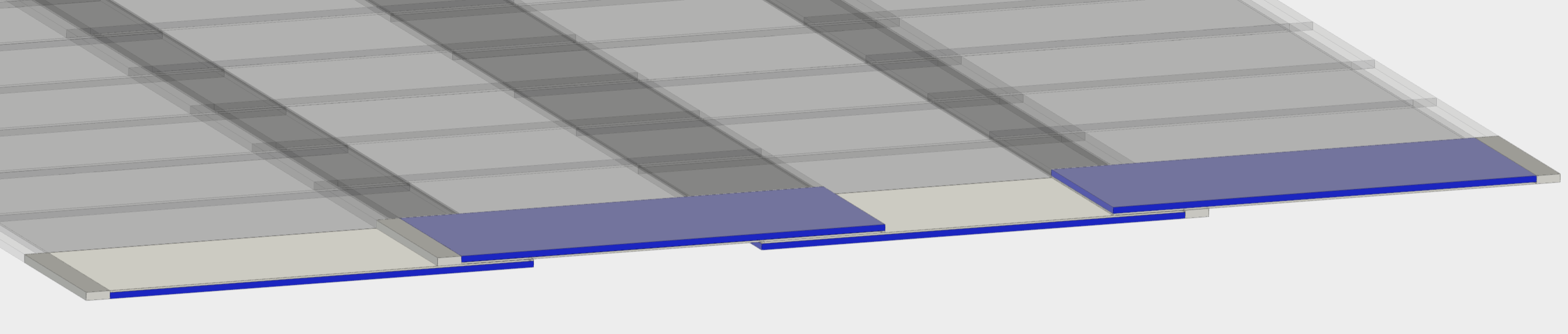}

    \\

    (c)

    %\vspace*{-5mm}
    \end{tabular}
    &
    
    \begin{tabular}{c}
    \includegraphics[width=1\linewidth]{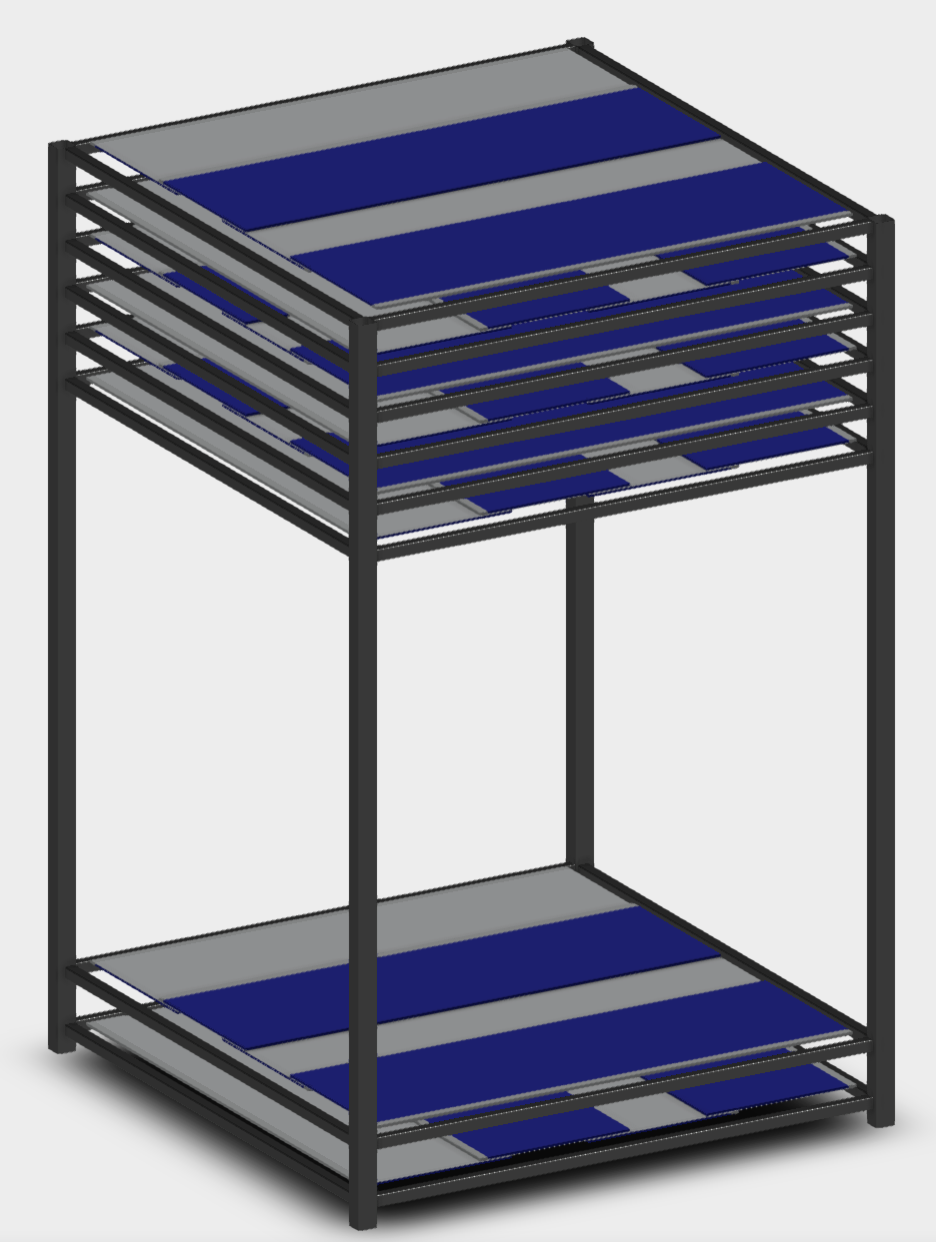}
    \\
    (d)
    \end{tabular}
    \end{tabular}
    \caption{Details of the \layers and their installation in \towers. (a) A single \sublayer of 8 \barasss joined side-by-side to make a sensor element with physical area of 2.8m $\times$ 8.96m, and sensor area (indicated with pink shading) of 2.3 $\times$ 8.96m. 
    (b) 4 \sublayers (blue and gray) are joined into a single \layer with sensor area $(9~\mathrm{m})^2$ and physical area of 9m $\times$ 9.3m.
    (c) Detail of the first \barass in each \sublayer 
    to show overlap along bar direction and alternating vertical offset.
    (d) \Tower with \ceilingtrackingmodule at top, comprising 6 \layers with 80cm spacing and alternating bar orientations, and two \floorvetolayers mounted at a height of 0.5m and 1m. 
    }
    \label{fig:modular}
\end{figure}

This benchmark layout informs R\&D and design efforts for every aspect of the MATHUSLA detector.
Details of the scintillator detectors and supporting electronics are discussed in Sections~\ref{s.detectordesign} and~\ref{s.electronicsdaqtrigger}. 
The modular nature is fundamental to the design of the trigger system, which drives data storage requirements (see Computing in Section~\ref{s.computing}) and 
also supplies a hardware trigger level signal to CMS to capture information on LLP production events.
The distinctive geometry of the MATHUSLA detector arrangement and its required large decay volume dictates specific Civil Engineering (Section~\ref{s.civeng}) and Safety (Section~\ref{s.safety}) requirements.

\subsection{Scintillating Detector Layers}
\label{Scintillator.plane}

The basic sensor building block of MATHUSLA is the \textbf{\barass}, shown in Figure~\ref{fig:FNAL_extrusions}. A single \barass comprises 32 scintillator \mbars of length \barlength, width \barwidth and thickness \barthickness, providing 2.35~m $\times$ 1.12~m area of sensor coverage with an approximate total physical size of 2.8~m $\times$ 1.12~m. 
Each \mbar is extruded with a hole at the center into which a $1.5$~mm diameter wavelength-shifting fiber (\WLSF) is inserted and connected at each end to a Silicon Photomultiplier (\SiPM).  The coordinate along the length of the bar is determined by the differential time measurement of the two ends of the bar, which has a resolution of $\sigma \sim \pm$ $15$~cm. The width of the bar determines the corresponding transverse coordinate with $\sigma \sim \pm 1$ cm. A $\sim 5$m long \WLSF is threaded through two nearby bars, with a $180^\circ$ bend at one end, so that \SiPM signals can be recorded only at one end of the \mbars
Each \barass requires 32  \SiPMs, one for each bar. The electronics readout box at one end of the bar assembly can be mounted flush with either the top or bottom of the bar assembly (from the perspective of Figure~\ref{fig:FNAL_extrusions}), allowing \barasss to be joined with minimal vertical gap in various configurations. 
The bars and electronics are attached to an aluminum strongback plate, resulting in a total thickness of $\sim$ 5cm and making each \barass self-supporting if mounted at the edges.

In Figure~\ref{fig:modular} we illustrate how \barasss are joined into larger sensor elements. 
Eight \barasss can be joined side-by-side to make \textbf{\sublayers} of width $\sim 896$~cm (Figure~\ref{fig:modular} (a)). Four \sublayers can then be joined end-to-end with small vertical off-sets to make a single contiguous \textbf{\layer}, providing hermetic surface-coverage over an area of $\sim9 \,\textrm{m} \times 9 \textrm{m}$ (Figure~\ref{fig:modular} (b), (c)).
Six \layers are then stacked into the ceiling tracking module with alternating bar orientations and 80cm separation between neighboring \layers, while \floorvetolayers are just \layers mounted near the floor in \towers. (Figure~\ref{fig:modular} (d)).
The wall tracker modules are identical, with structural adjustments to allow for vertical mounting.

The same principle is employed to assemble the 8 \wallvetolayers, except that \wallvetosublayer are made of 10 \barass joined end-to-end, and 5 of those make one \wallvetolayer. 
Similarly, the 
\floorvetostrips, which are mounted horizontally to cover gaps, or vertically to constitute \columndetectors, are each comprised of two tightly sandwiched sensor layers, each of which is simply $1/4$ of a \layer (2 \barasss wide and 4 long), resulting in a sensor area of 9m $\times$ 2.24m.

\newpage
%%%%%%%%%%%%%%%%%

%%%%%%%%%%%%%%%%%
\section{LLP Searches with the MATHUSLA Detector at LHC Point 5}
\label{s.detectoratcms}
%%%%%%%%%%%%%%%%%

We now discuss, in detail, how the MATHUSLA detector at the LHC Point 5 would search for and discover LLPs produced at the HL-LHC.

In Section~\ref{s.sboverview}, we provide an overview of LLP decay signals that MATHUSULA will reconstruct as displaced vertices and discuss various HL-LHC and cosmic backgrounds. In Section~\ref{s.geant} we describe a \textsc{Geant4}~\cite{GEANT4:2002zbu} simulation of the full MATHUSLA detector, a pipeline for hit digitization and track/vertex reconstruction, and simulations of LLP signals and the various backgrounds that are fed into the GEANT geometry.
We can then simulate a search for LLP decays representative of MATHUSLA's main physics target, with full backgrounds and realistic reconstruction and background veto strategies, which is summarized in 
Section~\ref{s.LLPsearch}. This demonstrates that the main physics target can be searched for with near-zero background and a signal efficiency of $\sim$ 50\% for LLP decays in the MATHUSLA detector volume.

Having confirmed MATHUSLA's ability to conduct background-free LLP searches with good efficiency for our physics target LLP signals, we then estimate its sensitivity across the parameter space of key LLP benchmark scenarios in Section~\ref{s.signalacceptance}. This demonstrates MATHUSLA's powerful new physics reach which robustly complements and greatly extends the discovery capability of the LHC main detectors and other proposed LLP experiments.

In all of the below, we assume that the $3000 \mathrm{fb}^{-1}$ of HL-LHC luminosity is accumulated in 5 years of total beam-on-time, over a real-time period of at least 10 years~\cite{hllhclumi}.
This means that for cosmic backgrounds, 5 years of data can be gathered with the beam off, which will greatly aid in characterizing these backgrounds, which are challenging to simulate, and verifying LLP search strategies to ensure they are in fact background-free. \textbf{This ability of MATHUSLA to conduct detailed studies of its dominant background without any possible signal contamination is a key feature in the robustness of the experiment, allowing it to claim a BSM LLP discovery with confidence in the event that anomalous displaced vertices are observed when the HL-LHC is running.} 

%%%%%%%%%%%%%%%%%
\subsection{Overview of Signal and Background}
\label{s.sboverview}
%%%%%%%%%%%%%%%%%

MATHUSLA is an air-filled detector at the surface above CMS. It is therefore a very radiation-quiet environment compared to the (HL-)LHC interaction points, which is what makes it such an ideal place to look for conspicuous decays of upwards-traveling LLPs,
However, several important backgrounds to LLP searches have to be accounted for, and they are generated by two sources of particle flux incident on MATHUSLA: cosmic rays (CRs), which are mostly muons but proton and neutron contributions are especially important; and muons from the LHC that have enough energy to penetrate the $\sim$ 100m of shielding separating MATHUSLA from the CMS chamber.

%%%%%%%%%%%%%%%%%
\subsubsection{LLP Signal} 
\label{s.dunno}
%%%%%%%%%%%%%%%%%

The signal from LLP decays is defined as a displaced vertex of at least two tracks, which are reconstructed using hits in the ceiling and back wall tracking layers and are required to travel upward or outward from the decay volume. By default, we also require that the decay vertex is situated strictly inside the decay volume of the detector. 
The main physics target signal,  $\mathcal{O}(10 - 100~\mathrm{GeV})$ LLPs  that decay hadronically, produces $\mathcal{O}(10)$ charged particles per decay, meaning the vertex can easily have similarly many tracks. This turns out to be crucial for rejecting cosmic ray backgrounds, but many vetos are needed in combination to eliminate all backgrounds (see Section~\ref{s.signalcriterionbackgroundvetostrategies} for a full list).
%

%%%%%%%%%%%%%%%%%
\subsubsection{Cosmic Ray Inelastic Backscatter}
\label{s.LPLPBGcosmicbackscatter}
%%%%%%%%%%%%%%%%%

The $\sim 5 \times 10^{13}$ CR particles incident on the floor, support structure, scintillator material, and the air in the detector during the HL-LHC run can undergo a variety of inelastic processes that result in the generation of a displaced vertex in the MATHUSLA decay volume.
Generation of cosmic ray DVs is entirely dominated by incident CR protons and neutrons, despite them only making up about one percent of the total incident CR flux.  
Much of this background can be  vetoed by requiring more than two tracks per vertex for primary physics target searches, but many other vetos have to be utilized in combination, including rejection of vertices associated with detector material, large numbers of hits before the vertex in time, vertices with incompatible orientation with respect to the IP, vertices featuring slow tracks, and vertices occurring together with additional hits or tracks from CR interactions (see Section~\ref{s.signalcriterionbackgroundvetostrategies}).
These processes are challenging to simulate, since only a tiny fraction of the very high CR rate gives rise to an upwards-traveling vertex.

In the following sections, we simulate this background by generating CR protons and neutrons incident on the detector using the \texttt{PARMA} \cite{sato2008development} 
and \texttt{CRY} \cite{CRY} generators.
We find no overwhelming preference for a specific part of the incoming proton distribution to source these backgrounds (high or low energy or incidence angle), necessitating large unweighted event samples. Due to computational limitations, the generated samples amount to about $1/30$ of the real incident rate. 
Our vetos eliminate all simulated background events.
To argue that this extends to MATHUSLA's full exposure, we  extrapolate the effect of all our cuts in combination to  provide convincing evidence that the overlapping nature of our different vetoes reduce this background to be $\ll 1$ event over the entire MATHUSLA runtime for our primary physics target LLP searches.

%%%%%%%%%%%%%%%%%
\subsubsection{Interactions and Decays of Muons from the HL-LHC} 
\label{s.LLPBGmuons}
%%%%%%%%%%%%%%%%%

Muons from the HL-LHC, produced primarily in electroweak processes and $b \bar b$ production, can penetrate into the MATHUSLA detector volume if they have sufficient momentum ($p_T \gtrsim 40 \gev$), and can then
 generate backgrounds to LLP searches through interactions with the detector material.
 The resulting secondary particles (mostly delta rays, i.e. knock-off electrons) can be separately reconstructed as a DV at the interaction location. This will also include rare contributions from muon trident processes.
 
Our \textsc{Geant4} simulations described below indicate that over the entire HL-LHC run, roughly $\sim 10^8$ muons from HL-LHC will leave at least one hit in the MATHUSLA detector, corresponding to a rate of about 1 Hz. Only about $0.5\%$ of those give rise to a reconstructible vertex. These background vertices can be efficiently rejected by the same strategies that veto CR backgrounds in our primary physics case LLP search. 
The CMS muon system L1 trigger could also be used as an additional veto for off-line rejection to reduce the size of this background by an additional order of magnitude. 
In our studies, we assume utilization of this CMS veto to reduce our muon simulation burden by a factor of 10, but LHC muons are so subdominant as a background compared to the (ultimately completely-rejected) CR neutron and proton backgrounds that a final realistic analysis is unlikely to rely on the CMS muon veto to achieve zero background.

A rare background process that is not contained in standard \textsc{Geant4} simulations is the rare 5-body muon decays ($\mu^-\rightarrow {\rm e}^- {\rm e}^+ {\rm e}^- \overline{\nu}_\mu \nu_{\rm e}$, with branching ratio $(3.4\pm0.5)\times 10^{-5}$~\cite{Tanabashi:2018oca}) that could be reconstructed as a DV.
This process was investigated by computing the 5-body decay kinematics of a muon at rest with Madgraph5~\cite{Alwall:2011uj}, and over-writing the relevant \textsc{Geant4} module to force all muons to decay accordingly. 
The spectrum of LHC muons
in MATHUSLA is steeply falling with energy, and the muons that actually decay are additionally weighted towards those with low momentum.
Preliminary simulations therefore predict $< 100$
five-body muon decays anywhere within the detectable volume of the experiment over the entire HL-LHC run; the rate at which these decays give rise to reconstructible tracks and vertices is low, since the electron final states rarely have momenta in excess of 50 MeV.
Muon decays with a reconstructible vertex will be from muons with such a high  momentum that the cone of the vertex points back to the original muon hit in the floor detector. Those few decays are therefore efficiently rejected by our veto strategies.

%%%%%%%%%%%%%%%%%
\subsubsection{Neutrino Scattering}
\label{s.LPLPBGneutrino}
%%%%%%%%%%%%%%%%

Neutrinos produced in atmospheric CR interactions are incident on the MATHUSLA decay volume, where they can inelastically scatter off nuclei in the air or detector material to generate a DV with two (or in rare cases more) tracks. 
This process was studied by 
generating CR neutrinos incident on air, steel, earth and detector material targets using GENIE~\cite{Andreopoulos:2009rq} and the FLUKA 3-D~\cite{battistoni2003fluka} atmospheric neutrino flux model. The final states of these interactions can then be inserted into the \texttt{GEANT4} simulation of the MATHUSLA detector.
The number of expected neutrino interactions resulting in the emission of charged particles can be estimated to be 
$< 6$ per year within the entire detector volume, resulting in $< 30$ events recorded throughout the HL-LHC run.
These few events can be effectively vetoed with the above-mentioned veto strategies.
Neutrinos from HL-LHC collisions contribute an order of magnitude fewer scatterings, which can be vetoed by similar means.

%%%%%%%%%%%%%%%%%%%%%%%%
%%%%%%%%%%%%%%%%%%%%%%%%
%%%%%%%%%%%%%%%%%%%%%%%%
%%%%%%%%%%%%%%%%%%%%%%%%
%%%%%%%%%%%%%%%%%%%%%%%%

%%%%%%%%%%%%%%%%%%%%%%%%
\subsection{GEANT Simulations} \label{s.geant}

We now outline the setup of our \textsc{GEANT4} simulations of MATHUSLA, as a slightly simplified implementation of the benchmark design shown in Figure~\ref{fig:MATHUSLAstructure}.
We then explain how an event is defined within the simulations, and how electronic noise, ambient CR hits, and per-hit-efficiencies are included.  Relevant details for the simulation of LLP signals and various backgrounds are also given.
This allows us to realistically reconstruct  tracks of simulated particles leaving some number of hits in the detector, and vertices comprising at least two tracks. With this setup, we can simulate a complete analysis for benchmark LLP searches in Section~\ref{s.LLPsearch}.

\subsubsection{Detector Geometry}

\begin{figure}
    \centering
    \includegraphics[width=0.8\columnwidth]{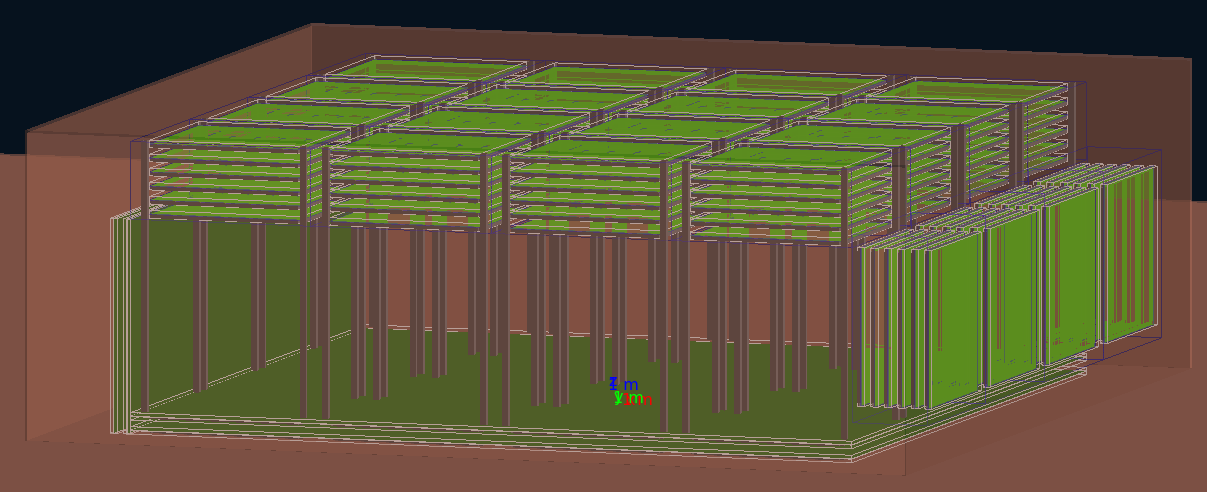}
    \caption{
    MATHUSLA geometry implementation in \textsc{GEANT}.
    }
    \label{f.geantgeometry}
\end{figure}

The \textsc{GEANT} implementation of the MATHUSLA detector geometry is shown in Figure~\ref{f.geantgeometry}. Six ceiling tracking layers are set 80~cm apart and start at a height of 11.8 m above a concrete floor. Six other wall tracking layers are placed at the back of the detector at a height of 1.85~m above the floor and extending 9 m upwards. The floor begins at ground level, 85.47 m above the CMS IP, and has a span of 42.8 m x 42.8 m. Both the floor and ceiling tracking layers begin 70 m along the beam axis away from the CMS IP, extending to 112.8~m away. The 1~m physical gap between \towers, and the thickness of the required vertical structural supports, results in 1.7~m gaps between the sensor areas of adjacent $(9\mathrm{m})^2$ horizontal tracking modules in the ceiling and vertical tracking modules on the back wall. Each tracking layer is made of scintillator bars with dimensions 1 cm $\times$ 4.5 cm $\times$ 450 cm arranged in alternating orientations (transverse to or along the beam axis) in each layer, with the lowest layer oriented transverse to the beam axis.\footnote{The different length of our simulated bars compared to the 2.3~m bars we eventually adopted in the \barasss of our physical design has no significant impact on our results.} Each layer also includes 0.3 cm of aluminum above and below each layer to represent the bar casing and mimic the material budget of the aluminum honeycomb support structure of the individual bar assemblies and \layers.
The front wall and floor are covered by two fully hermetic tracking layers with alternating bar orientation.  The first hermetic floor layer is placed on top of the iron floor, the second 80~cm above. The hermetic wall closest to the detector is constructed 1 cm away along the beam axis, while the second is, in turn, 0.8~m away from the first wall. Both walls have a height of 11.8~m and begin at ground level. 
The geometry includes vertical and horizontal steel support columns with realistic dimensions based on the detector design in Fig.~\ref{fig:MATHUSLAstructure}. The vertical columns are assumed to have a 14 inches square cross section with a thickness of 5/8 inches. The horizontal beams are assumed to have a 8 inches square cross section with a thickness of 1/2 inches. 
To include material effects from support columns without having to specify how the floor detector is made hermetic, the support columns `float' above the gapless floor sensor layer in this simplified geometry.

For the purpose of simulating HL-LHC backgrounds, the corresponding particles are placed at the HL-LHC collision point and propagated towards MATHUSLA through the CMS cavern and the separating rock. 
The CMS experiment is simulated as a cylindrical cavern 53m in length. The outer part of the cavern is a 1.28 m thick concrete shell, with an outer radius of 14.53m. An iron cylindrical ring of 20m length, 1.4 m thickness, and an outer radius of 8m is placed centered within the concrete shell to simulate the material budget of the CMS detector itself. Between the iron ring and the concrete shell, the cavern is filled with air. The access shaft is also included and is modeled as a hollow, vertical rectangle extending from the CMS cavern to ground level. It has a square cross section with an outer dimension of 20.5 m. It has an outer shell 1.5m thick made of concrete, with the rest filled with air. 
The separating rock is modeled using three layers. The first is sandstone situated at ground level and extending to a depth of 45.3m. This is followed by a layer of marl, extending to a depth of 63.55m. The lowest layer is a mixture of 50\% quartz and 50\% marl, which reaches a final depth of 100m in the simulation. 
Above the surface, air at ambient pressure and composition is included.

Our simulated geometry represents a slight simplification compared to the full detector design presented in Section~\ref{s.detectorintro}, but this should not significantly affect our conclusions. 
The staggered layering of vertical layers in the front wall to achieve hermetic coverage, and the more complicated geometrical arrangement of sensors in the floor and support columns to achieve full coverage in the presence of vertical structural supports, is not implemented. 
Note that the absence of column detectors wrapping the vertical supports in our simulation makes our assumed material veto capabilities pessimistic compared to the realistic geometry, where the column detectors will likely significantly aid in such a material rejection veto. 
The detector floor is also placed at surface level, rather than a few meters below grade. This has a very minimal effect on signal and background rates, and is done both for simplicity and to not make our simulations unduly dependent on fine details of the geometry that will likely change as the design is refined towards the final proposal through successive engineering studies. 

We also ensure that our results are robust by exploring 
less well-instrumented variations of our \textsc{GEANT} detector, which  can be well approximated by ignoring some hits in a given analysis, either randomly (to mimic realistic per-hit detection efficiencies or small gaps in the floor detector), or in certain layers (to simulate 5 or 4 tracking layers).
This gives us further confidence that our conclusions are indicative of the performance of a realistic MATHUSLA detector.

\subsubsection{Digitization and Event Definition}

Individual scintillator bars in the bar assemblies are assumed to give hit information with a spatial resolution of 2.25 cm (13 cm) transverse and longitudinal to the bar orientation, with a time resolution of 1 ns. Longitudinal resolution is implemented via Gaussian smearing, while transverse resolution is realized by forcing the coordinate to take the center coordinate of the closest bar. To mimic the light yield requirements of a realistic detector, a bar is assumed to detect a single digitized hit if the energy deposits of all hits within a single bar and within 20~ns exceed 600~keV. 
In the case of multiple hits, the longitudinal spatial coordinates and time coordinates of the hits are averaged, and Gaussian smearing which includes both the above native resolution and the spread of the different hits is applied to the final hit read-out coordinates. 
We consider various efficiencies to detect each hit: the default is 0.95, but we also run studies with efficiencies 0.9 and 1.0.
To simulate versions of this detector with fewer tracking layers, hits in one or two of the 6-layer tracking module are ignored for a given analysis. 
Ignoring hits in certain areas of the floor/wall veto layers mimics gaps in veto coverage.
We find that these variations do not significantly change our results. 

Cosmic rays are incident on the detector with a rate of $\sim$~300~kHz in each horizontal sensor layer. This is included in all studies by generating CR flux with realistic composition, angular distribution and momentum spectrum with the \texttt{CRY} generator and injecting it into simulations of LHC signals. 
We also include random electronic noise from SiPM dark count and other sources as random hits, with the default noise assumption being equal in overall rate across the detector to the CR rate, while more optimistic (0.1$\times$ cosmic) and pessimistic (10$\times$ cosmic) assumptions are also considered, again without significantly impacting results.

Unlike for the LHC detectors, where beam collisions dictate the time structure of each event, MATHUSLA is continuously monitoring and reading out hits in the entire detector. In practice, as described in Sec.~\ref{s.DAQ}, a simple upwards-traveling track trigger (as well as other possible triggers) will mark `interesting events' in the datastream, and all hits in the entire detector recorded within 600~ns of the trigger timestamps will be grouped into a single `event' and saved for later offline analysis. 
In our \texttt{GEANT} simulations of LHC signals or backgrounds, we add $\pm1000$~ns of random CR and electronic noise activity before and after the first hit in the detector, then include all hits within 600~ns of the first hit in the saved `event', which is then submitted for track and vertex reconstruction and further analysis.\footnote{The reason for simulating $\pm1000$~ns of CRs hitting the detector, despite the 600~ns timing window for an event, is to ensure that various 'partial tracks' from CRs that are in-transit across the detector are correctly included.} 
This results, on average, in the addition to the event of hits corresponding to 0.86 full CR tracks, plus a similar (1$\times$) level of electronic noise. 
In our simulations of pure CR backgrounds (inelastic processes of CR protons, and neutrino scatters), we include the electronic noise within the time window but not additional unrelated additional CR tracks in the time window. This has very little effect, apart from possibly slightly overestimating the reconstruction efficiency for displaced vertices resulting from these cosmic backgrounds.

\subsubsection{LLP Signal Simulation}
\label{s.signalsim}

MATHUSLA's main physics target is well-represented by the simplified model of exotic Higgs boson decays to two bosonic LLPs, which in turn decay hadronically.
This arises in many BSM scenarios, including many Hidden Valleys, the SM+S scenario below, but also in more complete BSM theories like the Fraternal Twin Higgs~\cite{Chacko:2005pe, Craig:2015pha, Curtin:2015fna} that solve the little Hierarchy Problem. 
The model has three parameters: the LLP mass, lifetime, and production rate in Higgs boson decays. The hadronic LLP decay is assumed to be to $\bar b b$, a similarly motivated decay mode to $gg$ gives extremely similar results (with very slightly higher signal acceptance due to the higher final-state multiplicity).

We use the pre-simulated LLP decay libraries produced by~\cite{Curtin:2023skh}, and insert LLP decay final states into the \texttt{GEANT} simulation of the MATHUSLA detector with a spatial distribution consistent with the assumed source process and LLP lifetime. Unless otherwise stated, we work in the long-lifetime limit $c \tau \gg 1000~\mathrm{m}$. 
These samples were generated in
MadGraph5 3.4.2~\cite{Alwall:2014hca} and showered in Pythia~8~~\cite{Sjostrand:2006za,Sjostrand:2007gs} to simulate SM Higgs production in gluon fusion and vector boson fusion, with the former implemented with an effective $ggh$ operator in the model. 
Jet matching with up to one extra hard jet and event reweighing to reproduce the  NLO+NNLL Higgs $p_T$ spectrum computed by HqT 2.0~ \cite{Bozzi:2005wk,deFlorian:2011xf} ensures the LLP momentum spectrum is accurate at the same order. The total Higgs production cross sections in gluon fusion (vector boson fusion) are taken to be 54.6 pb  (4.27 pb)~\cite{LHCHiggsCrossSectionWorkingGroup:2016ypw}, 54.6 pb for gluon fusion and 4.27 pb.
LLP decays are again showered in Pythia~8 to produce the final states that are inserted into the MATHUSLA detector simulation.

\subsubsection{Background Simulation}
\label{s.backgroundsim}

\vspace*{2mm}
\noindent 
\textbf{\emph{Cosmic Ray Protons:
}} \vspace{2mm}

\noindent The \texttt{PARMA} \cite{sato2008development} generator is used to simulate $1.04\times10^{11}$ protons incident on a 60~m $\times$ 60~m plane above MATHUSLA, representing about $1/32$ of the total rate over 5 years of HL-LHC beam-on time. The total flux is calculated at altitude of 511~m, which is roughly the altitude of Geneva. Simulating the full exposure is computationally very intensive and left for a future study, but various extrapolations allow us to utilize the current sample to make statements about the CR proton background at MATHUSLA with a high degree of confidence.

\vspace*{2mm}
\noindent 
\textbf{\emph{Cosmic Ray Neutrons:
}} \vspace{2mm}

\noindent The \texttt{CRY} \cite{CRY} generator is used to simulate $8\times10^{11}$ neutrons incident on a 60~m $\times$ 60~m plane above MATHUSLA, representing about $1/27$ of the total rate over 5 years of HL-LHC beam-on time. For the same reason, the simulation of the full exposure is left for a future study.

\vspace*{2mm}
\noindent 
\textbf{\emph{LHC Muons:
}} \vspace{2mm}

\begin{figure}
    \centering
    \includegraphics[width=0.45\linewidth]{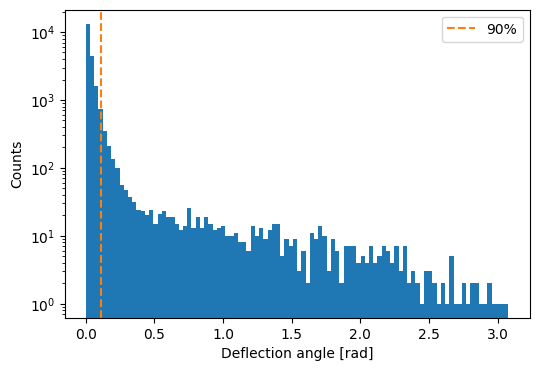}
    \includegraphics[width=0.45\linewidth]{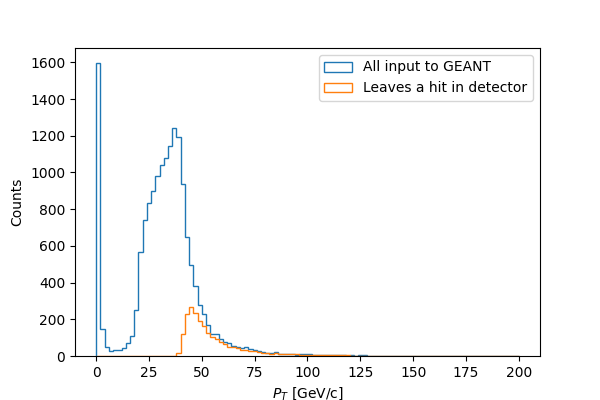}
    \caption{Muons from LHC. Left: LHC muon deflection angle for muons that leaves at least one hit in MATHUSLA; Right: pT distribution of simulated LHC muons, and the ones that leaves at least one hit in MATHUSLA.}
    \label{fig:ch2:lhc-muon-deflection}
\end{figure}

\noindent We first focus on electroweak muon production via $s$-channel on- and off-shell $\gamma^{(*)}, Z^{(*)}, W^{(*)}$ at the HL-LHC.  
The overall cross section for muons with $p_T > 10~$GeV and $|\eta| < 2.5$ is found to be 17.3 nb at lowest order and parton-level. This is multiplied by a $K$-factor of 1.3~\cite{ATLAS:2016fij} to include NLO effects, and used to normalize subsequent simulations with the same parton-level generator cuts. 
We then generate $6.1\times10^{9}$ muon production events in Madgraph~5 with showering in Pythia~8, and matching of up to one hard jet to accurately model the muon momentum spectrum. 
This represents about $1/10$ the total muon rate at MATHUSLA over the HL-LHC run; alternatively, this can be seen as the full rate of muons that reaches MATHUSLA after vetoing $90\%$ of muons that fly towards MATHUSLA using the CMS L1 Muon Trigger. 
Muons are then propagated in~\texttt{GEANT} towards MATHUSLA if their initial direction lies in the same hemisphere as the detector.
We find that only muons with $p_T > 35$~GeV  have any appreciable chance of reaching the detector, with the penetration probability approaching unity for muons pointed at the detector with $p_T \gtrsim 40$~GeV, see Fig.~\ref{fig:ch2:lhc-muon-deflection}.
The proximity of these thresholds to the kinematic edge near $m_W/2$ necessitates our careful simulation of the muon momentum spectrum.
We found that angular deflection  in the rock can be significant: muons that leave at least one hit in MATHUSLA undergo a deflection of, on average, 4.3~degrees compared to their initial direction at the LHC IP. This information informs some of the veto strategies we devise for rejecting LHC muons below.

Muons from $\bar b b$ are simulated in Madgraph 5 + Pythia 8, with their total production rate normalized to the FONLL HL-LHC prediction of $\sigma_{\bar b b [p_T > 20~\mathrm{GeV}, |\eta| < 3]} = 13300~\mathrm{nb}$. 
By propagating these muons through the \texttt{GEANT} simulation in an identical manner as above, we find that the number of muons from $\bar b b$ production that leave a hit in MATHUSLA is about 0.75 $\times$ the corresponding rate from electroweak production. Furthermore, the muons from $\bar b b$ that leave a hit in MATHUSLA are overall much softer and less likely to generate a vertex than the muons from electroweak processes.
Therefore, the impact of LHC muons from $\bar b b$ production on our backgrounds can be conservatively over-estimated by simply multiplying the electroweak LHC muon background  by a factor of 1.75, which we do in our analyses below.

\vspace*{2mm}
\noindent 
\textbf{\emph{Atmospheric Neutrinos:
}} \vspace{2mm}

\noindent The \texttt{GENIE} \cite{Andreopoulos:2009rq, Andreopoulos:2015wxa} generator is used to simulate 10 years of atmospheric neutrino exposure scattering off the material modeled in the \texttt{GEANT} simulation, including air, steel, earth and detector material. Neutrinos are generated based on FLUKA 3-D~\cite{battistoni2003fluka} atmospheric neutrino flux model. The number of neutrino interactions resulting in the emission of charged particles is roughly 31 within the detector's decay volume, and 1248 including a 1 m region outside the detector. Over 90\% of interactions outside the detector are from the earth material. All 1248 interactions are further simulated in the full \texttt{GEANT} simulation.

\subsubsection{Track and Vertex Reconstruction} \label{s.trackreconstruction}

Tracks are identified and reconstructed using a Kalman Filter \cite{fruhwirth1987application} based algorithm. Track identification starts by forming track seeds consists of a doublet of hits from two different layers with a  selection criterion of space-time interval between the two hits. For each seed, Kalman Filter is applied twice. The first round of Kalman filter is for finding the consistent hits and remove outliers. Search is performed layer by layer using the progressive nature of Kalman Filter to find hits that are consistent with the track seed. A minimum of 4 hits is required to keep the track after all layers are traversed. The second round of Kalman filter extracts the track parameters from the found hits. Uncertainties from multiple scattering \cite{lynch1991approximations,wolin1993covariance} is considered in the Kalman Filter, assuming deflection in the tracking layers corresponding to a 0.5~GeV muon.

\begin{figure}[htpb]
    \centering
    \includegraphics[width=0.48\linewidth]{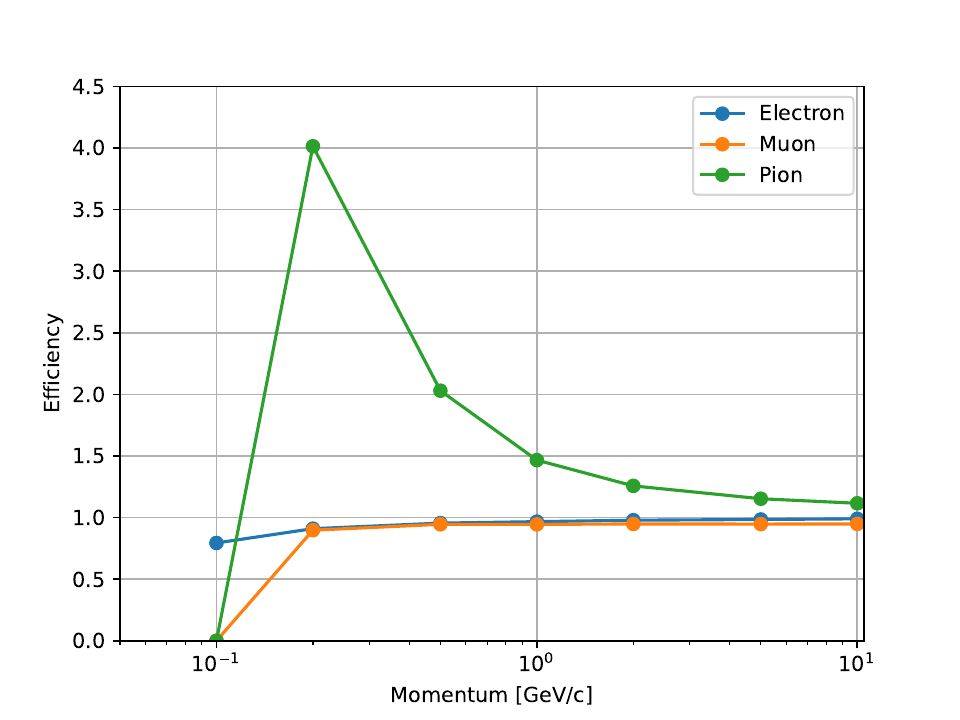}
    \includegraphics[width=0.48\linewidth]{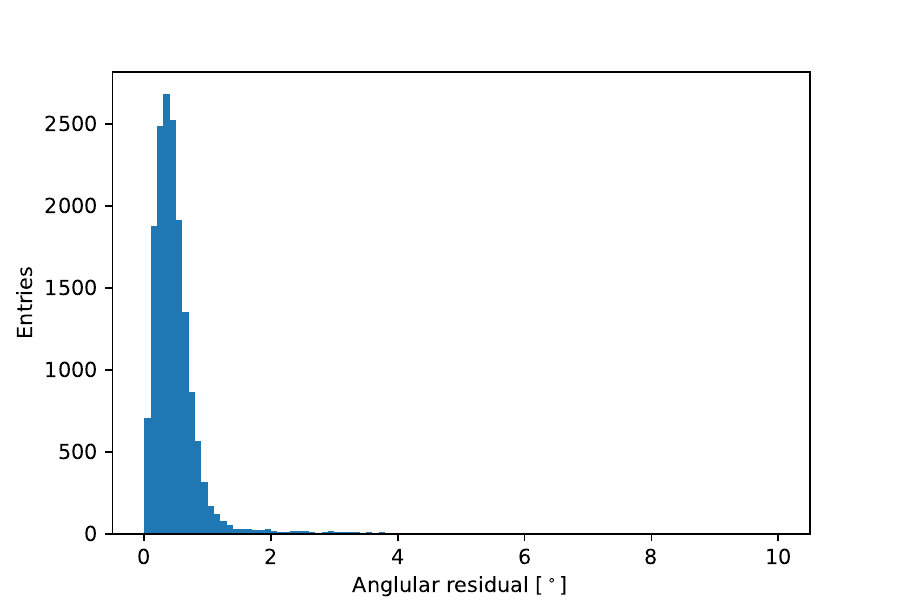}    
    \caption{Left: track reconstruction efficiency for electrons, muons and pions. Note that the greater-than-unity efficiency for pions is an artifact of defining pions  that decay in flight as `unreconstructible', but secondaries usually allow for the pion to be nevertheless reconstructed. %\DC{wordsmith, or add explanatory plots} 
    Right: distribution of the angular difference between the reconstructed and truth track for 10 GeV/$c$ muons.}
    \label{fig:ch2:single_track_eff}
\end{figure}

Figure \ref{fig:ch2:single_track_eff} shows the track reconstruction efficiency for muons, pions and electrons injected into the detector with with certain momenta and an uniform angular distribution. The track reconstruction efficiency is $>  90\%$ for charged particles with momentum above 0.5 GeV/$c$ that leave at least 4 hits in the detector.
For reconstructed tracks, the truth-level particle trajectory is generally reconstructed within 1 degree.

Vertex reconstruction also starts from vertex seeds made with pairs of reconstructed tracks. For each vertex seed, the remaining tracks are added to the vertex one by one until the increment of $\chi^2$ exceeds a set threshold. Then the vertices are constructed from tracks associated with each seed by finding the best 4D point with a least-squares fit. To assess the vertex reconstruction efficiency and precision of the reconstructed vertex position, we compared reconstructed vertices in all physics target LLP decay simulations to the truth-level LLP decay positions. 

\begin{figure}[htpb]
    \centering
    \includegraphics[width=0.5\linewidth]{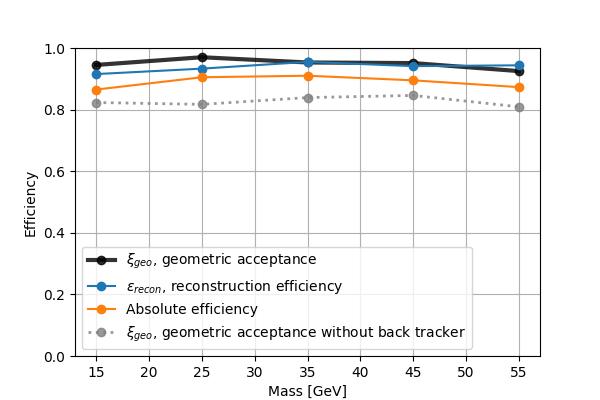}
    \includegraphics[width=0.4\linewidth]{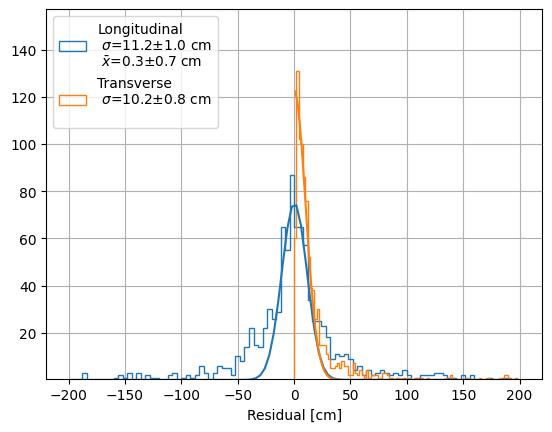}    
    \caption{Left: geometric acceptance, reconstruction efficiency and absolute efficiency for primary physics target. The dotted line shows the geometric acceptance if the back wall of MATHUSLA is not instrumented with tracking layers. Right: vertex position resolution for 25~GeV LLP.}
    \label{fig:ch2:hxx_efficiency}
\end{figure}

We define the \emph{geometrical acceptance} of the detector to be the fraction of LLP decays in the decay volume for which at least two LLP decay final states leave at least 4 hits each. 
For the primary physics target simulations of LLPs arising from exotic Higgs decays, the geometrical acceptance as a function of LLP mass is shown in Fig.~\ref{fig:ch2:hxx_efficiency} in the long-lifetime limit ($c\tau\sim1000$~m). The geometrical acceptance is near-unity, thanks to the tracking modules in the back wall of MATHUSLA. Without these rear trackers, about 15\% - 20\% of signal would be lost. 
For LLP decays that pass this geometric acceptance criterion, the vertex reconstruction efficiency is also shown in Fig.~\ref{fig:ch2:hxx_efficiency}. As expected, this is near unity for the high-multiplicity LLP decays of the primary physics target. The vertex position is found to be reconstructed with a precision of 10~cm (11~cm) transverse to (longitudinal to) the direction of LLP propagation, as shown in the right panel of Fig.~\ref{fig:ch2:hxx_efficiency}.

These reconstruction studies demonstrate that this MATHUSLA design has good acceptance and reconstruction precision for our targeted physics signals.

%%%%%%%%%%%%%%%%%%%%%%%%%%%%%%%%%
%%%%%%%%%%%%%%%%%%%%%%%%%%%%%%%%%
%%%%%%%%%%%%%%%%%%%%%%%%%%%%%%%%%
\subsection{Simulated Analyses for Benchmark LLP searches}
\label{s.LLPsearch}

We now use our signal and background samples to perform full simulated analyses corresponding to a realistic  search for our main LLP physics target signal in MATHUSLA. 
This demonstrates that a background-free search can be performed with an overall signal acceptance $\times$ efficiency of $\sim 0.5$. 
We therefore conclude that our MATHUSLA design can successfully reject backgrounds and discover LLPs produced at the HL-LHC. These realistic signal rates also inform the overall sensitivity estimates across the parameter space of LLP benchmark models in the next Section.

\subsubsection{Signal Definition and Background Rejection Strategies}
\label{s.signalcriterionbackgroundvetostrategies}

The basic criterion for an LLP signal in MATHUSLA is the reconstruction of a vertex in the decay volume (defined as being contained by the ceiling and rear tracking modules, the floor veto layers, and the front wall veto layers) that comprises at least two tracks. Furthermore, all tracks of the vertex must travel in the upwards direction (or exiting the detector through the rear, for tracks reconstructed in the rear tracking modules). This eliminates random crossings of CRs and/or LHC muons, and downward-traveling vertices from CR scattering processes.

HL-LHC muons, CR protons/neutrons and atmospheric neutrinos can generate background vertices which pass the above criteria. To reject these backgrounds, we veto MATHUSLA events that fail any of the following cuts. When referring to a vertex, these cuts always apply to the primary vertex (PV), defined as the vertex in the event with the most tracks.
\begin{enumerate}
    \item \textbf{PV association with sensor material:} We shrink the fiducial volume by 0.3~m from the walls, and we also exclude 1~m$^2$ regions around the vertical steel pillars. 

    \item \textbf{Hits before the vertex in time:}
    Background vertices in the decay volume are usually generated by external charged particles, which leave hits in either the veto layer or a tracking layer before the vertex is created.  
    Therefore we reject any event with more than 1 hit (anywhere in the detector) within a 100~ns time window before the PV.

    \item \textbf{Tracks with abnormal speed:} 
    Atmospheric neutrinos and CR protons with momenta near a GeV often produce non-relativistic protons or charged pions as final states. Therefore we veto any event in which any track in the PV has $\beta < 5/6$ or has an apparent speed much higher than $c$ (presumably a fake track).

    \item \textbf{PV orientation:} Vertices from LLP decays originate from the CMS IP, while cosmic backgrounds have random directions. Therefore we define a cut on the angle difference between the IP direction and the vertex cone edge. The IP direction is within the cone if this value is less than 0. But for LLP decays containing invisible particles, the boost of the LLP means the signal vertex can be oriented away from the CMS IP, i.e. it is possible the IP is not contained in the back-propagated vertex cone. Therefore we reject any event in which this value for the PV is lower than 0.3 rad, which balances signal efficiency and rejection power.

    \item \textbf{Number of tracks in PV:} This is a very powerful discriminator. Over $80\%$ of vertices from LLP decays, but $<0.2\%$ of vertices arising from cosmic proton and neutron backgrounds, have four or more tracks. Therefore we reject any event in which the PV has fewer than 4 tracks.

    \item \textbf{Fraction of downward tracks in each event:} Since CR backgrounds usually generate tracks that travel downward, we reject events in which too high a fraction of the tracks are downward-traveling. This cut is made on the downward fraction of tracks in the event, rather than the absolute number of downward-traveling tracks, to keep signal efficiency high -- since LLP decays may also generate some downward tracks when there are many decay products.

    \item \textbf{Fraction of hits that are inconsistent with PV:}  Neutron and proton scattering events tend to have more random hits that are not part of any reconstructed vertex. This results in a high fraction of hits that are inconsistent with the PV (based on maximum propagation speed $c$). We veto events in which this fraction exceeds a set threshold.

\end{enumerate}

\subsubsection{Benchmark Analyses}
\label{s.benchmarkanalysis}

For LLP signal, we simulated $h\to XX, X\to b\bar{b}$ decay at 5 different masses (15, 25, 35, 45, 55 GeV) of $X$. 
We applied the event selection criteria to the simulated backgrounds and signal. The result is shown in Table.~\ref{tab:cut_table}.

The signal efficiency ranges from 37\% to 53\%, decreasing with LLP mass due to the lower boost at higher mass. The backgrounds are completely rejected, although only the LHC muon and atmospheric neutrino samples have total exposure on par with MATHUSLA run time -- the cosmic proton (neutron) samples represent a total of 1/32 (1/27) MATHUSLA exposure. We use the ``ABCD method" to estimate the surviving cosmic proton + neutron background at full MATHUSLA exposure, by dividing the cuts into two groups that we expect to be largely uncorrelated.

\begin{table}[tpbh]
\resizebox{\textwidth}{!}{
    \begin{tabular}{l|rrrr|rrrrr}
    \hline
    \multicolumn{1}{r|}{\textbf{}}               & \multicolumn{4}{c|}{\textbf{Backgrounds}}                             & \multicolumn{5}{c}{\textbf{Signal}}   \\ \hline
    \multicolumn{1}{r|}{\textbf{}}               & \multicolumn{1}{l}{\textbf{\begin{tabular}[c]{@{}l@{}} cosmic \\ p \end{tabular}}} & \multicolumn{1}{l}{\textbf{\begin{tabular}[c]{@{}l@{}}cosmic \\ n \end{tabular}}} & \textbf{\begin{tabular}[c]{@{}r@{}}LHC \\ $\mu$ \end{tabular}} & \multicolumn{1}{l|}{\textbf{\begin{tabular}[c]{@{}l@{}}atmos. \\ $\nu$\end{tabular}}} & \multicolumn{1}{l}{\textbf{\begin{tabular}[c]{@{}l@{}}bb \\ 15GeV\end{tabular}}} & \multicolumn{1}{l}{\textbf{\begin{tabular}[c]{@{}l@{}}bb\\ 25GeV\end{tabular}}} & \multicolumn{1}{l}{\textbf{\begin{tabular}[c]{@{}l@{}}bb \\ 35GeV\end{tabular}}} & \multicolumn{1}{l}{\textbf{\begin{tabular}[c]{@{}l@{}}bb \\ 45GeV\end{tabular}}} & \multicolumn{1}{l}{\textbf{\begin{tabular}[c]{@{}l@{}}bb \\ 55GeV\end{tabular}}} \\ \hline \hline
    \textbf{Exposure (fraction of 5 years)}       & 1/32      & 1/27       & 1/1.75              & 2x   & \multicolumn{5}{c}{10000 events}                                                                                 \\ \hline \hline
    \textbf{All events with $\geq1$ vertex in volume} & 638739  & 218908  & 23689   & 24   & 8601&  9057&  9055&  8964&  8846   \\ \hline \hline
    \textbf{1. Fiducial volume}                       & 566577  & 200440  & 23056   & 22   & 8356&  8806&  8802&  8708&  8598   \\
    \textbf{2. \#Hits before PV}           & 55494   & 57386   & 17      & 0    & 7938&  8318&  8254&  8121&  7950   \\
    \textbf{3. Abnormal track speed in PV}                   & 42910   & 44924   & 17      & 0    & 7707&  8119&  8090&  7945&  7748   \\
    \textbf{4. PV to CMS direction}               & 4013    & 4630    & 15      & 0    & 7167&  7104&  6622&  5897&  4785   \\ \hline
    \textbf{5. \#Tracks in PV}                   & 10      & 4       & 0       & 0    & 5155&  5645&  5509&  4950&  4001   \\
    \textbf{6. Downward fraction of tracks}                    & 2       & 1       & 0       & 0    & 5033&  5522&  5357&  4811&  3876   \\
    \textbf{7. Hit fraction inconsistent with PV}                  & 0       & 0       & 0       & 0    & 4859&  5348&  5223&  4656&  3707   \\  \hline
    \end{tabular}
    }
\label{tab:cut_table}
\caption{Cut table for the benchmark $h\to XX$ hadronically decaying LLP search at MATHUSLA. Exposure for backgrounds are given with respect to the full MATHUSLA exposure of 5 years. The selection number matches the list in Sec.~\ref{s.signalcriterionbackgroundvetostrategies}. The cuts are divided into two groups: Cuts 1-4 do not explicitly depend on the number of tracks or hits that could be associated with the vertex, while cuts 5-7 directly depend on these properties of the vertex.}
\end{table}

The ABCD method requires two independent selections that together define the signal region (A).
By inverting either or both selections, three control regions (B, C, and D) can be defined. The assumption is that the number of background events $N$ in the four regions follows the relation $N_A/N_B=N_C/N_D$, and we can estimate $N_A$ from $N_B$, $N_C$, and $N_D$ because these regions have more background statistics. 
Here, we treat cuts 1-4 in Table~\ref{tab:cut_table} as the first selection, and cuts 5-7 as the second selection. The reason for this separation is that the former do not depend explicitly on the multiplicity of final states that can be associated with the vertex, while the latter explicitly depend on this multiplicity. These two cut groups should therefore have low enough correlation to allow for the ABCD method to give a reasonable estimate of the final background level.

With this division, $N_D$ is the total number of events before any selection, i.e. all events with $\geq 1$ vertex in  the volume; 
$N_B$ is the number of events left after applying cuts 1-4 to $N_D$; and $N_C$ is the number of events after applying cuts 5-7 to $N_D$. See Table ~\ref{tab:cut_table_inv}, where the order of the two cut groups has been reversed.

\begin{table}[tpbh]
\centering
\resizebox{0.75\textwidth}{!}{
    \begin{tabular}{l|rrrr}
    \hline
    \multicolumn{1}{r|}{\textbf{}}               & \multicolumn{4}{c}{\textbf{Backgrounds}}                             \\ \hline
    \multicolumn{1}{r|}{\textbf{}}               & \multicolumn{1}{l}{\textbf{\begin{tabular}[c]{@{}l@{}} cosmic \\ p \end{tabular}}} & \multicolumn{1}{l}{\textbf{\begin{tabular}[c]{@{}l@{}}cosmic \\ n \end{tabular}}} & \textbf{\begin{tabular}[c]{@{}r@{}}LHC \\ $\mu$ \end{tabular}} & \multicolumn{1}{l}{\textbf{\begin{tabular}[c]{@{}l@{}}atmos. \\ $\nu$\end{tabular}}} \\ \hline \hline
    \textbf{Exposure (fraction of 5 years)}       & 1/32      & 1/27       & 1/1.75              & 2x                                                                                                                                                                                                                \\ \hline \hline
    \textbf{All events with $\geq1$ vertex in volume} & 638739  & 218908  & 23689   & 24     \\ \hline \hline
    \textbf{5. \#Tracks in PV}                    & 1206&  216&  263&  2      \\
    \textbf{6. Downward fraction of tracks}                     & 54&  27&  248&  2      \\
    \textbf{7. Hit fraction inconsistent with PV}                   & 1&  0&  180&  0      \\ \hline
    \textbf{1. Fiducial volume}                       & 1&  0&  173&  0     \\
    \textbf{2. \#Hits before PV}           & 0&  0&  0&  0      \\
    \textbf{3. Abnormal track speed in PV}                   & 0&  0&  0&  0      \\
    \textbf{4. PV to CMS direction}               & 0&  0&  0&  0      \\
    \hline
    \end{tabular}
    }
\label{tab:cut_table_inv}
\caption{Cut table for the benchmark $h\to XX$ hadronically decaying LLP search at MATHUSLA, with the order of the two cut groups flipped.}
\end{table}

For protons, we estimate $N_A=0.2$; for neutrons, both regions C and D have zero counts; for muons, we estimate $N_A=0.2$. We therefore double the background from protons as a conservative estimate for the contribution of neutrons.
Therefore, the total background level for our benchmark LLP search with full MATHUSLA exposure is less than 1 event.

We repeated the analysis on a smaller sample in different configurations using only 5 or 4 tracking layers, combined with different detector efficiency of 90\% or 95\%. 
In the 4-layer setup, we require only 3 hits to form a track rather than 4 for the other setups. 
In the worst case (4 layers, 90\% hit efficiency),
signal efficiency remains above 17\%, while the background is affected more significantly, increasing by more than 10 times compared to our benchmark simulation above.
On the other hand, in the 5-layer setup has almost the same signal efficiency as the 6-layer benchmark, with backgrounds increased by less than a factor of 2. 
In all setups, the signal efficiency drops by less than 5\% when the detector efficiency is decreased from 95\% to 90\%. 
We also performed the study on the 6-layer setup with $10 \times$ the noise rate, and it has no significant effect on signal or background rejection efficiency.

We can therefore conclude that the ability of our six-layer detector design to conduct zero-background LLP searches with high signal efficiency is robust with respect to $\times 10$ increases in the noise rate, a doubling of the failure rate for detecting single hits (from 5\% to 10\%), and even the complete failure of a single tracking layer compared to our baseline assumptions.

%%%%%%%%%%%%%%%%%%%%%%%%%%%
%%%%%%%%%%%%%%%%%%%%%%%%%%%
%%%%%%%%%%%%%%%%%%%%%%%%%%%
%%%%%%%%%%%%%%%%%%%%%%%%%%%
%%%%%%%%%%%%%%%%%%%%%%%%%%%
%%%%%%%%%%%%%%%%%%%%%%%%%%%

\subsection{LLP Physics Reach}
\label{s.signalacceptance}

\begin{figure}
    \centering
   \includegraphics[width=0.8 \textwidth]{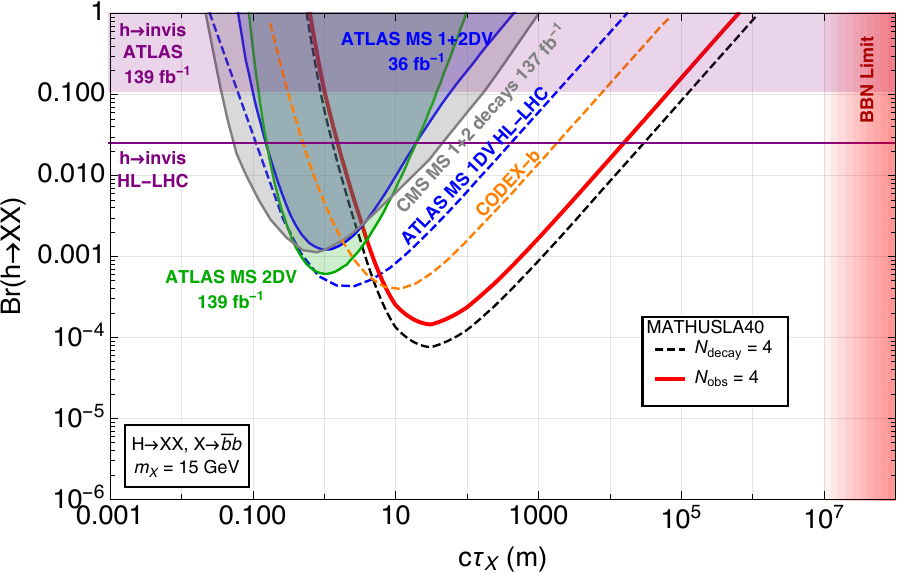}
    \caption{
    Sensitivity of the 40m MATHUSLA detector to hadronically decaying LLPs produced in exotic Higgs boson decays. The solid red curve shows the exclusion reach from a realistic background-free search for DVs with with 49\% signal reconstruction efficiency after background vetoes are applied. For comparison, the dashed black curve shows idealized reach for 4 decays in the decay volume.
    We also show the current $\mathrm{Br}(h \to \mathrm{invis})$ limit from ATLAS~\cite{ATLAS:2023tkt} (purple shading) and  the HL-LHC projection~\cite{Dainese:2019rgk} (purple line); current ATLAS constraints from searches for 1 or 2 DVs (blue shading) and 2 DVs (green shading) in the muon system~\cite{ATLAS:2018tup, ATLAS:2022gbw};
    current CMS constraints from searches for 1 or 2 LLP decays in the muon system~\cite{CMS:2024bvl} (gray shading);
    projections for an ATLAS 1DV search in the muon system at the HL-LHC~\cite{Coccaro:2016lnz} (blue dashed);
    and the idealized sensitivity of CODEX-b~\cite{Aielli:2019ivi} (orange dashed). 
    }
    \label{f.higgs}
\end{figure}

\begin{figure}
\centering
\includegraphics[width=0.7 \textwidth]{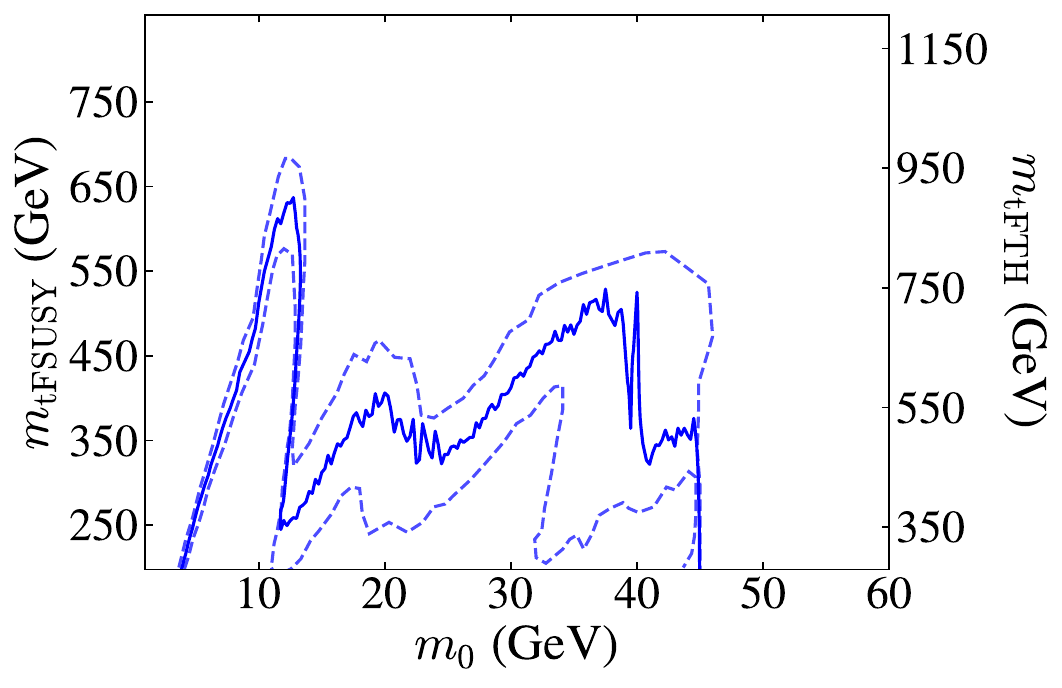}
\caption{
Reach of the 40m MATHUSLA design in a simplified parameter space of Neutral Naturalness, generated using the dark glueball Monte Carlo from~\cite{Batz:2023zef}. Dark glueballs, the lightest of which has mass $m_0$, are produced in exotic Higgs decays which undergo dark Lund-String hadronization. The effective higgs coupling to dark gluons, which also allows glueballs to decay, is generated by neutral top partners in the Folded SUSY~\cite{Burdman:2006tz} and Fraternal Twin Higgs~\cite{Craig:2015pha} models, with masses indicated on the horizontal axes. The solid blue curve shows the reach for 8 decays in the MATHUSLA decay volume, corresponding to the exclusion limit for 50\% reconstruction efficiency expected for near-background-free searches. The dashed curves represent theoretical uncertainties in this reach from unknown aspects of non-perturbative dark $N_f = 0$ QCD.}
\label{f.glueballreach}
\end{figure}

We can now use the geometry-only
\texttt{MATHUSLA FastSim} detector simulation code~\cite{MATHUSLAFastSim}
to compute the number of LLP decays in MATHUSLA's decay volume for the LLP benchmark models as a function of lifetime~\cite{MATHUSLAFastSim, Curtin:2023skh}. Applying the final reconstruction efficiency from Table~\ref{tab:cut_table} accounts for the effects of cuts and vetos required to make the search background-free, allowing us to make realistic estimates of MATHUSLA's physics reach. 
We use the same exotic Higgs decay events as the $\texttt{GEANT4}$ simulation.
The sensitivity of MATHUSLA is shown in \fref{f.higgs} for $m_{LLP} = 15~\mathrm{GeV}$. Higher masses are similar, with a shift of the sensitivity curve in lifetime corresponding to the average LLP boost.
The dashed line shows the contour of 4 decays in the MATHUSLA volume $(N_\mathrm{decay} = 4)$. Applying the realistic reconstruction efficiency for a background-free LLP search yields the solid red sensitivity curve $(N_\mathrm{obs} = 4)$.

The reach of MATHUSLA is compared to current and projected invisible Higgs boson decay bounds (purple line) and the proposed CODEX-b detector~\cite{Aielli:2019ivi} (orange dashed, not including realistic reconstruction efficiencies for a background-free search).
We also compare this to current bounds from LLP searches in the ATLAS and CMS muon systems~\cite{ATLAS:2018tup, ATLAS:2022gbw, CMS:2024bvl} (green, blue and gray shadings), and show the projected reach of the ATLAS search at the HL-LHC~\cite{Coccaro:2016lnz}, which is is  background limited for single LLP decays.

It is evident that the $(40~\mathrm{m})^2$  MATHUSLA design can probe LLP lifetimes and production rates 1-2 orders of magnitude beyond the reach of the HL-LHC main detector searches or the proposed CODEX-b detector.

The significance of this sensitivity is vividly demonstrated by considering MATHUSLA's reach for dark glueballs produced in exotic Higgs decays. In practical terms, this can be regarded as a slight elaboration on the minimal exotic Higgs decay simplified model, and is realized in many hidden valleys and, in particular, the 
Fraternal Twin Higgs~\cite{Craig:2015pha} models and Folded SUSY~\cite{Burdman:2006tz}  solutions to the little Hierarchy problem. Due to recent progress~\cite{Curtin:2022tou, Batz:2023zef} in modeling non-perturbative aspects of the dark shower and dark glueball hadronization, it is now possible to make quantitative predictions for the generated LLP signal arising from the production of all meta-stable dark glueball species. The sensitivity of MATHUSLA to these decays for the current 40m geometry is shown in Figure~\ref{f.glueballreach}, as a function of the dark glueball mass and SM-neutral top partner mass. MATHUSLA effectively probes neutral naturalness solutions of the little hierarchy problem in large parts of model's motivated parameter space with neutral top partner masses below a TeV.

%%%%%%%%%%%%

\newpage
%%%%%%%%%%%%%%%%%
\clearpage
\section{Detector Design}
\label{s.detectordesign}
%%%%%%%%%%%%%%%%%

%%%%%%%%%%%%%%%%%
\subsection{Detector Requirements}
\label{s.detectorrequirements}
%%%%%%%%%%%%%%%%%
MATHUSLA is intended as a discovery detector for new physics. It must therefore be capable of unambiguously identifying LLP signal candidates in the presence of near-zero backgrounds.  This necessitates well-understood detector performance and signal efficiency, the ability to calibrate and stably operate the detector, and the ability to record physics samples necessary to validate the physics performance and background rates. 
Conceptually, the detector consists of an empty decay volume located approximately 100~m from an LHC IP, instrumented with veto detectors on the upstream side and tracking detectors for the charged decay products of LLPs on the downstream side.

The basic performance requirement of the MATHUSLA detector is the capability to reconstruct vertices (occurring within the decay volume) from decays of LLPs produced in LHC collisions, and to differentiate these reconstructed vertices from backgrounds originating from both CRs and SM particles produced in LHC collisions. The location of the detector, on the surface roughly 100~m from the CMS IP, is dictated by availability of space situated as close as possible to the IP. Given this location, the physical dimensions are motivated by the cross sections of the relevant physics processes and the anticipated integrated luminosity of the HL-LHC, as well as practical limitations like building codes specifying the maximum height of buildings near CMS.  The challenge of MATHUSLA is to instrument the region surrounding the large decay volume in a cost-effective manner with detectors that meet the design requirements imposed by the physics goals.

To meet these requirements, the detector needs to be capable of track reconstruction in ``4D'' (i.e. including timing information) with a resolution sufficient to identify decay vertices over lever-arms of up to tens of meters, with vertex resolutions of a few cm in the transverse direction and tens of cm longitudinally.  This vertex resolution is motivated by the simulation studies of Section~\ref{s.detectoratcms}, allowing MATHUSLA to suppress backgrounds CR protons and LHC muons arising from inelastic interactions in the air or detector structure. Absolute hit timing needs to be sufficient to distinguish upwards-traveling from downwards-traveling tracks, and to be capable of associating hits in the lower floor veto \layers with tracks in the multi-tracking \layers  for the purpose of vetoing muons originating from the CMS IP.  Hit timing can also be used to associate MATHUSLA events with LHC bunch crossings, and hence CMS detector events, for the purpose of sending trigger requests to CMS in real time and correlating MATHUSLA and CMS events in offline analyses. 

The vertex resolution also needs to be sufficient to ensure that the reconstructed LLP candidate ``points-back'' to the CMS IP, to help distinguish signal events from, for example, randomly-oriented vertices produced by atmospheric neutrinos and CR protons.

%%%%%%%%%%%%%%%%%
\subsubsection{Tracking Resolution}
\label{s.trackingresolution}
%%%%%%%%%%%%%%%%%
The MATHUSLA detector is anticipated to consist of a series of large detector \layers
comprising narrow \mbars of plastic scintillator, oriented in the horizontal $x -y$ plane.  \mBars
in these \layers will be oriented either along the $x$ or $y$ axis, in an alternating 
manner.  Each \layer will therefore directly provide two spatial coordinates, i.e. $z$ plus 
$x$ or $y$ based on detected ``hits'' in individual \mbars, with the adjacent tracking 
\layers providing the third spatial coordinate.  Note that the third spatial coordinate in a hit bar can also be obtained based on differential hit timing from the \SiPMs at either end of the \WLSF, but with substantially lower resolution than the other two coordinates.

Simulation studies based on straw-man detector geometries are sufficient to determine 
the basic detector performance necessary to ensure adequate tracking and vertexing capabilities to
satisfy the signal reconstruction and background rejection requirements described in 
Section~\ref{s.detectoratcms}.  In particular, the $z$ and $(x,y)$ position resolution in
individual tracking \layers must be approximately $1$~cm, and the absolute hit time
resolution is required to be $1$~ns or better.   Based on this, the transverse dimension
of the scintillator \mbars needs to be $1~{\rm cm} \cdot \sqrt{12} \simeq 3.5~{\rm cm}$. 
The thickness of the \mbars in $z$ is not determined by the tracking resolution, but must
simply be large enough to provide sufficient scintillation light to ensure high hit efficiency and timing resolution. 

The detector concept foresees a total of six \planes of scintillator \mbars mounted in \ceilingtrackingmodules, and \walltrackingmodules in the rear, as well as at least two layers of veto coverage in the front wall and floor. 
The six \layers in the ceiling and rear wall comprise the
primary tracking system, each separated by 80cm in $z$.  
Therefore, ideal tracks will be
reconstructed from 
6 hits, which include 
three hits each with high-resolution along the $x$ and $y$ directions,
though we anticipate that good quality tracks can be reconstructed from as few as four hits.

%%%%%%%%%%%%%%%%%
\subsubsection{Timing}
\label{s.timingrequirements}
%%%%%%%%%%%%%%%%%

The difference of hit timing in adjacent vertical \planes of the detector is used to determine the up/down flight direction of particles traversing the detector, and is critical for rejection of CR backgrounds as well as for reducing hit combinatorics in track-finding algorithms.  Given the anticipated $\sim 80$~cm spacing between triggering \planes, we specify an absolute hit timing resolution of $\leq 1$~ns, such that the flight direction of a track candidate can be determined based on any pair of hits in adjacent \planes. In practice, hit timing in any scintillator \mbar is determined by measurements from two \SiPMs, which are located on opposite ends of a single \WLSF.  Consequently, this 1~ns time resolution, which can be determined from the resolution of the average (i.e. $(t_1 +t_2)/2$) of the two hit times, also determines the differential timing resolution (i.e. $t_1-t_2$), which is used to infer the hit position of the incident particle along the length of the scintillator \mbar.  For 1~ns timing resolution, we anticipate a hit position resolution of roughly $\sim 15$~cm.  Therefore, this differential timing can be used to provide longitudinal position information in the scintillator \mbars, albeit with a resolution that is an order of magnitude worse than in the transverse direction.  While this differential timing
information will not significantly impact the tracking resolution, it is anticipated to
aid in pattern recognition in the track reconstruction process.

%%%%%%%%%%%%%%%%%
\subsubsection{Vertex Reconstruction Resolution}
\label{s.vertexrequirements}
%%%%%%%%%%%%%%%%%

The geometry-only \texttt{MATHUSLA FastSim}~\cite{MATHUSLAFastSim} is used to investigate the impact of spatial tracking resolution on LLP signal acceptance for the exotic Higgs boson decay benchmark model of the main physics target (see Section~\ref{s.signalsim}), as well as for GeV-scale LLPs. %, as well as the RHN and SM+S benchmark models for GeV-scale LLPs (see Appendix~\ref{s.lowmassLLPs}).
These simulations do not perform explicit track and vertex reconstruction, but do merge two hits if they occur in the same scintillator bar. The effect of finite spatial resolution can then be overestimated by requiring that each track has at least 4 distinct (i.e. not merged) hits. 
For bar lengths of 1, 2, or 4~m, and bar widths of 4 or 8~cm, 
this requirement was found to decrease the geometric acceptance for LLP decays by at most 5\% relative to idealized simulations that did not include any spatial resolution criteria.  
The \mbar dimensions given in Sections~\ref{Scintillator.plane}~and~\ref{s.detectorbuildingblocks}, and resulting track and vertex spatial resolutions in realistic reconstructions, are therefore well-suited to MATHUSLA's physics objectives. 
This has been confirmed by the full \texttt{GEANT4} simulations of Section~\ref{s.detectoratcms}.

%%%%%%%%%%%%%%%%%
\subsection{Detector Building Blocks}
\label{s.detectorbuildingblocks}
%%%%%%%%%%%%%%%%%

This was introduced in Sections~\ref{s.detectorintro} and~\ref{Scintillator.plane}, but we repeat the main details here.
The $\sim 40~{\rm m} \times 40~{\rm m}$  footprint of the MATHUSLA detector is divided into
16  identical, $\sim$~\moduleside~$\times$~\moduleside \towers each with 4 vertical supports. 
Each \tower contains 6 identically constructed \layers in its \ceilingtrackingmodule, see Figure~\ref{fig:modular}. 
\Walltrackingmodules mounted on the rear wall are identical their ceiling equivalents, with modified structural supports to allow for their vertical orientation.

The \towers include two \floorvetolayers mounted by its vertical supports 0.5m and 1m above the floor, which are identical to the \layers and also arranged with alternating bar orientations. \Floorvetostrips are horizontally mounted at a height of 1.5m to cover the gap between \towers, and vertically mounted on the vertical support columns of the \towers to form a \columndetector that instruments the intersection of any two gaps. All of this constitutes a \floorveto that is hermetic with respect to downward cosmics. 
The front wall of MATHUSLA is instrumented with two layers of four $(11.2~\mathrm{m})^2$ \wallvetolayers.
This \wallveto completes the \veto system and makes it hermetic with respect to LHC muons as well.

Each $(9~\mathrm{m})^2$ contiguous \layer is made up of four identical \sublayers arranged end-to-end with some overlap, where each provides~\moduleside$\times$~\barlength of sensor coverage.
Each \sublayer is made up of  eight \barass units with 32 scintillator \mbars that are
\barlength~long~$\times$~\barwidth~wide. 
\Wallvetolayers are constructed similarly, out of 5  \wallvetosublayers comprising 10 \barass each, while \floorvetostrips are just two \sublayers sandwiched together.
Each scintillator \mbar is paired with a second \mbar located 16~\mbars away, and the pair is threaded with a common \WLSF, as shown in Figure~\ref{fig:FNAL_extrusions}. A timing coincidence between the two \SiPMs on the \fiber ends indicates the passage of a charged
particle, while suppressing false signals from electronic noise. 
The time difference between the two \SiPMs provides a longitudinal 
position along the fiber.

%%%%%%%%%%%%%%%%%
\subsubsection{Scintillator \mBars and Wavelength-shifting Fibers}
\label{s.scintillatorbars}
%%%%%%%%%%%%%%%%%

Extruded plastic scintillator provides a cost-effective solution for instrumenting the large MATHUSLA decay volume, and thus the primary detector element of MATHUSLA is a scintillator \mbar. 
The scintillator material is based on commercial polystyrene doped with PPO ($~1\%$) and POPOP ($~0.02\%$).  
Due to the relative ease of fabrication, the \mbar cost is dominated by material costs for commercial polystyrene pellets, and the total material need for MATHUSLA is estimated to about 160 tons.

\begin{figure}
\centering
{
\includegraphics[width=.7\textwidth]{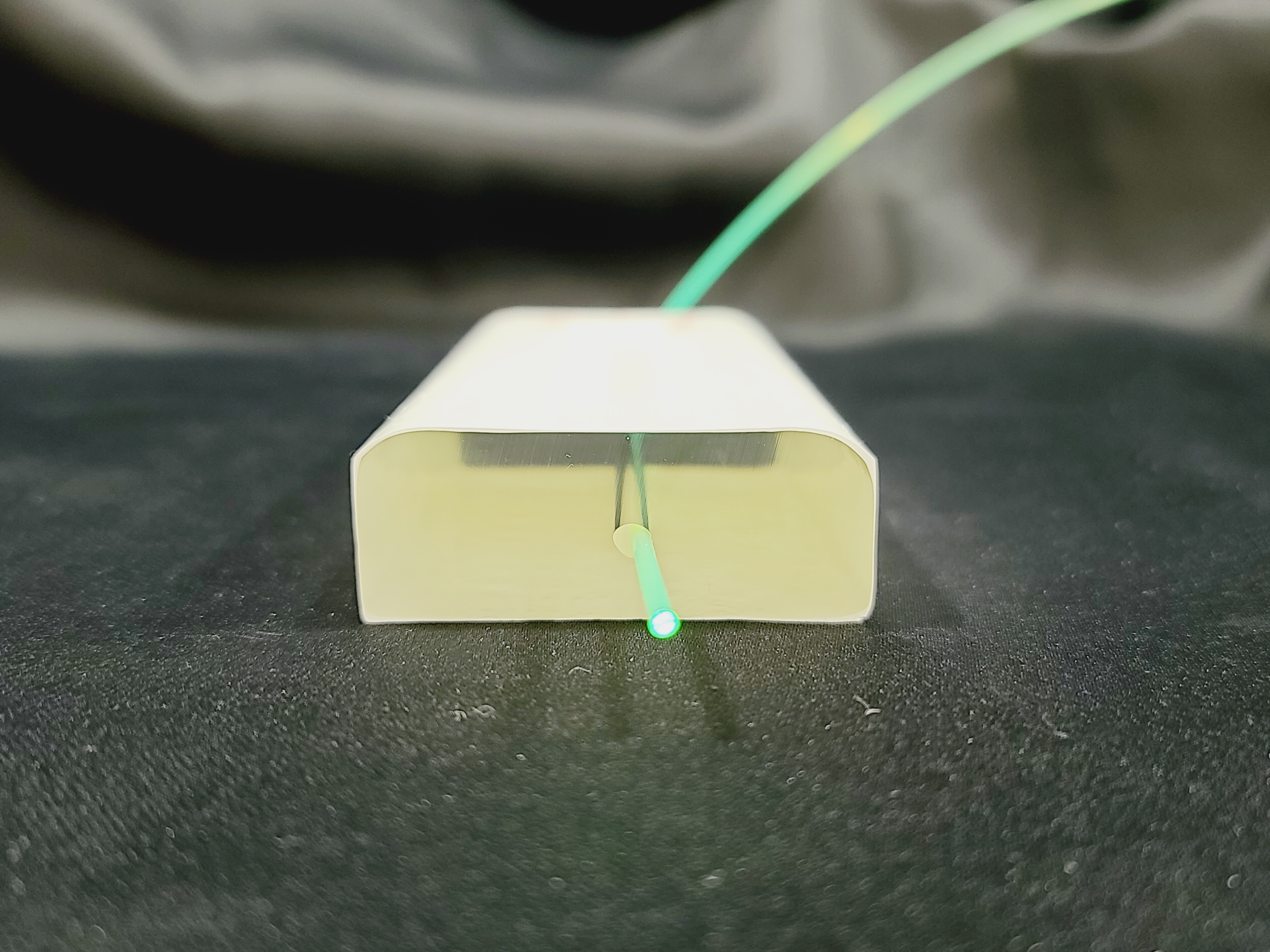}
\caption{\label{fig:Extrusion} 
A scintillator bar extrusion as made at Fermilab. A white co-extruded cladding and a central co-extruded hole can be seen. Through the hole is a \WLSF. Fiber ends will be instrumented with \SiPMs.}
}
\end{figure}

The scintillator plastic is produced using an extrusion process.
Such extruded scintillator produces scintillation light similar to more expensive cast organic scintillator, but with a relatively short attenuation length (typically $\sim 10$~cm).  
This is not a limitation for our choice of $\sim$ cm \mbar width. 
To maximize absorption in the embedded \WLSF, the exterior of the bars are coated with reflective material.  
Particles passing through the scintillator produce light, which is wavelength shifted and propagated along the \fiber to \SiPMs at each end, where a light yield threshold is applied and timing information is recorded. Figure~\ref{fig:Extrusion} shows a representative scintillator \mbar that was produced in an extrusion process with a co-extruded hole and white co-extruded reflective cladding. 

The efficiency of a \WLSF for capturing scintillation light is typically a few percent, depending on the fiber diameter and dopants. Attenuation of light within the fibers is also a significant consideration given the size of the detector.  For MATHUSLA, the transverse \mbar width ($\sim 3.5$~cm) is dictated by the desired position resolution of $\sim 1$~cm, while the thickness of the \mbars determines the amount of scintillation light produced by the passage of an ionizing particle, which in turn limits the light output from the \WLSF.  As the total material cost scales linearly with the \mbar thickness, it is desirable to minimize the bar thickness while retaining the desired hit efficiency and timing performance.  Studies have demonstrated that sufficient light yield can be obtained using $1 - 2$~cm thick \mbars, but there is a cost optimization to be made between the \mbar thickness and the \WLSF diameter. The \mbar length, which dictates the dimensions of the smallest mechanical assembly unit (a 32-\mbar, \barass module) is limited by the maximum tolerable attenuation in the \WLSF, which limits the fiber length to approximately $5 - 6$~m.  Although scintillation light must be detected at both ends of a fiber, \SiPMs will be mounted only on one side of each \sublayer, and consequently the 5~m fiber limits the scintillator \mbar length to approximately $2.35$~m. This allows each fiber to loop back through a second \mbar (see Figure~\ref{fig:FNAL_extrusions}), such that all fiber ends are along the same \sublayer edge.  The edge that contains the fiber loops adds $\sim 30$~cm to the \sublayer length, and these loops will be contained in, and protected by, the aluminum cladding that is foreseen to surround the \barasss for light tightness, structural support, and fire protection.

\subsubsection{Scintillator Fabrication}
\label{s.RDscintfab}

\begin{figure}
    \begin{center}
        \includegraphics[width=0.75
        \textwidth]{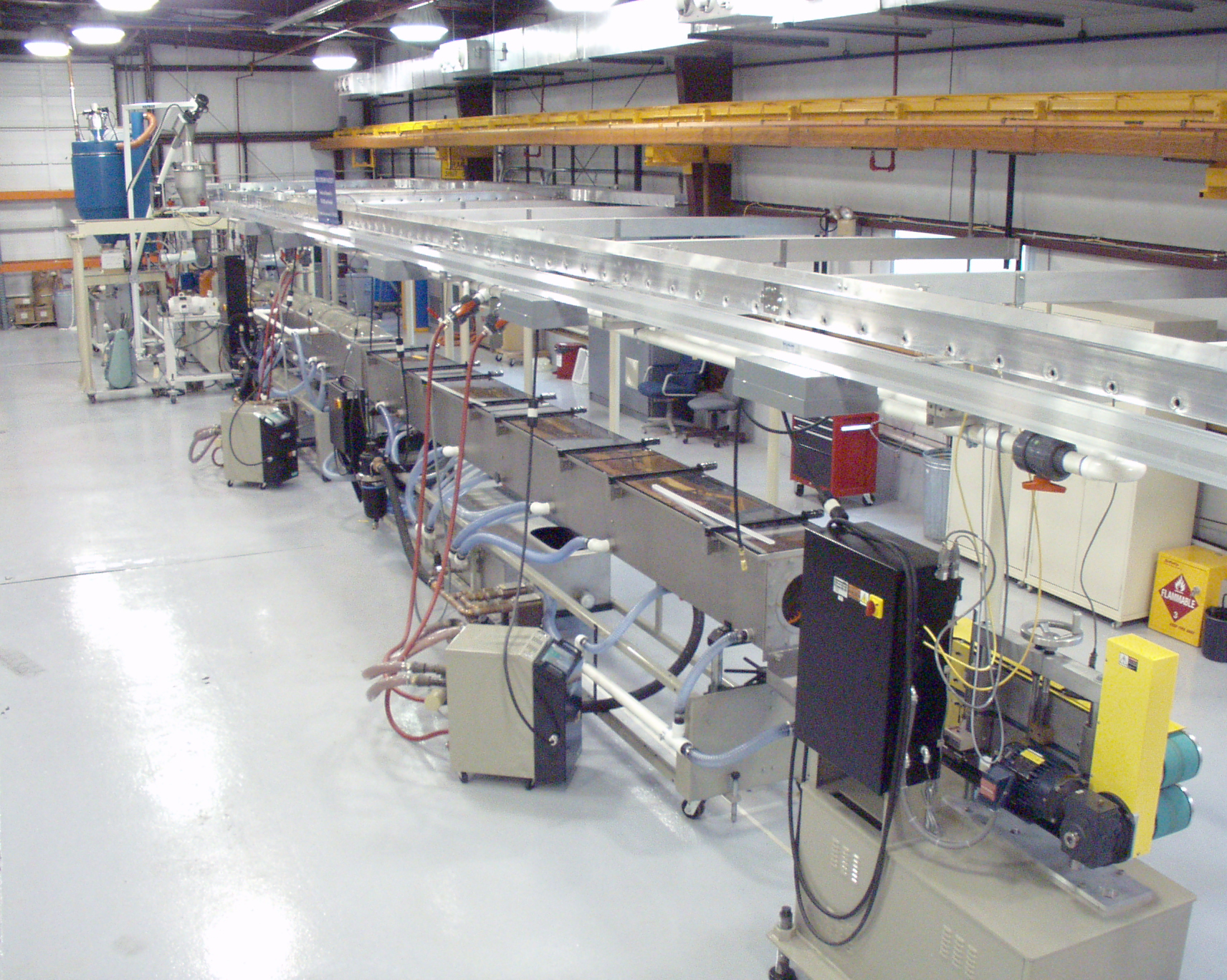}
        \caption{The scintillator extrusion facility at Fermilab.}
   \label{fig:Extrusion_line}
   \end{center}
\end{figure}

The extruded scintillator can be fabricated at the NICADD Fermilab extrusion facility at FNAL.  This facility has a long history of making similar extrusions, most recently for the Mu2e CRV detector~\cite{Mu2e:2014fns}.  Fig.~\ref{fig:Extrusion_line} shows the extrusion machine.  The FNAL extruder is capable of extruding about 75~kg/hour. Consequently, extrusion of 100~tons of scintillator would take about 1300~hours, or approximately 3~months of production at 24~hours-per-day operation.

\newpage

%%%%%%%%%%%%%%%%%
\subsubsection{Silicon Photomultipliers}
\label{s.RDSiPM}
%%%%%%%%%%%%%%%%%

Characterization studies of the individual detector components are being performed, and are detailed in Appendix~\ref{s.detectorRD}. A minimum ionizing particle traversing the nominal, $1$~cm thick scintillator bar produces a total light output of $\mathcal{O}(10)$ - $\mathcal{O}(100)$ photons from each end of a nominal, $\sim 5$~m \WLSF, depending on the incident angle and position along the bar. It has additionally been demonstrated that a $1$~ns timing resolution can be achieved with this light yield (see \fref{fig:timing} in Appendix~\ref{s.detectorRD}), with an appropriate choice of \WLSF. The timing resolution is being driven primarily by the decay time of the \WLSF fluors and the relatively low photon statistics. In this regime, the arrival time of the first detected photon is expected to provide the best measure of the hit time, hence it is desirable to use photodetectors with a fast signal rise time and sensitivity to single photon signals. \SiPMs are well matched to this application and have been shown to not be the limiting factor in the timing resolution of a system.  Although the performance of \SiPMs from a number of vendors (Hamamatsu, Broadcom, ONSemi/SensL) has been studied, for the purposes of this document only Hamamatsu devices are considered.

\SiPMs provide an effective and economical solution to the detection of light from blue-green wavelength shifting optical fibers with sensitivity to light in the range of $270 - 900$~nm with peak sensitivity at 450~nm.  Hamamatsu S14160 or S13360 series \SiPMs have photon detection quantum efficiencies in excess of 0.5 in the peak emission range for 
Saint Gobain BCF-92XL  WLSFs. Due to the relatively low light yield from the \WLSF in the MATHUSLA default configuration, it is desirable to maintain the highest possible photon detection efficiency.

\SiPMs are two dimensional arrays of silicon avalanche photodiodes (``pixels'') connected in parallel, such that the output signal represents the sum of the signals from all activated pixels.  Consequently, they can act as photon counters, with sensitivity down to the single photon level. 
\SiPMs are operated at a voltage slightly above the photodiode breakdown voltage, such that an incident photon can trigger breakdown in individual photodiode pixels, which results in a current pulse proportional to the number of activated pixels.
Spontaneous breakdown can also occur within individual photodiode pixels, resulting in dark noise.  Since pixel breakdown is generally uncorrelated, this noise is mostly comprised of single-pixel pulses, with an exponentially decreasing rate of larger amplitude pulses.  For example, Hamamatsu S13360-3050 \SiPMs quote a dark noise rate of $0.5$~Mc$/$s under nominal operating conditions. 
The dark noise rate potentially imposes a lower threshold on the amplitude of triggerable signals.  Since interesting signals are those in which there is a coincidence between \SiPM signals at the two ends of a $ \sim5$~m \WLSF within a $\sim 50$~ns time window, the effective requirement is simply that the coincidence rate from dark noise be lower than the rate resulting from CRs within the scintillator bar.  This rate has been studied and has been found to be easily achievable for commercially available $3\times3$~mm \SiPMs, for example Hamamatsu S14160 or S13360 series MPPCs (\SiPMs),  by requiring the signal to be consistent with three or more photo-electrons.  Signals from minimum ionizing particles traversing the scintillator bars have been measured to result in $\mathcal{O}(10 - 100)$ \SiPM photoelectrons, hence the loss of signal efficiency resulting from this noise threshold requirement on the coincident signals is $\ll 1\%$.

The dark noise rate also increases with increasing overvoltage, and generally scales with the active area of the SiPM.  Consequently, using smaller \SiPMs further reduces the potential coincidence rate.  Given the anticipated \WLSF diameter of $1.5$~mm, a \SiPM of diameter $1.8$~mm or $2$~mm would be optimal.  Hamamatsu has previously produced $2$~mm S13360 and S14160 \SiPMs, but it is anticipated that, given the $\sim 10^5$  \SiPMs required for the full MATHUSLA detector, a custom order is feasible.

\SiPMs are commercially available in a number of sizes and with a variety of pixel pitches.  For the purposes of MATHUSLA, the pixel size has been shown to not impose a limitation; however, for a given size \SiPM, smaller pixel sizes result in more dead material between pixels and hence a lower \SiPM photon detection efficiency (PDE).  Given the relatively small number of photons exiting the \WLSF, it is desirable to have the highest possible PDE. Although the number of pixels imposes an upper limit on the effective dynamic range of detected signals, this is not expected to be a significant limitation
given the $\mathcal{O}(100)$ photons.  For example, a $2\times 2$~mm \SiPM with a $50~\mu$m pixel pitch would have $\sim 1600$~pixels, and larger pitch (e.g. $75~\mu$m) is also possible.  A PDE of $50\%$ is quoted for S13360-3075 and S1416-3050 SiPMs, with $75~\mu$m and  $50~\mu$m pixel pitches, respectively, hence this value is realistically achievable. Additionally, the photoelectron signal is larger for larger pixel size, which supports favoring the largest available pixel size. 

\SiPMs display a significant temperature dependence on the gain, and hence it is important to either regulate the temperature or apply a temperature correction to the overvoltage to maintain a constant gain. 
Based on temperature excursions observed in CERN buildings comparable to the envisioned MATHUSLA experimental hall 
and the \SiPM temperature correction coefficient from Hamamatsu, a voltage control range of~1.0 V will adequately control the gain. 
This is compatible with the anticipated level of voltage trimming already foreseen to compensate for device gain variations within individual \SiPMs. 
Direct control of \SiPM temperatures has been considered, but temperature correction using sensors integrated into the front-end electronics is sufficient for the experiment's needs. 

%%%%%%%%%%%%%%%%%
\subsection{Detector Assembly}
\label{s.RDassembly}
%%%%%%%%%%%%%%%%%

\subsubsection{\Barass}
The basic unit of the scintillator layers is a ``\barass'' that provides a sensor coverage of \barlength $\times$ 1.12~m (32 bars wide), with physical dimensions close to 2.8~m $\times$ 1.12m. Figure~\ref{fig:FNAL_extrusions} shows an overview of the \barass. 
The \barasss are foreseen to be assembled and tested at collaborating universities and institutions before shipment to CERN. Once at CERN, eight \barasss are connected to form an approximately \barlength$\times$ \moduleside \sublayer of scintillators. Four such \sublayers comprise one full scintillator \layer or \floorvetolayer. \Wallvetolayers and \floorvetostrips are assembled in an analogous manner.

The pre-polished fibers are loaded into the holes in the scintillator, with about 5~cm of excess on the \SiPM end. Printed circuit boards (PCBs) are attached to one side of each \barass. Each 2~mm square \SiPMs is mounted normal to a PCB, and 3D printed spring-loaded clamps are also mounted on the PCB. The 5~cm length of fiber is used to thread the fiber through the clamp allowing spring-loaded force against the \SiPM to achieve a stable optical coupling.
A bracket attaching the PCB(s) to the \barass will also serve as a stiffener to prevent the board from flexing during handling and assembly. 
Each PCB has a header to connect to a readout daughter board. After fabrication, a daughter board is temporarily connected to the \barass for local quality control testing before shipment to CERN.
At CERN, the \barasss are received in the assembly area of the experimental hall at P5. 

The \barasss are made by gluing \mbars  to a rigid sheet of aluminum hex panel strongback and covering them with thin aluminum sheet on the other three sides. This makes them fire-resistant  and structurally self-supporting when mounted on hardpoints at their edges. Each bar assembly will therefore have a weight close to $\sim$ 50 kg.

\subsubsection{Scintillator \layers}
Once at CERN, eight \barasss are joined with additional structural support 
to become a ``\sublayer''. Four \sublayers are used to form the complete 9~m$\times$9~m scintillator \layer. 
The complete scintillator \layer has pick points for crane or hydraulic lifting, and will weigh about 4 tons. These complete scintillator \layers are craned into the proper location and permanently connected to the \tower or \walltrackingmodule structure. Installation of horizontal and vertical \floorvetostrips, and \wallvetolayers, proceeds similarly.

\begin{figure}
    \begin{center}
        \includegraphics[width=0.9\textwidth]{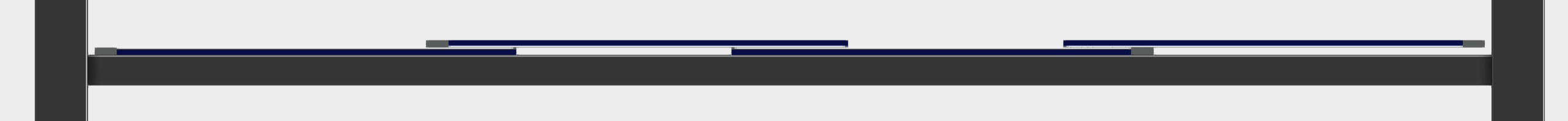}
        \caption{Cross section of a scintillator \layer in a \tower, showing the 4 \sublayers. Note the staggering between \sublayers, which prevents dead areas. 
        Cables will be routed to the edge of the module and up towards the DAQ on the roof.
        }
   \label{fig:4subplanes}
   \end{center}
\end{figure}

\begin{figure}[th!]
    \begin{center}
        
        \begin{tabular}{c}
        \includegraphics[width=0.8\textwidth]{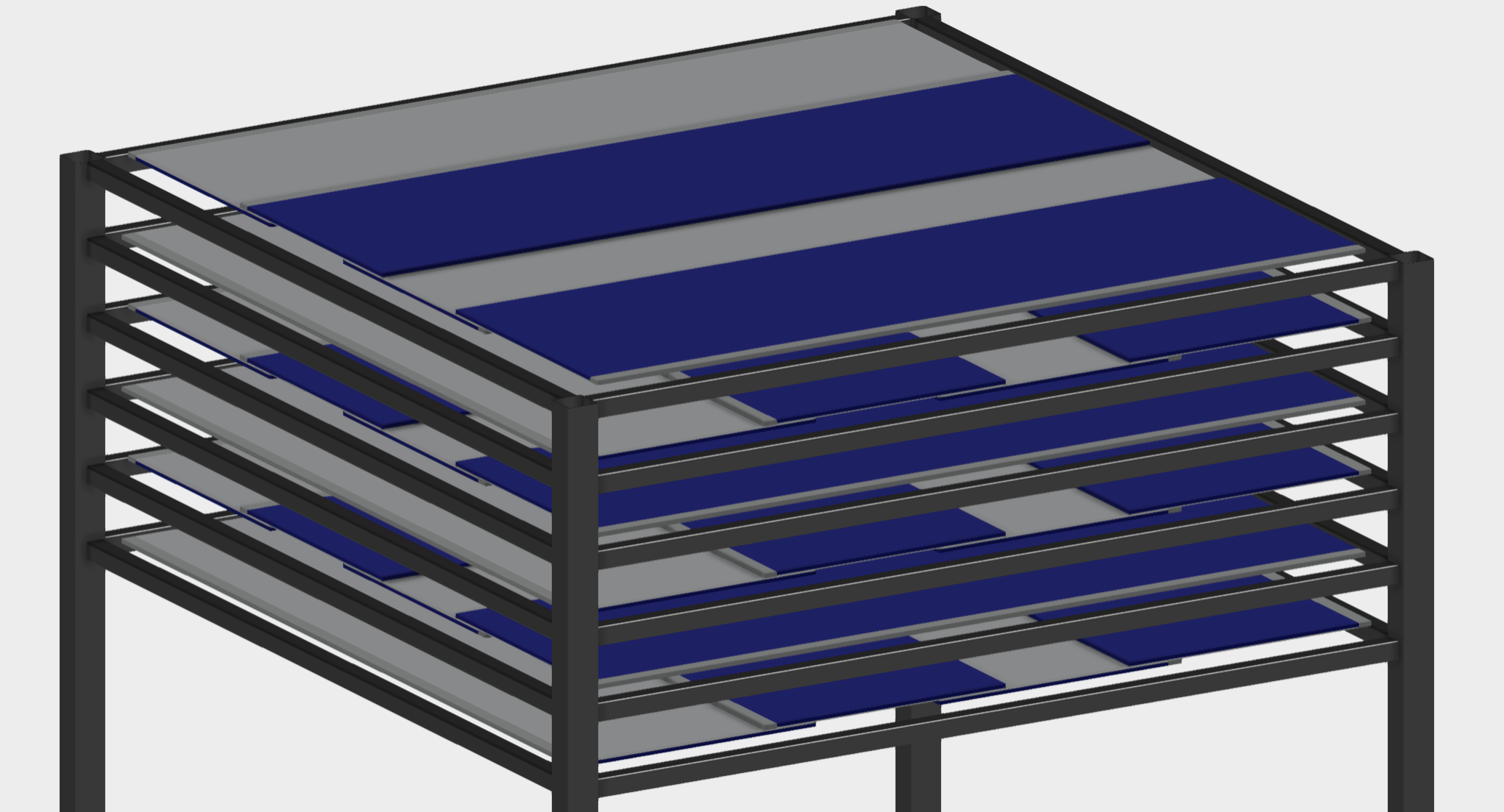}        

        \\

        \includegraphics[width=0.8\textwidth]{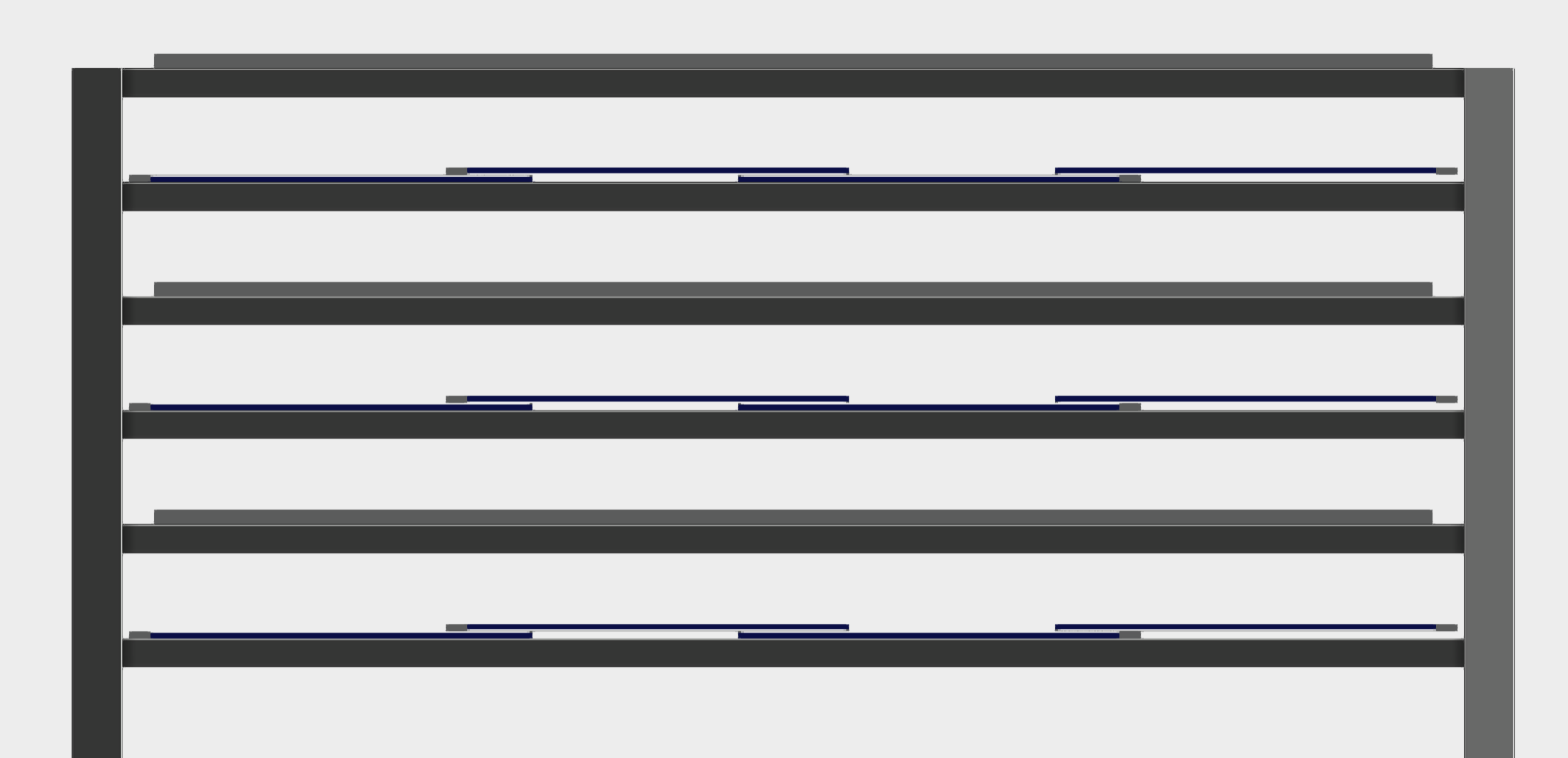}  
        
        \end{tabular}|
        \caption{Isometric view (top) and side view (bottom) of a \ceilingtrackingmodule comprising 6 complete  \layers, showing the four \sublayers.
        }
   \label{fig:planeiso}
   \end{center}
\end{figure}

Maintenance, if necessary, can be performed at whatever time is optimal for data taking, not limited to beam stops. 
The cross section of a scintillator \layer, as well as the honeycomb support plate, is shown in Figure~\ref{fig:4subplanes} . 
Cables will be routed from the \sublayer to the edge of the \tower and upwards to the data acquisition (DAQ) electronics mounted at the top of the \tower, see Section~\ref{s.DAQ}. 
Figure~\ref{fig:planeiso} shows an isometric view of the scintillator \layer, with the 4 \sublayers visible. Note that each \sublayer is light-tight and completely encased in aluminum.
%

%%%%%%%%%%%%%%%%%
\subsubsection{Structural Support}
\label{s.structural}
%%%%%%%%%%%%%%%%%

The honeycomb substrate on which the \barasss are mounted will provide significant stiffness to minimize sag. An entire $9~\textrm{m} \times 9~\textrm{m}$ detector layer will have attachment points for crane pickup to move the detector layer for installation in a \tower. The \tower has mounting bars that will support the \layers and \floorvetolayers, as well as horizontal and vertical \floorvetostrips to cover gaps between modules. The position and number of connection points will be informed by finite element analysis to ensure that the \layer sag is within tolerance. 

%%%%%%%%%%%%%%%%%
\subsubsection{Detector Installation and Integration}
\label{s.integration}
%%%%%%%%%%%%%%%%%

Construction will start at the front of the detector, closest to CMS, and proceed toward the back.
Each MATHUSLA \tower will be built in place in the experimental hall. A set of \tower vertical and horizontal support structures are first assembled at their final locations. The four \towers at the front each include additional support structures for their two respective \wallvetolayers. The four \walltrackingmodules at the rear of the detector are installed last.

Once the support structure for a \tower or \walltrackingmodule is in place,  utilities are dressed to the structures, which include DAQ low voltage and DAQ communications fibers. The \SiPM bias voltages will be generated locally, and no cooling infrastructure is foreseen. 

The various scintillator detector layers are built and tested in the adjacent assembly area of the experimental hall, then craned and attached to their respective support structures (see Section~\ref{s.civengrequiremenets} for a discussion on the use of cranes in light of the civil engineering constraints of the experimental hall).
Sensor layers would therefore be installed from the bottom-up.
After each detector layer is attached, it is connected to the utilities and read out by a portable DAQ test system to verify functionality. Following successful tests, the next layer is  installed. Once each \tower is completed, DAQ servers to read out the \ceilingtrackingmodules and associated sections of \wallveto and \floorveto are mounted on top of the \tower and on the floor respectively.

\Towers are very much self-contained, and therefore a phased startup of operation is possible. Meaningful physics data can start to be taken as soon as the first \tower is installed.

%%%%%%%%%%%%%%%%%
\subsection{Environmental Conditions}
\label{s.enviroment}
%%%%%%%%%%%%%%%%%
\subsubsection{Temperature Fluctuations}
\label{s.tempfluct}

Like other CERN above-ground buildings, the hall containing the MATHUSLA detector would have to be equipped with basic climate control systems to limit the temperature variation throughout the year to a reasonable range that is compatible with optimal detector performance. 

For example, if the maximum temperature variation througout the year is 15$^\circ$C, 
the only component sensitive to these temperature fluctuations is the \SiPM. The \SiPM has a $\sim $ 40 mV$/^\circ$C breakdown voltage change, and corresponding gain change for a fixed bias voltage. There are several ways to compensate for this variation, both actively~\cite{SIPM_active} and passively~\cite{SIPM_passive}. Compensating for this effect is discussed in Section \ref{s.RDSiPM}.
%%%%%%%%%%%%%%%%%
%%%%%%%%%%%%%%%%%
\subsubsection{Humidity}
\label{s.humidity}
%%%%%%%%%%%%%%%%%

Ducted air circulation will be required due to the vertically stratified geometry of the detector. 
The relative humidity in Geneva is generally about 80\%, and somewhat lower in the summer. To improve electronics and scintillator stability, the humidity of the hall interior will be controlled. 
This will be done by exchanging $\sim$5\% of the air per hour and requires an industrial desiccant dehumidifier. Suitable units are available from a number of manufacturers.

\subsection{Calibration}
\label{s.calibration}
%%%%%%%%%%%%%%%%%

Calibration is envisaged both to ensure the correct functioning and the gain of readout channels.  Three distinct levels of calibration are possible: electronic (via charge injection), \WLSF-level (via light injection into fibers within \sublayers) or physics-level (via offline or near-line analysis of cosmic ray data). The last of these can be performed offline using triggered data, while the other two will require dedicated calibration runs, some of which can potentially be conducting in a ``rolling'' manner across all detector channels without disrupting physics data taking.  Ideally, it is anticipated that it will be possible to calibrate the full detector several times a day, e.g. to be able to track possible gain variations due to temperature excursions.

The most likely sources of failure or gain drift are anticipated to be due to \SiPM gain changes due to temperature, bias voltage, or aging effects; \WLSF performance problems due to damage, aging, or optical coupling to the \SiPMs; and electronic connection problems.  It is therefore desirable to monitor both the light output from fibers and the \SiPM photoelectron gain. In principle, the \SiPM gain can be tested by performing a scan of the dark noise rate as a function of the trigger threshold in the vicinity of the single-photoelectron peak. However, pulse height measurement capability is also being considered for the readout electronics, either via a dedicated ADC, or Time over Threshold (ToT) measurement.  If this functionality is available, then gain calibration of the full system can be performed by measurement of a \SiPM ``finger plot'' when light is injected into \WLSF. 
We therefore anticipate using an 
LED pulser
to monitor the fibers and \SiPMs.

Gain calibration can also be performed directly from cosmic ray data simply by comparing hit rates in individual channels or by recording pulse amplitudes with the DAQ. However, the relatively low rate of cosmic rays in individual \mbars precludes using this as a ``real-time'' calibration system, and instead would more likely be useful, if needed, as an offline calibration of physics data. 
\newpage
%%%%%%%%%%%%%%%%%

%%%%%%%%%%%%%%%%%
\section{Electronics, DAQ \& Trigger}
\label{s.electronicsdaqtrigger}
%%%%%%%%%%%%%%%%%

The MATHUSLA plans for electronics, trigger, and data acquisition (DAQ) are based on design concepts that are very well-understood, to provide reliability and scalability. The front-end electronics are similar to commercially available circuits with optimizations for MATHUSLA. The trigger is similar to many recent experiments: a hardware-based Level-1 trigger (L1T) with a short and well-defined latency, and a high-latency software-based high-level trigger (HLT). Data acquisition using commercial off the shelf (COTS) hardware is parallelized by \tower and \walltrackingmodule to ensure functionality, scalability, and performance from the installation of the first detector \tower to the last \walltrackingmodule. Furthermore, the design plans can be readily adapted to accommodate detector layout changes that may occur. 

%%%%%%%%%%%%%%%%%
\subsection{Front-end electronics}
\label{s.frontend}
%%%%%%%%%%%%%%%%%

The front-end ASICs attached to each \SiPM feature two comparators: a low-threshold corresponding to roughly 0.5 photoelectrons (p.e.) for timing purposes and a high-threshold corresponding to roughly 1.5 p.e. A hit is defined as the coincidence of the two \SiPM signals, where at least one exceeds the higher threshold, while the time of the hit is given by the average times of the two low-threshold signals. The logical OR of the two high-threshold signals is used to reject much of the noise rate while maintaining high efficiency for real hits. The hit rate arises from two main sources: the transit of charged particles and dark noise.

\paragraph{Charged particle transit}
The hit rate due to charged particles is dominated by cosmic rays. For two scintillator \mbars connected by a \WLSF, the hit rate is $\sim$30 Hz.

\paragraph{Dark noise}
Hits can also arise from the random coincidence of near simultaneous noise in two \SiPMs. The goal is to maintain a dark rate at the same level or lower than the cosmic rate of 30~Hz. The \SiPMs are separated by roughly 37~ns, determined by lab measurements with realistic choices of \WLSF technology and length ($\sim5$~m). This means that two dark counts within 37~ns of each other can mimic a signal. For the goal of achieving a 30~Hz  background rate, each \SiPM can tolerate an individual noise rate of $\sim$20 kHz. The \SiPMs have two counting thresholds; the low threshold (0.5 p.e.) has a noise rate around $\sim$100 kHz while the high threshold (1.5 p.e.) has a noise rate of $\sim$500 Hz. Coincident noise hits can occur if a low threshold hit is within a 74~ns window centered on the initial high threshold hit. If the trigger is set for a high threshold hit on one \SiPM with a coincident low threshold hit on the other \SiPM within $\pm37$~ns the random noise coincidence rate is 7.4~Hz, well below the 30~Hz budget. 

%%%%%%%%%%%%%%%%%
\subsection{Aggregator}
\label{s.concentrator}
%%%%%%%%%%%%%%%%%

The signals from each ASIC are transmitted along a PCB to an FPGA, one for each \layer, \wallvetolayer and \floorvetostrip.
The signals from \trackingmodules in the ceiling or rear wall need to be processed quickly as they also constitute the trigger system, and are routed to FPGAs on the roof of the respective \tower or \walltrackingmodule. 
The signals from the \veto can be aggregated into separate FPGAs on the floor or the front wall and collected for read-out. 
Due to interconnecting bandwidth limitations and CMS latency requirements, the aggregation must be done quickly and in a compact representation. 
Feasibility has been demonstrated in a  preliminary design.

%%%%%%%%%%%%%%%%%
\subsection{DAQ}
\label{s.DAQ}
%%%%%%%%%%%%%%%%%

All hits, i.e. the coincidence of both \SiPMs from a fiber, are read out and stored temporarily in a disk buffer maintained by different DAQ servers assigned to different portions of the detector.
Assuming 20 bytes per hit, and applying a factor of $\times 2$ to the cosmic ray rate to  include dark count, each \barass produces a datarate of about 40 KB/sec. 
The \ceilingtrackingmodule and \walltrackingmodule would each be assigned one of 20 DAQ servers accepting their datarate of 7MB/sec. 
The \floorveto can be organizationally broken up into sections associated with each \tower~-- for \towers in the interior, this includes its two \floorvetolayers, 2 horizontal \floorvetostrips, and one \columndetector, producing 6 MB/sec. The rate is lower for \towers on the periphery, meaning one of the DAQ servers monitoring the \floorveto of a \tower at the front of MATHUSLA could additionally be assigned the entire \wallveto, which only produces 2MB/sec. 
Overall, this modularity allows for scaling and staging.

Ignoring the minor variations in datarates, this means the entire MATHUSLA detector outputs 36 datastreams each under 7MB/sec, or about 0.6 Tb/day. This rate is easily accommodated by COTS servers, and readily available storage devices can buffer one or more days of hits.

The HLT system uses the buffers to collect all hits from the entire detector within a  $\pm 0.5~\mu$s time window around each L1T time stamp  (roughly double the speed of light time across the detector), and sends them to permanent storage. 
This Permanently Saved Data (PSD) will constitute at most 0.1\% of the full detector data rate.

An additional server acts as  accumulator for the PSD selected by the HLT, accepting an average data rate of 0.3 MB/s from the whole detector. 
This data, at most about 8 TB per year, is then sent to CERN centralized storage for long-term archival, remote access and offline analysis.

%%%%%%%%%%%%%%%%%
\subsection{Trigger}
\label{s.trigger}
%%%%%%%%%%%%%%%%%

The MATHUSLA trigger is spit into two primary layers, a Level-1 (L1T) trigger which prompts the detector readout, and a High Level Trigger (HLT) which will further refine the selection and send data to permanent storage.

This trigger design is based on a detailed design study that was conducted when the benchmark MATHUSLA geometry was $(100~\mathrm{m})^2$, with 100 tower modules utilizing only ceiling trackers. Given that the current proposal has near-identical detector components (with the exception of adding \walltrackingmodules in the rear, which are electronically identical to the \ceilingtrackingmodules) and significantly smaller size, the same design  would also work for the 40m geometry presented in this report. 
The \walltrackingmodules can be integrated into this trigger scheme, but the exact technical details will require additional studies. These investigations will also re-visit the original trigger principles to ensure they are still optimal for the current design.

\subsubsection{Level-1 trigger}
The Level-1 trigger is formed from a coincident signal from the six \layers in the ceiling or rear wall of MATHUSLA. 
With a separation of 0.8 m between adjacent layers, a hit time resolution of 1~ns ensures a negligible probability of a downward going track masquerading
as an upward-going track due to Gaussian timing fluctuations. We search for spatially aligned hits compatible with an upward going track in a \ceilingtrackingmodule and its eight nearest neighbors that form a 3$\times$3 grid, where upward vs downward direction is given by the relative timing of hits. 
It is expected that upward particles exiting the rear wall of the detector can be similarly triggered on by searching for spatially aligned hits compatible with tracks going upwards and outwards in groups of 3 adjacent \walltrackingmodules.

Using hits from a given \ceilingtrackingmodule and its eight neighbors in a 3$\times$3 pattern, tracks can be inclined at zenith angle of $\arctan(30$m$/5$m$) = 80^\circ$.\footnote{Requiring at least one track in the DV of an LLP decay to have 4 hits confined within a 3$\times$3 group of \towers only reduces the geometric signal acceptance by at most a few percent~\cite{Curtin:2023skh}, which has negligible impact on the physics goals of MATHUSLA.}
The identical algorithm\footnote{\Towers at the edge of the MATHUSLA array differ only in receiving no hits from their nonexistent neighbors.} is applied in parallel to all \trackingmodules to ensure that trigger latency is insensitive to whether there are only a few  \towers and \trackingmodules during early installation or once the full area  has been instrumented. In order to eliminate duplicates, tracks that cross multiple \trackingmodules are assigned to the \trackingmodule where it crosses at a reference height (or distance from the rear wall, for the \walltrackingmodules). Upward-going tracks from \emph {all} \trackingmodules are used to search for a 4-dimensional vertex. The spatial position of the vertex must be consistent with the MATHUSLA decay volume to be accepted.

MATHUSLA intends to implement a number of Level-1 triggers for physics and calibration purposes. The primary MATHUSLA L1T is the 4-dimensional, i.e. space and time, overlap of two or more upward going tracks, which we will refer to as "LLP L1T". MATHUSLA will also implement a single upward-going track trigger. This is expected to be predominantly high-$p_T$ muons from LHC collisions, which are expected to have an incident rate of $\mathcal{O}(\mathrm{Hz})$ on the detector.
They are useful to validate MATHUSLA and for calibration and crosscheck with CMS. A random trigger will also be included in MATHUSLA L1T to monitor background conditions. The rate of cosmic rays on the full MATHUSLA detector
is 300~kHz. 
Therefore, each random trigger will contain an average of
%2.8 
$\lesssim 0.5$ 
cosmic rays in a 
%$\pm 1~\mu$s 
$\pm 0.5~\mu$s 
window. 

Finally MATHUSLA is investigating a number of other interesting physics triggers. The most important examples are 
potential triggers from CMS. Their timestamps will be written to the trigger stream in order to select hits for permanent storage.
The logical-OR of all these L1T signals constitute the "Global L1T". They are recorded to a L1T disk buffer, which is common to the entire MATHUSLA detector. The recorded information includes the type of trigger and a universal time stamp.

\subsubsection{High Level Trigger}
\label{s.HLT}
Using the timestamp information from the two sets of disk buffers (hits and L1T), the HLT system selects hits that are within a
%$\pm 1~\mu$s
$\pm 0.5~\mu$s
window of each Global L1T and writes them to archival storage for Permanently Saved Data (PSD), where the window size is derived from the  time of flight (at the speed of light) from one corner of MATHULSA to another. This ensures that all hits coming from an LLP decay are captured.

\subsubsection{CMS Trigger}
\label{s.CMStrigger}

While the trigger and DAQ described so far is sufficient to discover an LLP signal if one exists, understanding such a new physics signal would benefit from having information from the CMS detector in order characterize the event's 4$\pi$ topology~\cite{Barron:2020kfo}. We therefore generate a "CMS L1T" signal, transmitted to CMS where it will be treated as a force-accept trigger for the CMS high-level trigger. Since the propagation speed of the candidate LLP is a priori unknown, it is not possible to pin-point a specific bunch crossing (BC). The range of BCs will depend on the slowest propagation speed that we want to include in MATHUSLA acceptance. 

The primary implementation challenge is forming the trigger within the latency requirements of the CMS trigger system of $\sim 9~\mu$s. Architecturally, the trigger is broken into two phases.  The first is the local track-finding within each of the 16 groups of 3$\times$3 (or fewer) \ceilingtrackingmodules. (Future trigger design studies will extend this to the  \walltrackingmodules.)  
The second is the collection and vertexing of all tracks from all the \towers into one global vertex finder. A detailed accounting of all the latencies, such as signal transit times and aggregation times, leaves $\sim 2.5~\mu$s to do the track-finding and vertexing.  A prototype of the track-finding, which is the most compute intensive, indicates this can be done in $2.0~\mu$s using a large FPGA, thus leaving $0.5~\mu$s for the vertexing. 

Based on discussions with experts in CMS, the plan is for MATHUSLA to provide a stream of CMS L1T signals corresponding to the BC range we wish to see recorded. There are several constraints on such a stream. The earliest signal must remain within the CMS Level-1 trigger latency. In fact, this is the only pertinent requirement on MATHUSLA Global L1T latency. The latest signal corresponds to a LLP propagating at the speed of light from CMS IP to the LLP decay vertex in MATHUSLA. The sum of all such signals must be sufficiently low rate to not have noticeable impact on CMS operations. We envision "CMS L1T" to be the logical-OR of the MATHUSLA "LLP L1T" and a heavily prescaled version of single upward track in MATHUSLA. For the former, we plan to send the longest stream of BC signals that can be accommodated in order to maximize acceptance of slow LLP. And for the latter, we plan to request one or two BC's since single upward going particles are expected to be predominantly high-energy muons. They are crucial to establish the relative position and timing between CMS and MATHUSLA. And since CMS will likely trigger on such high-energy muons anyway, this is expected to contribute very little net rate for CMS.

An important opportunity is being provided by the CMS scouting buffer, which is part of the Phase-2 CMS Level-1 trigger upgrade~\cite{CERN-LHCC-2020-004}. 
MATHUSLA's call to the CMS trigger system would therefore include a request to save some part of this scouting buffer to permanent storage.
This would allow for the study of production events for very non-relativistic LLPs, albeit with less information than the full event record written by the hardware trigger.

%%%%%%%%%%%%%%%%%
\subsection{Slow Controls and Monitoring}
\label{s.monitoring}

Individual SiPM voltages need to be monitored and trimmed to correct for gain and temperature variations and in response to rolling calibrations which are anticipated to take place several times per day. Temperature and bias voltage determine the operating conditions of the detector, hence these quantities will be actively monitored and tracked in time, and this functionality will be integrated into the front end boards.   
The hall environmental conditions, e.g. temperature and humidity, will also be monitored but this functionality is independent of the MATHUSLA DAQ system.

\newpage
%%%%%%%%%%%%%%%%%
%%%%%%%%%%%%%%%%%
\section{Computing}
\label{s.computing}
%%%%%%%%%%%%%%%%%

MATHUSLA generates about 8 TB/year of permanently saved data, as described in Section~\ref{s.DAQ}.
This data, in the form of all hits within the whole detector in the time window around a L1T signal, will be sent to CERN centralized data storage for long-term archival and remote access.

The MATHUSLA collaboration expects to perform various types of analyses on this data.
The dominant computing workload will likely be taken up by data quality monitoring, alignment, calibration, and background simulations. 
Reconstruction of tracks and vertices in the full detector can then be performed offline using the permanently saved data from the detector.
This will enable the most important objective of MATHUSLA, once the collaboration understands the detector's characteristics: the search for BSM LLP decays.

For these analyses, the relevant data can be accessed remotely and transferred to various computing facilities at CERN and other participating institutions.
Due to the relative simplicity of the detector, its data, and associated analyses, as well as the expected low rate of LLP events, the computing requirements for these analyses are comparatively modest, especially when compared to many other high-energy physics analyses.

We are studying the best way to implement public sharing of MATHUSLA data. Our goal is to make a usefully formatted subset of the data publicly accessible, possibly with some time delay to facilitate initial analyses by the MATHUSLA collaboration, allowing independent research groups to conduct both LLP searches and unforeseen physics investigations.

\newpage
%%%%%%%%%%%%%%%%%
%%%%%%%%%%%%%%%%%
%%%%%%%%%%%%%%%%%
\section{Engineering}
\label{s.civeng}
%%%%%%%%%%%%%%%%%

Civil Engineering (CE) and infrastructure typically represent a significant proportion of the overall implementation budget for projects at CERN. For this reason, CE studies are of critical importance to ensure a viable cost-efficient conceptual design. This chapter provides information on the engineering design concept for the MATHUSLA detector structure, and discusses several civil engineering aspects that were studied to evaluate a suitable position of the new facility to house the detector on CERN-owned land near CMS.

\subsection{Location}

\begin{figure}
    \begin{center}
        \includegraphics[width=0.7\textwidth]{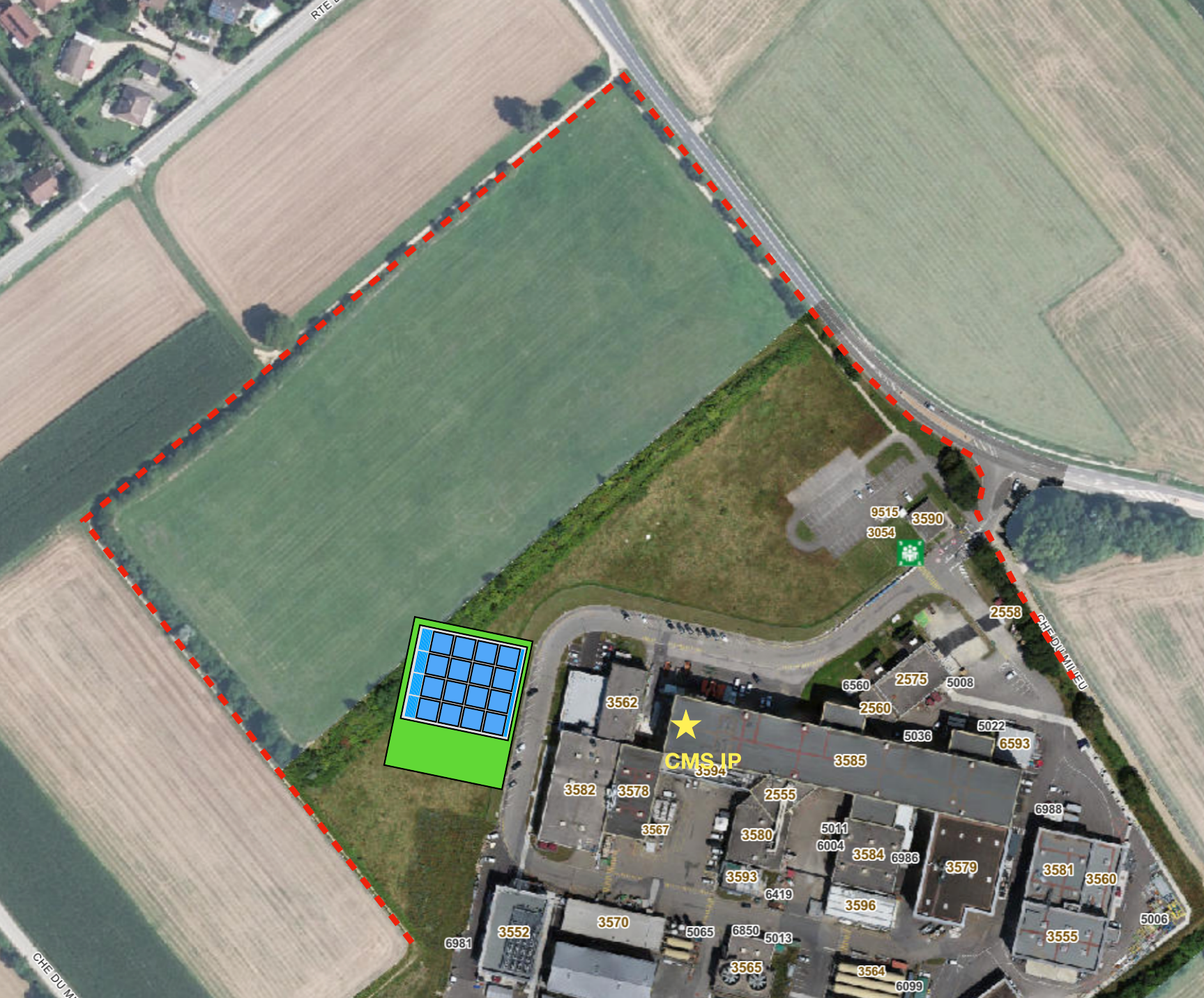}
        \caption{Proposed location of MATHUSLA and existing CMS site, the dashed red line indicating the CERN non-fenced boundary.}
   \label{fig:CE_1}
   \end{center}
\end{figure}

The Civil Engineering studies were based on the requirement of implementing the MATHUSLA at CERN near one of the LHC's interaction points. 
Following an initial study of the existing LHC infrastructure and existing geological conditions, the CMS Interaction Point 5, on CERN's French site, was identified as the most suitable location to have the proposed facility entirely located within existing CERN land. The proposed location and its relation to the CMS IP5 is shown in Figure ~\ref{fig:CE_1}. However, we note that the precise location could be shifted within the proximity of CMS without significantly impacting MATHUSLA's physics performance~\cite{Alkhatib:2019eyo}, should later engineering studies identify compelling reasons to change the location of the experimental hall.

\subsection{Geology and Hydrology}

\begin{figure}
    \centering
    \includegraphics[width=0.6\textwidth]{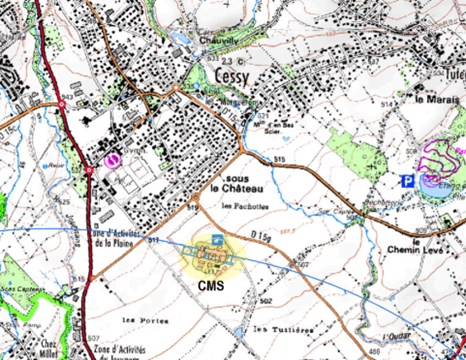}
    \caption{Figure showing the location of the CMS Interaction Point 5 in relation to Cessy village. 
    }
    \label{fig:CE_2}
\end{figure}

The proposed location of the detector building is situated within the Geneva basin, at the foot of the Jura Mountains, in the commune of Cessy, France. The terrain is about 0.5~km southeast of the Cessy village (see Figure~\ref{fig:CE_2}). the main land use in the surrounding area is characterised by agricultural land, wooded area and villages. The terrain slopes gently away from the Jura Mountain towards Lake Geneva in a southeast direction. Cessy village is located between approximately 560~m and 520~m above mean sea level (msl) and the proposed location is at an average of 510~m above msl.

Ground investigations have previously been carried out in this area for the civil engineering works of the CMS and HL-LHC projects.\footnote{
%Ref. 1 
GADZ HL-LHC and 
GIBBS-SGI-Geoconsult joint venture (1996), LHC civil engineering consultancy services package 02; 
%
%Ref.2 
GADZ SA-High Luminosity LHC Point 5 Geotechnical report.} 

\begin{figure}
    \centering
    \includegraphics[width=0.95\textwidth]{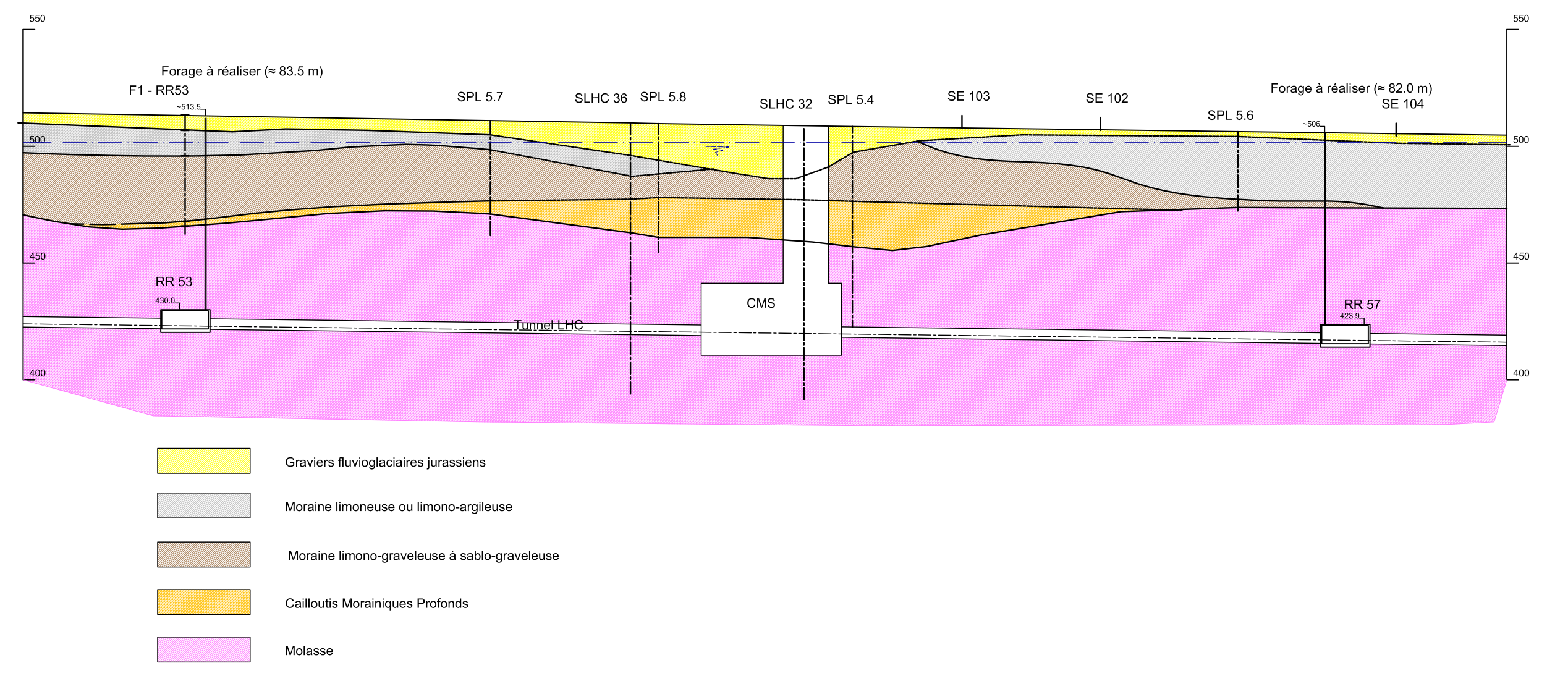}
    \caption{Stratigraphic profile at CMS } 
    \label{fig:CE_3}
\end{figure}

In general, the studies showed two main geological materials characterising the area: moraines and molasse, as shown on Figure~\ref{fig:CE_3}. 

The moraines are the youngest deposits compromising fluvio-glacial material, boulder and gravel moraine and fluvial out-wash material, all deposited during the Wurmian glacial period. The formations are very heterogeneous; however three main hydro stratigraphic units can be discerned that control the hydrogeology in the area. 

The fluvio-glacial deposits comprise of sandy and silty gravels of Jurassic and Alpine rock types. The units are compact to very compact, with clasts generally between 3-5~cm and occasionally up to 20cm. Sandy gravel are considered to be generally permeable and form a shallow aquifer (upper aquifer) at approximately from 22~m on the north side to 10~m on the south side of the CMS complex. On the contrary, the silty gravels are characterised by a low permeability. 

A second, deeper aquifer (lower aquifer) is found at approximately 30~m depth in the fluival outwash deposits. The material consists of silty to sandy gravels with clasts varying from 5 to 20~cm. This unit varies in thickness from 12~m to 21~m over the area and is most likely fed by underground inflow from the Jura Mountain and leakages from the upper aquifer. The aquifers are separated by a semi-impermeable layer with a thickness of 5-10~cm on the north side and 19-21~m on the south side of CMS.

The lower aquifer is located near the boundary with the molasse. This material consists of a complex alternating sequence of marls, sandstones and formations of intermediate composition with an average hydraulic conductivity less then $1 \times 10^{-8}$~m/s. The Molasse is thus considered impermeable.

\subsection{Engineering Concept of MATHUSLA Detector structure}
\label{s.engconcept}

In designing the engineering concept for the detector structure shown in Figure~\ref{fig:MATHUSLAstructure}, various ways of supporting the \ceilingtrackingmodules were considered. The number of vertical support columns, in particular, can be regarded as a free parameter. In principle, it is desirable to minimize the number of  columns, since they make gaps in the floor detector and provide material interaction targets for cosmic ray and LHC muon backgrounds, which has to be addressed by wrapping them in \columndetectors.
On the other hand, reducing the number of vertical supports requires the horizontal support beams in the ceiling to be much stronger to support the $(9~\mathrm{m})^2$ \layers, weighing more than 3 tons each, over spans of 20m or more. This can easily make the amount of material in the ceiling assembly so large that it starts to interfere with the propagation of charged particles between \ceilingtrackingmodules, therefore seriously hampering MATHUSLA's ability to reconstruct particle tracks and vertices. Furthermore, since about a meter of clearance between \trackingmodules is required for maintenance access, thicker horizontal beams also result in overall larger gaps, further reducing detector efficiency. 
Therefore, since vertical supports constitute a very small fraction of the overall detector volume that can be instrumented to further reduce its negative effects on LLP searches, it was decided to minimize the required horizontal support strucuture by maximizing the number of vertical support columns, given $(9~\mathrm{m})^2$ \layers in the \trackingmodules. This led to the chosen design in Figure~\ref{fig:MATHUSLAstructure}.

Having vertical supports every $\sim$ 10m has the further advantage of organizing the detector into self-sufficient \towers, and  makes design of the hermetic  \floorveto layout relatively straightforward, utilizing the \floorvetostrips to instrument gaps and vertical support columns. 

We note that hermiticity of the \wallveto with respect to LHC muons requires the inner \wallvetolayer to be no further than about a meter from the front \towers and their associated \floorvetolayers. However, given that these sensors can be supported by structural supports between the two \wallvetolayers, we do not anticipate this to present any serious design problems.

\subsection{Civil Engineering}
\label{s.civengrequiremenets}

The geometry of MATHUSLA40, and therefore the requirements on its enclosing facility, were chosen to minimize infrastructure costs while maintaining physics reach. 
MATHUSLA, with its overall footprint of $\sim$ 50m $\times$ 45m and height of 16m, alongside a neighboring space for assembly with a width of $\sim$ 20m can be accommodated by a $\sim$ 50m $\times$ 70m experimental hall equipped with a ceiling crane system for detector assembly.

\begin{figure}[]
\begin{center}
\includegraphics[width=\textwidth]{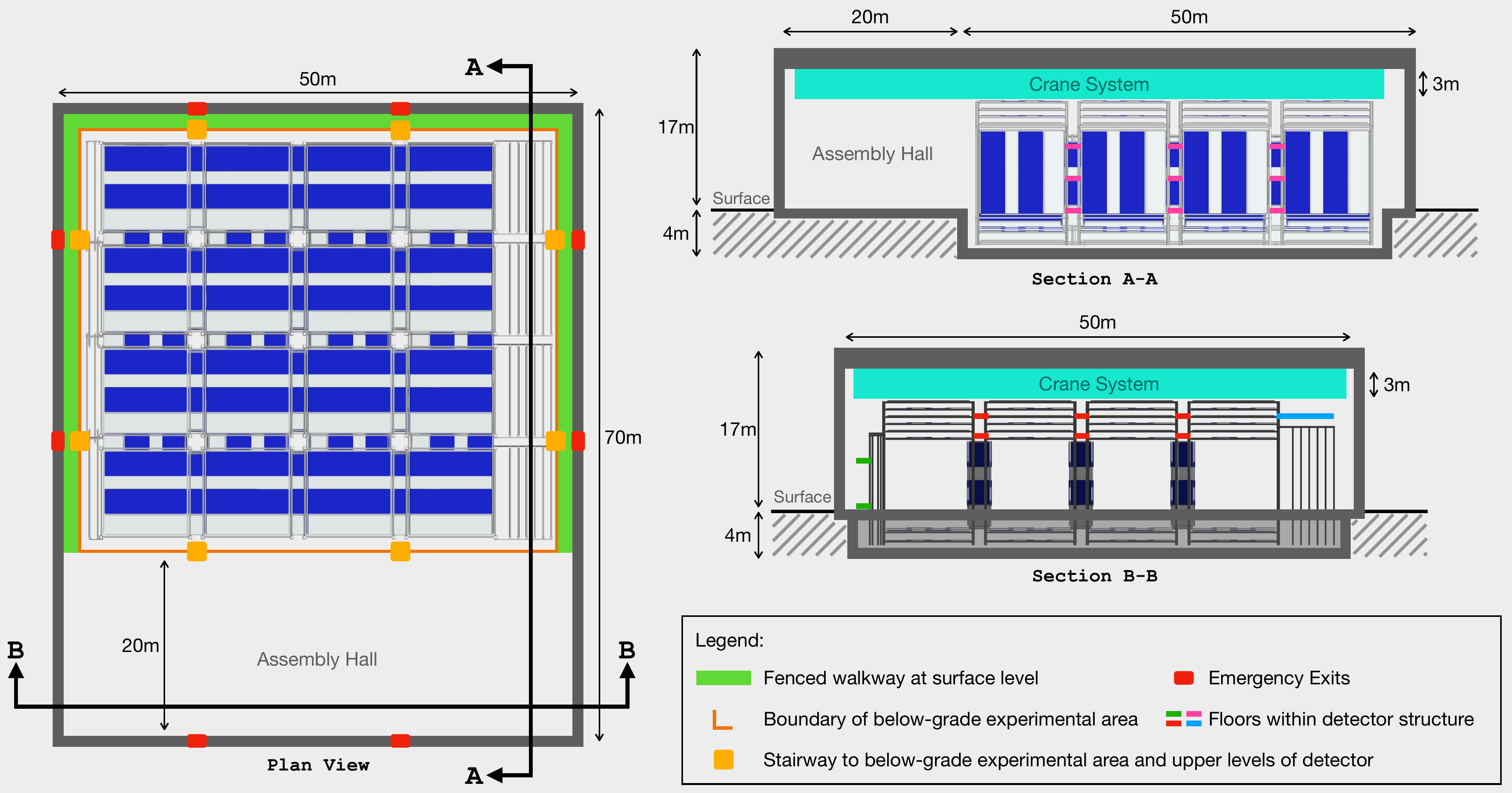}
\end{center}
\caption{
Sketch of the civil engineering concept for the MATHUSLA experimental hall (all indicated dimensions approximate). The total footprint of the building is 50m $\times$ 70m, with a total exterior height of 17m. The detector (footprint 40m $\times$ 45m not counting access stairs, total height 16m) is situated in a below-grade experimental area (within orange boundary in top view), requiring at most 4m of excavation. This ensures sufficient vertical space for an interior crane system utilizing low-headrooom hoists (cyan) below the roof. Detector components are prepared for installation in the adjacent 20m wide assembly hall, then lifted into place with the cranes. 
Several emergency exits (red rectangles), and stairways  to the below-grade experimental area and the upper levels of the detector (orange squares), provide means of egress for emergency evacuation. Red, green, blue and pink lines in the Section A-A and B-B views indicate where floors could be placed within the detector structure for installation, testing and servicing of individual \layers, \wallvetolayers and the \walltrackingmodules respectively, as well as overall \tower and detector integration. (The blue floor might have hatches to access the \walltrackingmodule below.)
} 
\label{fig:civil_engineering_concept}
\end{figure}

Since vertical space is at a premium, the ceiling crane system will utilize low-headroom hoists, four of which can be synchronized to pick up a single horizontal \layer from the assembly area using hardpoints at the corners, lift the \layer as high as possible below the ceiling, and move it into place for installation in the detector structure. Vertical \wallvetolayers can be similarly transported, except that after being picked up in the horizontal orientation, the four hoists would move to orient the layer near-vertically for installation in the front wall. 
The relatively low weight of the sensor layers (about 4 tons for the \layers and 7 tons for the \wallvetolayers) allows entry-level systems like the FX Crane System by R\&M or the DVR Crane System by Demag to fulfill these requirements.
These systems can be configured to only require about 1m of vertical space for the crane components themselves. Factoring in horizontal support girders, which would have to be sizable to bridge the 20-40m spans of the detector (depending on precise number and location of structural vertical supports in the building interior, which can be placed within MATHUSLA at the corners of some \towers, surrounded by \columndetectors), as well as the vertical extent of a suspended horizontal \layer itself, we conservatively assume that the total vertical height between the underside of the roof and the top of the MATHUSLA40 detector structure that is required for the crane system is at most 3m.

The vertical extent of the roof structure can be estimated by comparison to similar buildings at CERN. For example, Warren trusses can support the roof across the required horizontal span with a vertical height requirement of about 2m, especially if vertical structural supports are included at the corners of some \towers as mentioned above.

In total, we are therefore led to a conservative estimate of 21m for the total vertical extent of the MATHUSLA40 experimental hall in the area occupied by the detector structure.
This can be accommodated by excavating the floor of the experimental hall on which the detector structure is placed to a depth of about 4m below grade.
A sketch of this conceptual design for the MATHUSLA experimental hall is shown in Figure~\ref{fig:civil_engineering_concept}.

\subsection{Cost Estimate}
\label{s.civengcost}
A clean-sheet design study for the MATHUSLA experimental hall will be undertaken in the future, but its total infrastructure cost can be roughly estimated by rescaling the itemized cost estimate for a previous 100m MATHUSLA design that was prepared by CERN engineers, with different aspects of the cost scaling with either the total building area, circumference, vertical height, or detector area. 
This leads us to estimate a total building cost of $\sim$ 21.5M CHF, taking into account the significantly increased cost associated with building on 'made ground' which comprises the identified site. A conservative estimate for the support infrastrucuture required by the experiment (electrical, HVAC, etc) would add another 25\%.
The accuracy of the estimate is considered Class 5 - Concept, which could be 20-50\% lower or 30-100\% higher (in line with AACE International's best practice recommendations as has been used for previous CERN project). The study is at an early feasibility stage so until the projects requirements are further developed, it is suggested that the maximum range be adopted i.e -40\% to +75\% for CE costs.

\newpage
%%%%%%%%%%%%%%%%%
\section{Safety}
\label{s.safety}
%%%%%%%%%%%%%%%%%

A preliminary hazard analysis was made following the Launch Safety Discussion process of the CERN HSE Unit Project and Experiment Safety Support activity.  This is documented in EDMS 2766696,  which serves as an interface document to the CERN HSE Unit and will be used to further develop the project safety requirements in detail.   

At this stage, two major areas of concern for the construction and operation of the MATHUSLA detector have been identified:

\begin{enumerate}

    \item \textbf{Working from height: } MATHUSLA's air-filled LLP decay volume has a vertical height of about 11 meters, with the entire detector structure extending up to 15m. As discussed in Section~\ref{s.RDassembly}, each 9~m $\times$ 9~m scintillator \layer will be assembled on-site in the assembly hall, then moved by crane to vertical support structures that will have been built-in-place for each \tower. Integration and commissioning of each detector-\layer necessitates the presence of (temporary or permanent) floors that can be installed between \towers at varying heights as needed.  Floors can also be installed at various heights between the \walltrackingmodules for easy access, and adjacent to the \wallveto, see Figure~\ref{fig:civil_engineering_concept} (side and back view).
    The personnel performing the installation will follow standard CERN safety procedures for operating at height, including appropriate training and the wearing of fall protection and other personal protective equipment. 
    
    The floors will greatly simplify installation procedures for the  \layers. Appropriate PPE would be worn during all such work.  \Layers are designed to be fully integrated and serviced from the gaps between \towers, 
    and it should never be necessary to access the center region of a \tower directly. 

    Stairways will be permanently in place to reach the various floors from the periphery of the detector, see Figure~\ref{fig:civil_engineering_concept} (top view).

    \item \textbf{Scintillator Fire Safety:} 
    MATHUSLA plans to use $\sim$ 160 tons of polystyrene based scintillator. The material is flammable and great care needs to be taken to avoid fire. We plan on the following design features to mitigate this concern. 
    \begin{enumerate}
        \item Equipment near the scintillator will not present ignition risks. For example, front-end electronics are low power, and conduction via printed circuit boards will keep them well below the danger level. 
        \item Higher-power electronics, such as FPGAs, are in their own metal enclosures. 
        \item All scintillators will be jacketed in aluminum skins to reduce the risk of accidental heat sources and ignition.
        \item 
        The striated layout of the MATHUSLA detector makes fire control systems in the ceiling of the detector hall ineffective. Each  \layer in each \tower will therefore be outfitted with a self-contained local fire protection system. This local system will include smoke detectors; temperature sensors; at least one wide-angle IR camera for remote monitoring; and a suitable fire suppression system.
        This will be implemented with the advice and approval of CERN HSE Unit.

        \item Emergency egress: The layout of the detector will be designed to provide routes that are as short as practically possible to safety egress points. 
        Stairs to various parts of the detector structure provide rapid access to emergency exits on the periphery of the housing building,      see Figure~\ref{fig:civil_engineering_concept} (top view), with the floor situated at grade and freely traversable along the gaps between \towers. 
        For the worst case we foresee approximately 40~m  path-length, including traversal of stairs, to the exits from any point on the floor, or from any working space between the \trackingmodules to an access point on the perimeter. We note that this is in line with the normal maximum path of 40 meters for the LHC P5 site. 
        %e will work with the CERN HSE unit to address this issue. 
        
 \item The Project will produce, in collaboration with the concerned safety actors (e.g., DSO / LEXGLIMOS) of the relevant CERN Departments, a fire risk assessment of the planned content of the facility (activities, equipment and storage), to be later added to the safety file of the facility, and showing the adequacy of the fire safety measures proposed.
       
    \end{enumerate}

\end{enumerate}

We list other areas of concern below, but the associated hazards are judged to be of low severity or completely absent.
\begin{enumerate}    

    \item Mechanical safety. MATHUSLA's operation will involve no moving parts. During installation, cranes will be used to lift tracking \layers into place in the \towers and vertical \walltrackingmodules and the \wallveto. 

    \item Electrical safety. Custom electronics will have appropriate current limitations. The readout electronics will be low voltage, and will not provide a shock hazard. 
    
    \item SIPM high voltage - bias voltage. The SiPM will be biased at about 45~V. The current per SiPM will be limited to less than 1~$\mu$A.
    
    \item Seismic activity. In consultation with CERN civil engineering, the hall and detectors will be designed in accordance with the relevant seismic activity accelerations accelerations and European Standards. Document EDMS No. 1158454 details parameters needed for the design.

    \item Gas hazards. MATHUSLA's operation does not rely on the handling of any explosive or non-explosive gas. 

    \item Cryogenic or vacuum safety. MATHUSLA's operation will not employ any cryosystems or vacuum systems. 

    \item Chemical safety. MATHUSLA's operation will not employ any hazardous chemicals except for those that may be a part of the fire suppression systems. 

    \item Non-ionizing radiation safety. MATHUSLA's operating environment is the surface adjacent to CMS, radiation hazards are therefore completely absent.

\end{enumerate}
\newpage
%%%%%%%%%%%%%%%%%
%%%%%%%%%%%%%%%%%
\section{Cost and Schedule}
\label{s.costschedule}
%%%%%%%%%%%%%%%%%

We estimate the detector cost in Canadian Dollars based on experience from R\&D and quotes from vendors. Note that this is separate from the cost of the MATHUSLA experimental hall given in Section~\ref{s.civengcost}.

For the physical construction of the detector structure in the experimental hall, and installing the sensor modules in their appropriate positions in the structure using the ceiling crane system, we conservatively assume a labour cost that is 100\% of the material cost of the structure. The salary is calculated assuming eight postdocs (annual salary \$60k each), eight graduate students (annual stipend \$30k each) and one part-time project manager (annual salary contribution \$100k) for seven years. The travel cost assumes one trip to CERN (\$5k) each year, for each post-doc and each grad student. The annual maintenance and repair cost is estimated at 2\% of the sum of detector fabrication costs and readout electronics costs.

\begin{table}[hpbt]
\centering
\begin{tabular}{lr}
\hline
\multicolumn{1}{r}{}          & \textbf{Total {[}k\$ CAD{]}} \\ \hline
\textbf{Detector Fabrication}  & \textbf{\$19,179}        \\
\hspace{2mm} Scintillator bars               & \$4,914                  \\
\hspace{2mm} WLS fibers                      & \$2,859                  \\
\hspace{2mm} SiPMs                           & \$2,008                  \\
\hspace{2mm} Aluminum casing                   & \$5,556                  \\
\hspace{2mm} Glue                           & \$2,342                  \\
\hspace{2mm} Assembly \& test equipments      & \$1,500                  \\
\textbf{Electronics}           & \textbf{\$10,798}         \\
\hspace{2mm} Frontend                       & \$996                    \\
\hspace{2mm} Data acquisition               & \$5,378                  \\
\hspace{2mm} Cables                          & \$1,793                  \\
\hspace{2mm} Miscellaneous other components                           & \$1,992                  \\
\hspace{2mm} Server \& network             & \$640                    \\
\textbf{Detector installation} & \textbf{\$7,590}     \\
\hspace{2mm} Support structure material cost                & \$2,100     \\
\hspace{2mm} Physical installation labour cost                & \$2,100     \\
\hspace{2mm} Sensor shipping \& handling         & \$3,112     \\
\hspace{2mm} Sensor assembly \& testing    & \$278     \\
\textbf{Detector operations \& project management}    & \textbf{\$6,300}     \\
\hspace{2mm} Salaries                       & \$5,740     \\
\hspace{2mm} Travel                         & \$560     \\ 
\textbf{Maintenance \& repair}    & \textbf{\$1,477}     \\ \hline  
\textbf{Total}                 & \textbf{CAD \$45,344}     \\
\textbf{With 35\% contingency}   & \textbf{CAD \$61,214.}     \\
\hline
\end{tabular}
\label{tab:cost}
\caption{Summary of cost estimation for MATHUSLA-40 in CAD. The total including contingency corresponds to $\approx$ 40M €, or 38M CHF, at the time of writing.}
\end{table}

In the immediate future, MATHUSLA partner labs will engage in the additional R\&D needed to finalize some outstanding aspects of the detector design, including a sufficiently fast trigger system to supply a CMS L1 signal. 

Within 1-2 years of CERN approving the MATHUSLA experiment, international partner labs (likely in Canada, Europe and/or Israel) will begin setting up assembly lines for the construction and quality-testing of \barasss. In their first year of operation, these assembly lines will produce all the \barasss needed for a \emph{single} tower module and optimize their processes for rapid subsequent scale-up in output.

These \barasss would then be shipped to CERN and installed in a first \tower over the coming months. This \tower could be situated in the MATHUSLA experimental hall if it is available, but if the hall is still under construction, the first \tower could be installed elsewhere near ATLAS or CMS, and then disassembled and re-installed in the experimental hall when it is completed. 

The first \tower will collect cosmic ray alignment and calibration data for up to a year, during which time partner labs continue to produce  bar assemblies for the other 15 \towers, the \wallveto and the \walltrackingmodules.
Starting in 2029 (at the earliest), the first module will collect LLP search data during HL-LHC runs. Installation of  additional modules will proceed over the course of about a year while existing modules take data as they are completed.

\newpage
%%%%%%%%%%%%%%%%%
\appendix
\newpage
%%%%%%%%%%%%%%%%%
%%%%%%%%%%%%%%%%%

%%%%%%%%%%%%%%%%%
\section{Detector R\&D}
\label{s.detectorRD}
%%%%%%%%%%%%%%%%%

%%%%%%%%%%%%%%%%%
\subsection{Scintillator Timing and Light Yield Studies}
\label{s.RDWLSF}

\begin{figure}
\centering
{
\includegraphics[width=.7\textwidth]{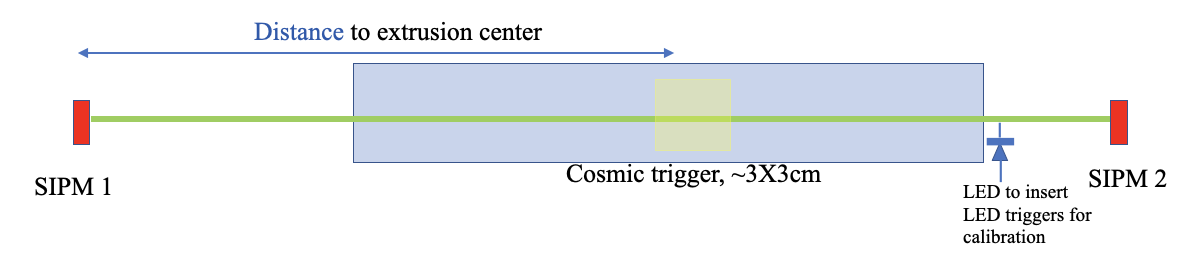}
\caption{\label{fig:cosmicSetup} 
The configuration of the cosmic ray timing and light yield test stand. The blue rectangle represents the extruded scintillator, while the green line represents the wavelength shifting fiber. The excitation point is measured as the distance to the extrusion center from SiPM 1.}
}
\end{figure}

We made an extensive study to understand light yield and timing for the scintillator - extrusion - SiPM system. Figure~\ref{fig:cosmicSetup} shows the basic measurement configuration. A 5~m long fiber was threaded through a shorted piece of extrusion, typically 50~cm long for convenience. (Separate studies showed that this was long enough to achieve the asymptotic light yield, due to finite photon attenuation length in the scintillator.) On each end of the fiber were SiPMs. The SiPMs were read out with PSI DRS4 waveform digitizers operating at 5 gigasamples per second (5~GS/s). Cosmic rays (triggered by a coincidence of small trigger counters) create light in the two SiPMs. The leading edge of the SiPM pulse is measured. Then the difference in time between the two SiPMs is determined.

\begin{figure}[tbh]
    \begin{center}
        \includegraphics[width=0.9\textwidth]{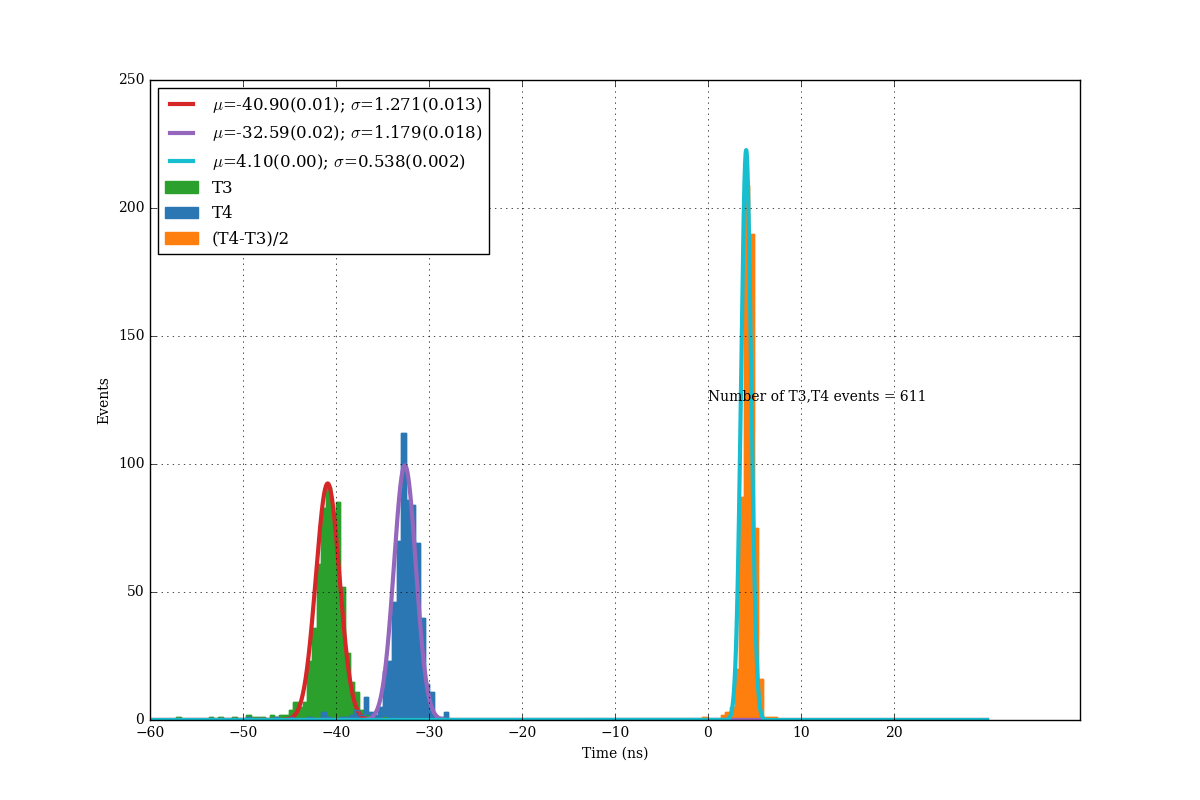}
        \caption{Timing measurement for a 5~m long fiber through a $1\times 4$~cm extrusion. This location is at 250~cm along the fiber, equidistant from the two SiPMs. Time distributions (relative to the cosmic trigger start time) are shown for the two SiPM channels (Chan 3, Chan 4). Also shown is the difference, (T4-T3)/2. We note this difference divided by 2 is our figure of merit for timing. The factor of 2 comes from the observation that different points along the fiber separated by $\delta$ have a $+\delta$ increase in distance from one SiPM, and a $-\delta$ decrease in distance from the other. The timing resolution of 0.538~ns corresponds to about 9~cm rms position resolution, well within MATHUSLA requirement.}
   \label{fig:timing}
   \end{center}
\end{figure}

Figure~\ref{fig:timing} shows a typical result. The SiPM pulse arrival times are shown in green or blue for the 2 SiPMs on the fiber ends. The difference (divided by 2) is plotted in orange. 

\begin{figure}[ht]
    \centering
    {
    \includegraphics[width=.7\textwidth]{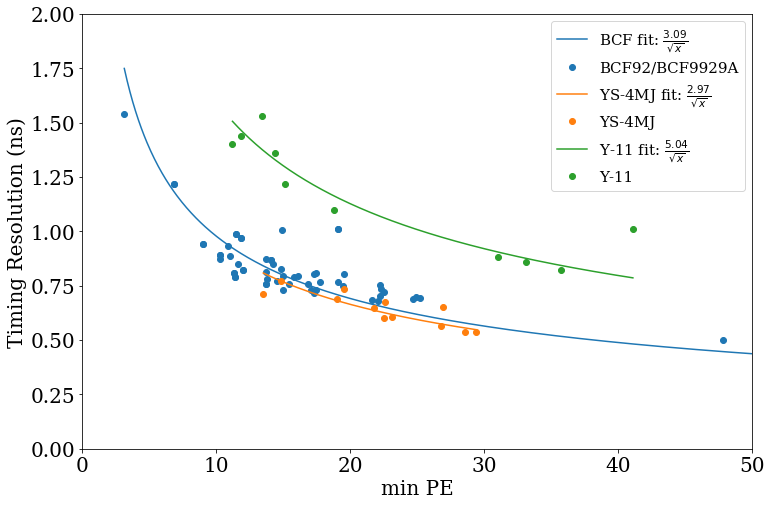}
    \caption{\label{fig:timeResolution} 
    The timing resolution versus the minimum average light yield (SIPM1 or SIPM2) for BCF92 or BCF9929A, Y-11, and YS4-MJ fibers. Each point is a separate cosmic ray measurement consisting of about 1000 triggered events.}
    }
    \end{figure}

Fig.~\ref{fig:timeResolution} depicts timing measurements for multiple cosmic ray runs. Each run had about 1000 events. Three different fiber types are shown in different colors. These fibers differ due to the different WLS fluors used in their manufacture.
We plotted the timing resolution as described in Figure~\ref{fig:timeResolution} vs. the minimum light yield for the BCF92 (and BCF9929A which is the same fluor, different concentration), Y-11, and YS4-MJ fibers. The minimum light yield is the smaller of the two average light yields recorded by the two SiPMs. Also plotted is a fit with form ${\rm A}/\sqrt{\rm x}$.   We see that for a given fiber type the time resolution varies by $1/\sqrt{{\rm n~(PE)}}$. Faster fiber fluors (YS-4 and BCF92) yield better timing results than slower Y-11, as expected.

\subsection{Scintillator and Fiber Baseline Choices}
\label{s.RDscintbaseline}    
Table~\ref{tab:extrusionfiberresults} shows measured light yield and timing resolution for a $1\times 4\times 50$~${\rm cm}^3$ extrusion with an approximately 5~m long 1.5~mm diameter fiber. Two fiber types are shown. The BCF9929A fiber is 5~m long and the YS-4 fiber is 5.6~m long. Both have the approximate length of the fiber required for MATHUSLA. Measurements were taken at the fiber middle and 50~cm from the end. These are illustrations of configurations that satisfies our design criteria. Thus our baseline scintillator is extruded scintillator with co-extruded single hole and white cladding. The profile is 3.5~cm by 1~cm thick. The fiber chosen is 1.5~mm diameter fiber. We note that fibers from San Gobain (BCF9929A) and Kuraray (YS-4) both satisfy our requirement.

\begin{center}
\begin{table} [ht]
\centering
\caption{Light Yield and Timing For Different Excitation points, 1.5mm diameter, 5.0m or 5.6m fiber, 1x4 cm extrusion profile. \label{tab:extrusionfiberresults}}
\begin{tabular}{ |c|c|c|c|c| }
 \hline
 Fiber Type & position along fiber & Avg. LY (SiPM 1) & Avg. LY (SiPM 2) & Timing res. (ns)\\ 
 \hline
  BCF9929A & 250 / 500 & 22.2 & 22.5 & 0.70 \\ 
BCF9929A & 50 / 500  & 37.9 & 13.7 & 0.81 \\ 
  YS-4 & 50 / 560 & 31.1 & 14.8 & 0.79 \\
  YS-4 & 260 / 560 & 29.4 & 31.7 & 0.54 \\
 \hline
\end{tabular}
\end{table}
\end{center}

\subsection{Test-Stands}
Detailed studies are currently ongoing using test-stands at the University of Victoria (UVic) and University of Toronto (UofT). 
These test stands have already demonstrated  combinations of extruded scintillator, WLSFs, and SiPMs that achieve the required light yield and timing performance. 
They also suggest optimization strategies for the readout system -- using commercial, off-the-shelf equipment combined with inexpensive custom-made electronics -- and calibration schemes for the large number of readout channels.

\subsubsection{University of Victoria Test-Stand}
\begin{figure}[htbp]
    \centering
    \includegraphics[width=0.50\linewidth]{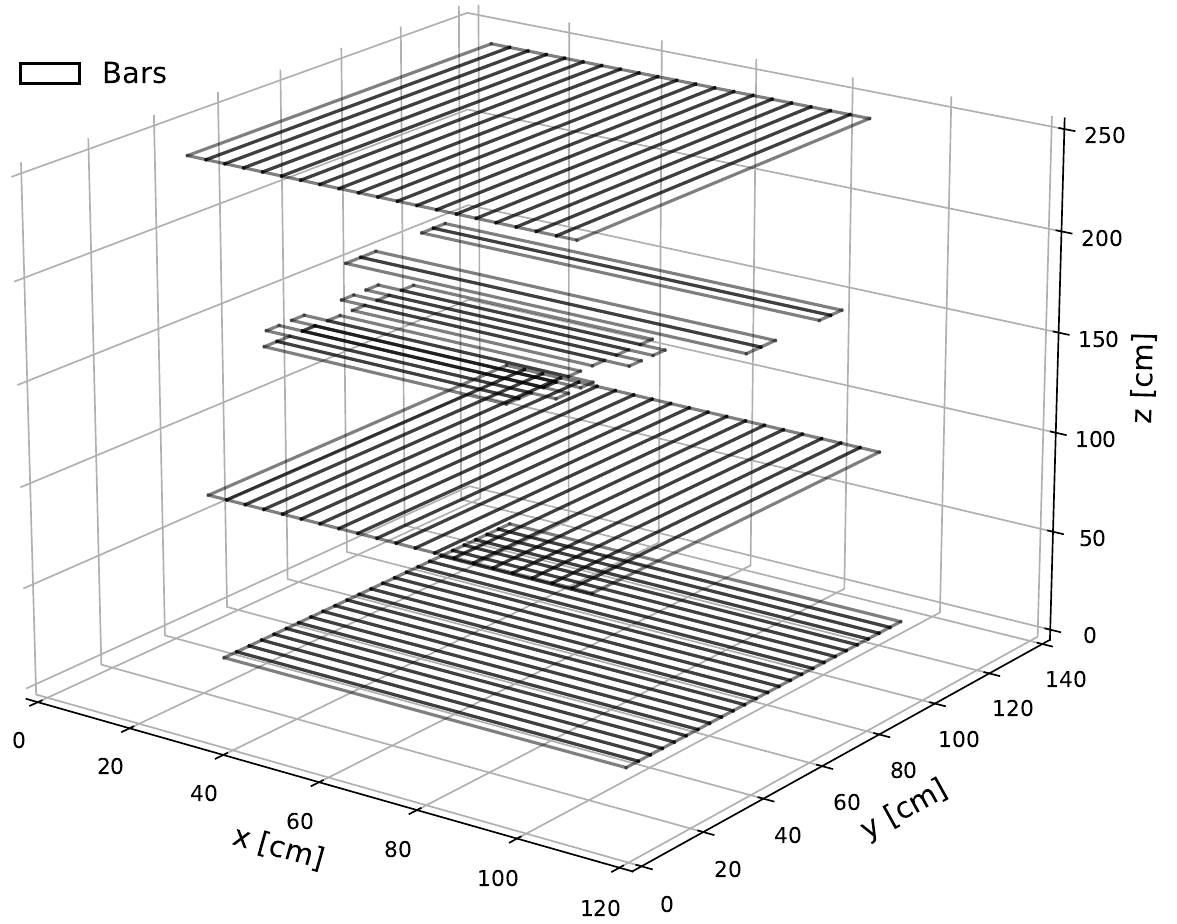}
    \caption{The UVic MATHUSLA prototype detector coverage. The third layer has partial coverage due to limited material and available readout channels.}
    \label{fig:uvic_reco}
\end{figure}

The primary purpose of the Victoria setup is to study the performance of the nominal wavelength shifting fiber. The setup comprises 4 layers of scintillator of alternating orientation consisting of 80 scintillator bars arranged into $80\times100$~cm$^2$ individual layers. The scintillation light produced by cosmic muons is collected in WLSFs 6~m in length, with each end of the fibre connected to a 64 channel Hamamatsu S13661-3050 MPPC SiPM array. 
The lowest and top layers thread each fibre through two bars, while the second layer has four bars per fibre. The third layer features three non-nominal styles of scintillator bar: two have the fibre threaded through a groove on the top, one centred and one offset, and the third has a hole through the bar but horizontally offset. 
After the instrumentation of three full layers, the remaining SiPM channels are connected to the third layer providing partial coverage; some fibres pass through four bars and others only two based on the relative distance to the SiPM. 
The uppermost, fourth, layer again features only two bars per fibre. A geometric representation of the scintillator layout is shown in Figure \ref{fig:uvic_reco}.\\
The SiPM array is powered and readout through a CAEN DT5202 unit. When a the baseline current is exceeded in two paired adjacent channels (0+1, 2+3, etc..) the array is triggered and records all light collected from -50ns to 300ns around the trigger event. An offline trigger requires at least 3 of the 4 layers to be hit providing a significant data sample with which to study the tracking and overall performance. \\
 
\begin{figure}[htbp]
    \centering
    \includegraphics[width=0.50\linewidth]{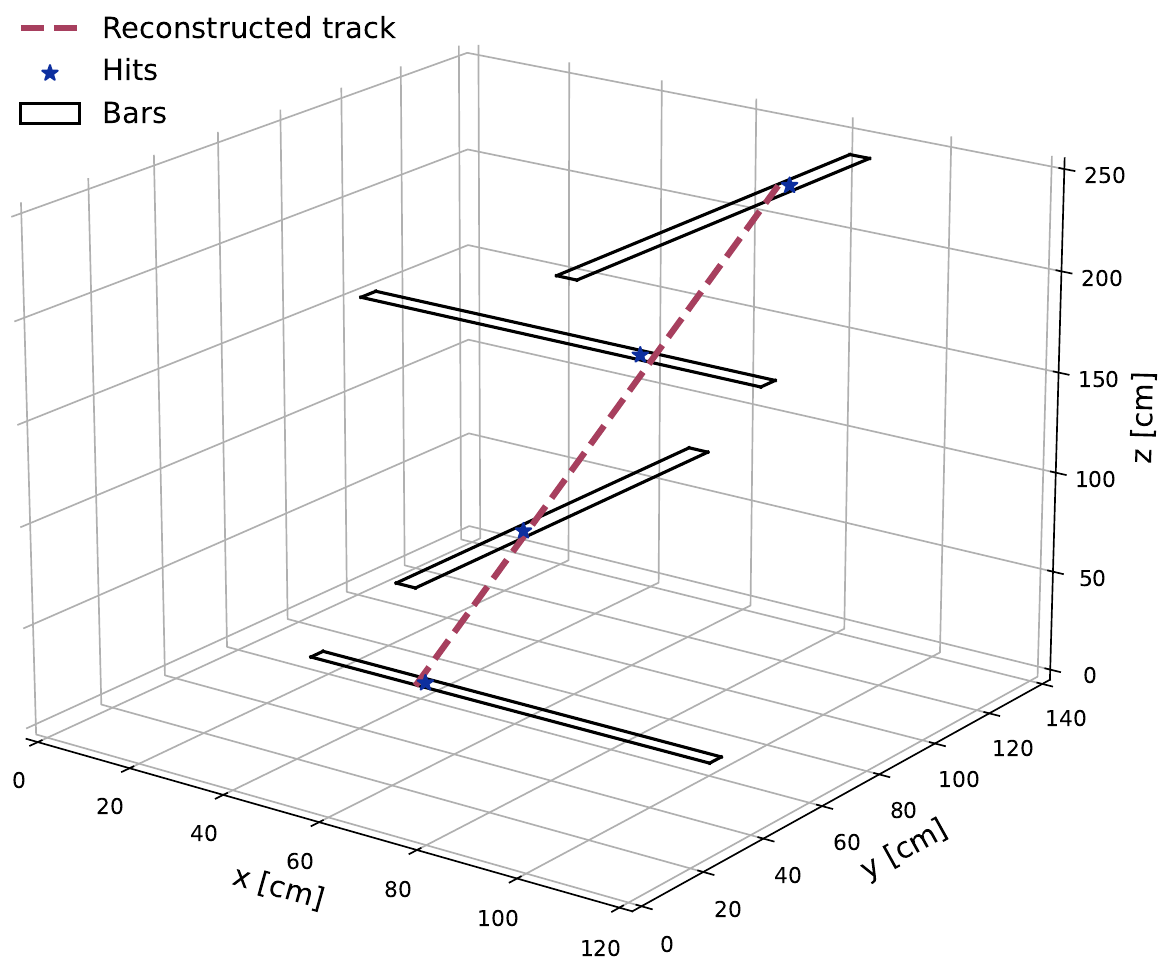}
    \caption{A reconstructed track passing through all four layers of the prototype, with hits reconstructed in each of the four layers. %DESCRIBE BETTER HERE
    }
    \label{fig:uvic_trk}
\end{figure}

\begin{figure}
    \centering
    \includegraphics[width=0.80\linewidth]{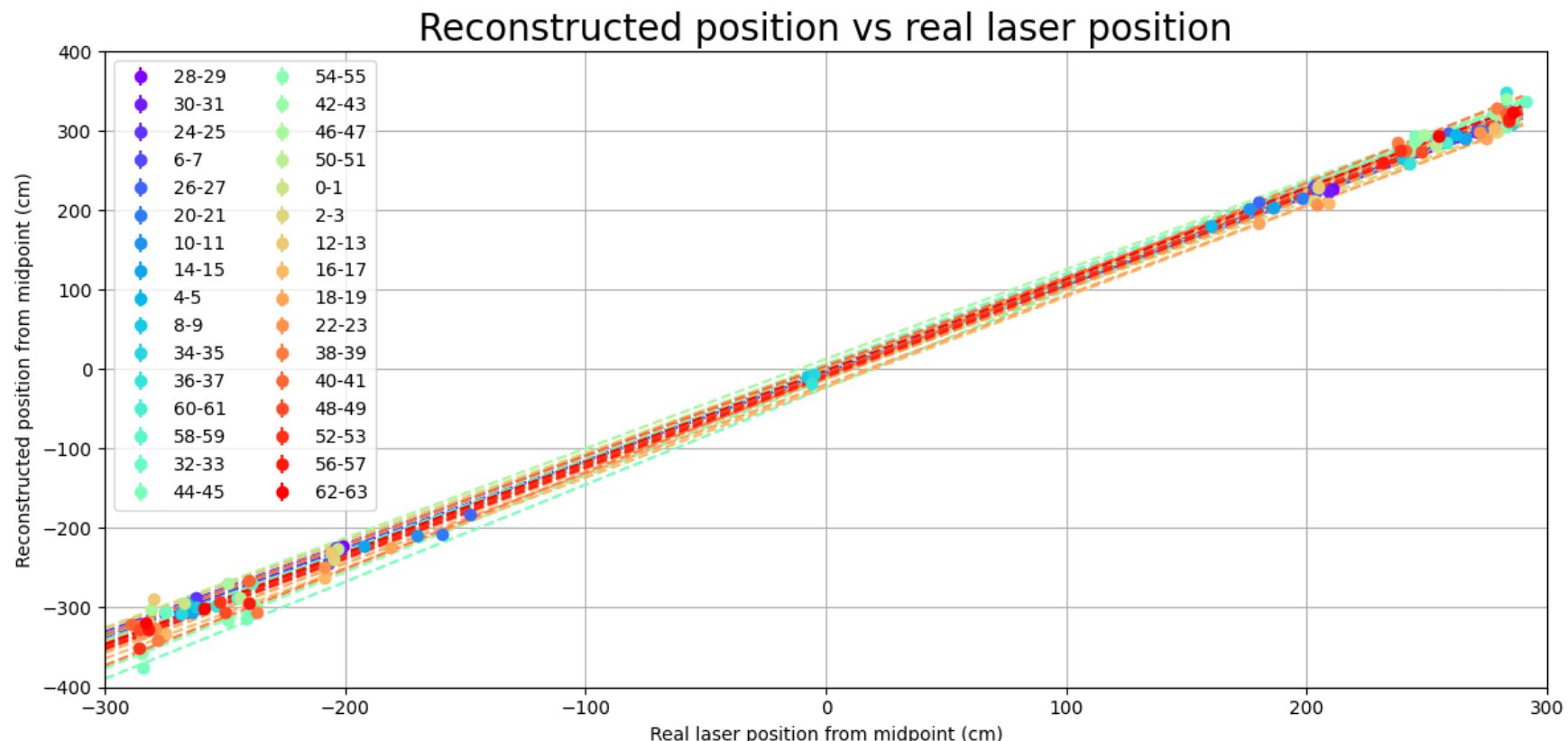}
    \caption{Reconstructed hit position plotted as as a function of the known light injection location into each of the 32 WLSFs installed in UVic test stand. Each point represents a location at which the fibre was pulsed with a laser, and a linear fit is performed to each fibre. Reconstructed locations are determined by a gaussian fit and have an uncertainty equal to the standard deviation. A small number of points were removed from the final index of refraction calculation; these fibre connection locations were difficult to access inside the test stand, resulting in poor laser-fibre connections and thus sub-par data.
    }
    \label{fig:bennett's plot}
\end{figure}

Recorded cosmic ray tracks will be used to benchmark full detector reconstruction. Reconstructed tracks will be used to determine the detection efficiency for the different components, different geometries, and as a function of incoming particle angle. 
An example of a reconstructed cosmic ray candidate with a recorded hit in every layer is shown in Figure~\ref{fig:uvic_trk}\\
\clearpage
\paragraph{R\&D studies}
The following R\&D studies are ongoing at the UVic test stand.
\subparagraph{Position Reconstruction Correction}
The MATHUSLA position reconstruction relies on the time of arrival difference between the two ends of a fibre. With the use of a MHz laser pulser, light can be injected at known locations along the fibre and the position reconstructed. As the light does not take a direct path down the fibre, and rather bounces around internally, a bias is observed in the reconstruction. The bias reconstructs the position closer to the nearer of the two ends of the fibre. This can be corrected by an effective shift to the index of refraction. Figure \ref{fig:bennett's plot} shows the position measurements for 32 channel pairs measured on the same WLSF. The slope represents the needed multiplicative correction to the index of refraction, which is 1.127.
%with a standard deviation of XXXX. \DC{missing number}
\subparagraph{WLSF Aging}
After 8 months of being placed in the third layer of the test setup, some WLSFs which lacked SiPM channels to connect to were recovered from the setup in order to have more full length fibres for testing in a small darkbox. Small deviations/kinks were observed in the recovered fibres corresponding to the areas of maximum curvature. While it is unclear if these formed during installation, slowly while remaining under stress, or during recovery, it highlights the need to evaluate fibre performance after installation and potentially over time. The fibres with small deviations were found to have different timing resolution performance compared to those without. 
\subparagraph{WLSF Quality Assurance}
One of the planned studies for the Victoria setup is a robust testing design for monitoring fibre performance. This includes investigating ways to insert light into the WLSFs via lasers or LEDs mounted to the scintillator or radioactive sources.
\subparagraph{CAEN performance}
The CAEN DT5202 readout system detected trigger rates comparable to the Toronto setup. However, the number of multi-layer hits was significantly lower. The cause of the discrepancy is not known, and a PETsys TOFPET2 ASIC evaluation kit was purchased to understand the source of the discrepancy. This study is ongoing.
\paragraph{Future plans}
Future studies planned for the setup include studies of the timing and track resolution achievable with cosmic rays. The design of the setup also allows for new WLSF compounds and scintillators to be replaced and compared for performance. Temperature testing of the SiPM performance has been carried out but requires further analysis to capture effects deviating from the expected results provided by the datasheet. We also note our events are comparatively low multiplicity compared to MATHUSLA, simulated noise is being considered to study the effects on the detector resolution.

\begin{figure}[h]
\begin{center}
\includegraphics[width= 0.35\textwidth]{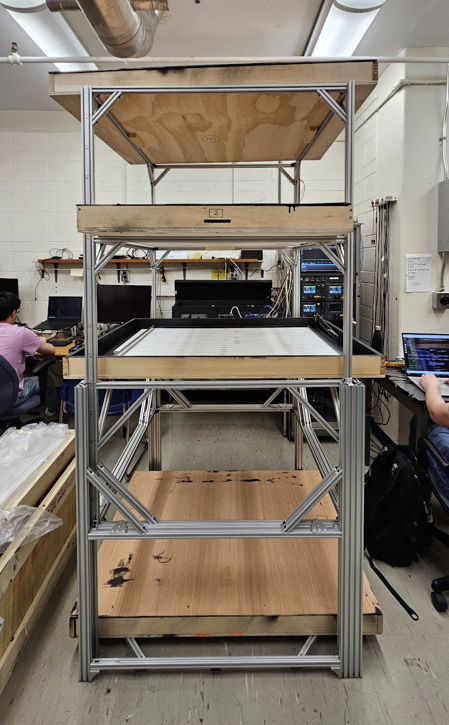}
\hspace{10mm}
\includegraphics[width= 0.55\textwidth]{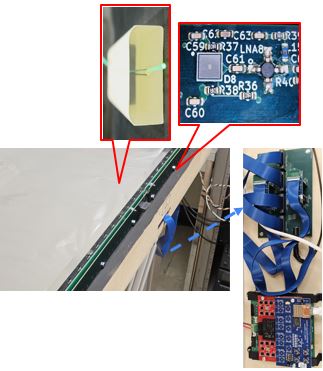}
\end{center}
\vspace*{-8mm}
\caption{Left: The UofT test-stand, with each of the four layers encapsulated in its own light-tight wooden box. Right: The SiPMs touching the fiber ends are all mounted on a PCB on one side of the layer, connected to the readout unit via a cable. }
\vspace*{-2mm}
\label{f.UofT-teststand}
\end{figure}

\begin{figure}[h]
\begin{center}
\includegraphics[width= 0.45\textwidth]{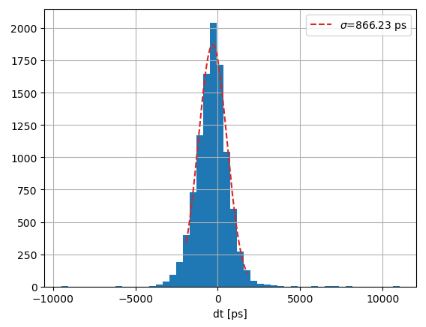}
\includegraphics[width= 0.53\textwidth]{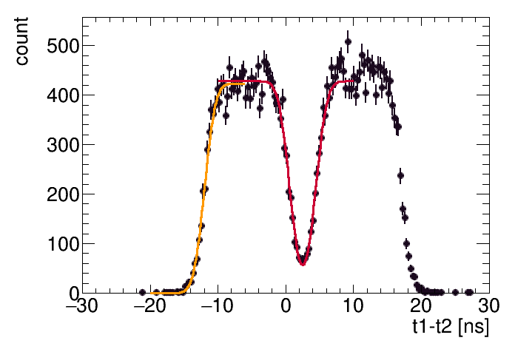}
\end{center}
\vspace*{-8mm}
\caption{Left: In CR events, timing resolution in individual readout channels of the UofT test-stand was found to be 0.87 ns on average. Right: All coincidences between the two SiPMs on the fiber ends were recorded, and the time difference in each pair of SiPM hits ($t_1-t_2$) calculated. The absolute value of $t_1-t_2$  depends upon the event position longitudinally in the bar, while the sign depends upon which bar (in the pair traversed by the WLSF) segment the hit occurred in. The resolution on $t_1-t_2$ was found to be 1.15 ns, which corresponds to 9.1 cm resolution on longitudinal hit position. }
\vspace*{-2mm}
\label{f.UofT-teststand-timing}
\end{figure}

\begin{figure}[h]
\begin{center}
\includegraphics[width= 0.5\textwidth]{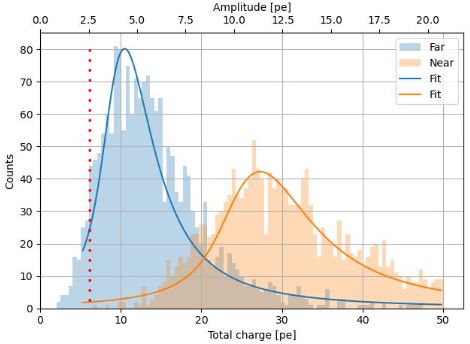}
\includegraphics[width= 0.45\textwidth]{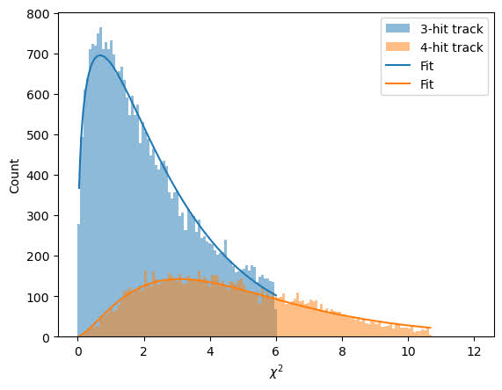}
\end{center}
\vspace*{-8mm}
\caption{Left: In CR events in the UofT test-stand, average charge/hit was 35 photo-electrons in the nearest SiPM channel (orange), and 14 in the SiPM channel at the far end of the same WLSF segment (blue). Right: Goodness-of-fit for the CR tracks reconstructed using a Kalman filter; note some tracks had hits in only 3 layers (blue) rather than all 4 layers (orange) due to their angle of entry, but were still well-reconstructed.}
\vspace*{-2mm}
\label{f.UofT-teststand-results}
\end{figure}

\subsubsection{University of Toronto Test-Stand}

  The main goal of this test-stand is to establish the modularity required for the 40 m layout.
  Rather than routing fibers from all scintillator bars to a multi-channel SiPM array, which makes for lengthy fiber paths that present light-tightness challenges, a SiPM is mounted at one end of each bar with each fiber looping through two bars (Figure~\ref{fig:FNAL_extrusions}) as in the 40 m design. This allows each layer of the test-stand to be encapsulated in its own light-tight unit (Fig.~\ref{f.UofT-teststand}) with all fiber ends and SiPMs along the same layer edge.  The SiPMs are mounted on a PCB with preamplifiers, to maintain the signal-to-noise ratio when connecting to the readout unit with a long cable.
  
   All 4 layers use Saint-Gobain BCF92XL (1.5 mm diameter) fiber, Hamamatsu S14160-3050HS (3 mm) SiPMs, and 1 m x 4 cm x 1 cm scintillator bars. The layers have alternating orientation of the bars. Each layer has an area of  1 m x 1 m; the WLSF segments are each 237~cm long, less than half the length of the nominal MATHUSLA design used in the UVic test stand. 

  The readout unit, a TOFPET2 ASIC evaluation kit from PETsys electronics, is able to power and read 256 SiPM channels. The ASIC features a multi-threshold trigger that maintains good timing resolution with the low threshold while suppressing dark current with the high threshold. Events are triggered on each SiPM channel, and  offline processing selects coincidences between the two SiPMs on the fiber ends. Timing calibration is performed on all the readout channels simultaneously.

  In a set of CR events, the test-stand demonstrated sub-ns timing resolution in individual SiPM channels, and longitudinal hit coordinate resolution of better than 10 cm (Figure \ref{f.UofT-teststand-timing}). With the trigger threshold set to 3.5 photo-electrons, where average charge/hit in the nearest SiPM channel was 35 photo-electrons (left panel of Figure \ref{f.UofT-teststand-results}), the average event rate was measured to be 38500/hr. After tracks with hits in 3 or 4 layers of the were reconstructed using a Kalman filter (panel of Figure \ref{f.UofT-teststand-results}), this yielded a CR track detection \& reconstruction efficiency of $\sim 90 \%$, based on track rate predictions from simulations.

  Forthcoming investigations will explore alternative preamplifiers and cables with comparable performance but reduced cost. The Toronto test stand will also be used to conduct R\&D for the MATHUSLA trigger system.

\newpage
%%%%%%%%%%%%%%%%%

\section{Supporting Documents}
\label{s.supportingdocuments}
%%%%%%%%%%%%%%%%%

\begin{itemize}
    \item 2016 Original MATHUSLA concept proposal by Chou, Curtin and Lubatti~\cite{Chou:2016lxi}.
    \item 2018 MATHUSLA Physics Case document~\cite{Curtin:2018mvb}.
    \item 2018 Letter of Intent, presented to the LHCC~\cite{MATHUSLA:2018bqv}.

    \item 2019 Masters project quantifying how LLP decay acceptance depends on the position of the MATHUSLA detector near CMS~\cite{Alkhatib:2019eyo}.
    
    \item 2019 Input  by the MATHUSLA collaboration to the update process of the European Strategy for Particle Physics~\cite{MATHUSLA:2019qpy}.
    \item 2020 Update to the Letter of Intent~\cite{MATHUSLA:2020uve}.
    \item 2020 Analysis of data collected in 2018 by the MATHUSLA test stand operated above the ATLAS interaction point~\cite{Alidra:2020thg}.
    \item 2020 Analysis by Barron and Curtin demonstrating how correlated event information from MATHUSLA and CMS can determine LLP production mode and underlying parameters~\cite{Barron:2020kfo}. Extends earlier analysis by Curtin and Peskin, which demonstrates how geometry of LLP decay vertex can be used to measure event-by-event LLP boost~\cite{Curtin:2017izq}.
    \item 2022 Contribution by the MATHUSLA collaboration to the North American Snowmass process~\cite{MATHUSLA:2022sze}. 

    \item 2023 Analysis by Curtin and Grewal containing zero-background sensitivity estimates for various LLP benchmark models, and studying how various features of the detector geometry impact signal acceptance.~\cite{Curtin:2023skh}. Also presents the \texttt{MATHUSLA FastSim} code~\cite{MATHUSLAFastSim}.

    \item 2023 Study of the performance of MATHUSLA as an instrument for studying the physics of cosmic rays by measuring extensive air showers~\cite{Alpigiani:2023qjb}. 
\end{itemize}

\newpage
%%%%%%%%%%%%%%%%%

%%%%%%%%%%%%%%%%%
\section{Glossary}
\label{s.glossary}
%%%%%%%%%%%%%%%%%

\begin{itemize}

\item {\bf \mBar}: A single bar is a \barwidth~$\times$~\barthickness bar, \barlength long, made of extruded plastic scintillator. 

\item {\bf \WLSF}: A 1.5~mm wavelength shifting fiber.

\item {\bf \SiPM}: Silicon photomultiplier.

\item {\bf \Barass}: A mechanical unit consisting of 32 
scintillating bars.  Its active sensor area is 112cm $\times$ 235cm, with a physical area of approximately 112cm $\times$ 280cm (the extra length accounts for the electronics readout box containing the SiPMs and space for bending the fibres). Each \barass has a backing of aluminum strongback for structural stability, making it self-supporting when attached at the edges.

\item {\bf \Sublayer}: A unit comprising 8 
\barass modules joined side-to-side, making it a 280~cm $\times$ 896~cm plane.

\item {\bf \Layer}: An assembly of 4 \sublayers, joined along the \barass direction with some overlap, to form a single sensor plane with an active sensor area of about 897 cm $\times$ 896 cm.

%four \sublayers that form an $\sim$ \moduleside$\times$\moduleside unit. It takes into account an overlap of 15~cm of the active (scintillating) area between any two neighboring \sublayers). The active area of each detector \layer is about 895~cm $\times$  896~cm.

\item {\bf \Trackingmodule:} A stack of 6 \Layers, arranged with alternating bar orientation separated from each other by  80cm, placed either horizontally on the ceiling, or vertically on the back wall. 

\item {\bf \Ceilingtrackingmodule:} A tracking module that is mounted on the ceiling.

\item {\bf \Walltrackingmodule:} A tracking module that is mounted on the back wall.

\item {\bf \Plane}: All the tracking \layers at a given height (or at a given horizontal distance from the the rear wall) across different \trackingmodules constitute one \plane.

\item {\bf Trigger layers}: Only \layers in \trackingmodules are used for triggering.

\item {\bf \Wallvetosublayer:} Similar to a \sublayer, but comprising 10 \barass joined end-to-end, making it a 280cm $\times$ 1120cm plane.

\item {\bf \Wallvetolayer:} Similar to a \layer, but comprising 5 \wallvetosublayers, giving it an active sensor area of about 11.18m $\times$ 11.2m.

\item {\bf \Wallveto:} A double-layer of \wallvetolayers (4 per layer with some overlap) covering the wall of MATHUSLA closest to CMS to aid in rejecting LHC muons.

\item {\bf \Floorvetolayer:} Identical to the \layers in the tracking modules, but placed 0.5 and 1m above the floor with perpendicular bar orientations.

\item {\bf \Floorvetostrip:} Two \sublayers sandwiched closely together, mounted 1.5m %\DC{check height} 
above the floor to cover the gap between \floorvetolayers.

\item {\bf \Columndetector:}
Four \floorvetostrips are mounted vertically, on the vertical support columns of two neighboring \ceilingtrackingmodules, with some additional sensor coverage beyond the width spanned by the two columns, to wrap around the intersection of two gaps between \towers. 
This instruments the height of the support columns from the \floorvetostrips to near the \ceilingtrackingmodules, negating the gap created by the support columns in the \floorveto, and enhancing material veto capabilities for inelastic cosmic ray interactions in the columns.

\item {\bf \Floorveto:} The \floorvetolayers, \floorvetostrips and \columndetectors  together comprise the \floorveto, which is hermetic for LHC muons and downward cosmics.

\item {\bf \Veto:} The \wallveto and \floorveto together comprise the veto system of MATHUSLA to aid in rejecting various LHC and cosmic backgrounds. These are not used in triggering.

\item {\bf \Tower}: A $\sim$15~m 
%\DC{check} 
tall assembly of a \ceilingtrackingmodule in the ceiling and two \floorvetolayers near the floor, supported by four vertical support columns. The \floorveto gaps between \towers are covered by \floorvetostrips and the \columndetectors.

\item {\bf Decay volume}: The volume of the detector below the \ceilingtrackingmodules (area $\sim 40~\mathrm{m} \times 40~\mathrm{m}$), starting near the top of the \floorveto to bottom of the \ceilingtrackingmodules ($\sim 11$~m).

\item{\bf 4D reconstruction}: For tracks and vertices, including timing information

\item{\bf Neutral LLP}: Electromagnetically-neutral, Beyond the Standard Model long-lived particle.

\item{\bf Displaced Vertex (DP)}: A vertex which originates in the MATHUSLA decay volume

\item{\bf Tracklet}:
A pseudo-track involving only a few points, such as the CMS IP and hits in MATHUSLA's floor/wall detectors.

\item{\bf Track}: The algorithmically determined line constructed from multiple hits detected across MATHUSLA \layers. Analogous to a charged particle

\item{\bf Vertex}: A convergence point of multiple tracks

\item{\bf LHC}: Large Hadron Collider

\item{\bf CERN}: European Organization for Nuclear Research

\item{\bf DAQ}: Data acquisition

\item{\bf HL-LHC}: High luminosity LHC.

\item{\bf LLP}: Long-lived particle

\item{\bf CMS}: Compact Muon Solenoid, a LHC experiment

\item{\bf BSM}: Beyond Standard Model

\item{\bf ATLAS}: A Toroidal LHC Apparatus, a LHC experiment

\item{\bf BBN}: Big Bang Nucleosynthesis

\item{\bf SHiP}: Search for Hidden Particles, a proposed CERN experiment~\cite{Alekhin:2015byh}.

\item{\bf SHADOWS}: Search for Hidden And Dark Objects With the SPS, a proposed CERN experiment~\cite{Baldini:2021hfw}

\item{\bf CR}: Cosmic ray

\item{\bf IP}: Interaction point of $pp$ collisions

\item{\bf SM}: Standard Model

\item{\bf RHN}: Right-handed neutrino

\item{\bf NLO}: Next-to-leading order accuracy

\item{\bf NNLL}: Next-to-next-to-leading logarithmic accuracy

\item{\bf PBC}: Physics Beyond Colliders, a CERN study group to investigate physics opportunities beyond the LHC collider complex~\cite{Beacham:2019nyx}. 

\item{\bf FASER2}: An upgrade to FASER, the Forward Search Experiment~\cite{Feng:2022inv}.

\item{\bf PCB}: Printed circuit board

\item{\bf PE}: photo-electron

\item{\bf ADC}: Analog to digital converter

\item{\bf ToT}: Time over threshold

\item{\bf L1T}: Level 1 Trigger

\item{\bf HLT}: High level trigger

\item{\bf COTS}: Commercial off the shelf

\item{\bf ASIC}: application-specific integrated circuit

\item{\bf FPGA}: Field-programmable gate array

\item{\bf BC}: Bunch Crossing

\item{\bf PSD}: Permanently Saved Data by the MATHUSLA experiment.

\item{\bf CE}: Civil engineering
\end{itemize}
\newpage
%%%%%%%%%%%%%%%%%

%%%%%%%%%%%%%%%%%
\section{MATHUSLA organization structure}
\label{s.organization}
%%%%%%%%%%%%%%%%%

The MATHUSLA collaboration organizational structure consists of a management board and spokesperson.  The current board members are:
\begin{itemize}
    \item David Curtin (University of Toronto, \texttt{dcurtin@physics.utoronto.ca})
    \item Miriam Diamond (University of Toronto, \texttt{mdiamond@physics.utoronto.ca})
    \item Erez Etzion (Tel Aviv University, \texttt{ereze@tauex.tau.ac.il})
    \item Steven H. Robertson (University of Alberta, \texttt{srobert6@ualberta.ca}) 
    \item Heather Russell (University of Victoria, \texttt{hrussell@uvic.ca})
\end{itemize}
Erez Etzion is the current spokesperson for the MATHUSLA experiment. 
Different subgroups work on their respective R\&D efforts independently and report their progress at weekly MATHUSLA meetings on Zoom, at which various collaboration issues are discussed. In addition, the governing board meets on the order of once per month.

\newpage
%%%%%%%%%%%%%%%%%

%%%%%%%%%%%%%%%%%
\bibliography{references}

\providecommand{\href}[2]{#2}\begingroup\raggedright\begin{thebibliography}{10}

\bibitem{Curtin:2018mvb}
D.~Curtin et~al., {\it {Long-Lived Particles at the Energy Frontier: The
  MATHUSLA Physics Case}},  {\em Rept. Prog. Phys.} {\bf 82} (2019), no.~11
  116201, [\href{http://arxiv.org/abs/1806.07396}{{\tt arXiv:1806.07396}}].

\bibitem{MATHUSLA:2018bqv}
{\bf MATHUSLA} Collaboration, C.~Alpigiani et~al., {\it {A Letter of Intent for
  MATHUSLA: A Dedicated Displaced Vertex Detector above ATLAS or CMS.}},
  \href{http://arxiv.org/abs/1811.00927}{{\tt arXiv:1811.00927}}.

\bibitem{MATHUSLA:2020uve}
{\bf MATHUSLA} Collaboration, C.~Alpigiani et~al., {\it {An Update to the
  Letter of Intent for MATHUSLA: Search for Long-Lived Particles at the
  HL-LHC}},  \href{http://arxiv.org/abs/2009.01693}{{\tt arXiv:2009.01693}}.

\bibitem{MATHUSLA:2022sze}
{\bf MATHUSLA} Collaboration, C.~Alpigiani et~al., {\it {Recent Progress and
  Next Steps for the MATHUSLA LLP Detector}},  in {\em {Snowmass 2021}}, 3,
  2022.
\newblock \href{http://arxiv.org/abs/2203.08126}{{\tt arXiv:2203.08126}}.

\bibitem{Chou:2016lxi}
J.~P. Chou, D.~Curtin, and H.~J. Lubatti, {\it {New Detectors to Explore the
  Lifetime Frontier}},  {\em Phys. Lett.} {\bf B767} (2017) 29--36,
  [\href{http://arxiv.org/abs/1606.06298}{{\tt arXiv:1606.06298}}].

\bibitem{Alimena:2019zri}
J.~Alimena et~al., {\it {Searching for long-lived particles beyond the Standard
  Model at the Large Hadron Collider}},  {\em J. Phys. G} {\bf 47} (2020),
  no.~9 090501, [\href{http://arxiv.org/abs/1903.04497}{{\tt
  arXiv:1903.04497}}].

\bibitem{Alimena:2021mdu}
D.~Acosta et~al., {\it {Review of opportunities for new long-lived particle
  triggers in Run 3 of the Large Hadron Collider}},
  \href{http://arxiv.org/abs/2110.14675}{{\tt arXiv:2110.14675}}.

\bibitem{Beacham:2019nyx}
J.~Beacham et~al., {\it {Physics Beyond Colliders at CERN: Beyond the Standard
  Model Working Group Report}},  {\em J. Phys. G} {\bf 47} (2020), no.~1
  010501, [\href{http://arxiv.org/abs/1901.09966}{{\tt arXiv:1901.09966}}].

\bibitem{ATLAS:2012av}
{\bf ATLAS} Collaboration, G.~Aad et~al., {\it {Search for a light Higgs boson
  decaying to long-lived weakly-interacting particles in proton-proton
  collisions at $\sqrt{s}=7$ TeV with the ATLAS detector}},  {\em Phys. Rev.
  Lett.} {\bf 108} (2012) 251801, [\href{http://arxiv.org/abs/1203.1303}{{\tt
  arXiv:1203.1303}}].

\bibitem{ATLAS:2018tup}
{\bf ATLAS} Collaboration, M.~Aaboud et~al., {\it {Search for long-lived
  particles produced in $pp$ collisions at $\sqrt{s}=13$ TeV that decay into
  displaced hadronic jets in the ATLAS muon spectrometer}},  {\em Phys. Rev. D}
  {\bf 99} (2019), no.~5 052005, [\href{http://arxiv.org/abs/1811.07370}{{\tt
  arXiv:1811.07370}}].

\bibitem{ATLAS:2019jcm}
{\bf ATLAS} Collaboration, G.~Aad et~al., {\it {Search for long-lived neutral
  particles produced in $pp$ collisions at $\sqrt{s} = 13$ TeV decaying into
  displaced hadronic jets in the ATLAS inner detector and muon spectrometer}},
  {\em Phys. Rev. D} {\bf 101} (2020), no.~5 052013,
  [\href{http://arxiv.org/abs/1911.12575}{{\tt arXiv:1911.12575}}].

\bibitem{ATLAS:2022gbw}
{\bf ATLAS} Collaboration, G.~Aad et~al., {\it {Search for events with a pair
  of displaced vertices from long-lived neutral particles decaying into
  hadronic jets in the ATLAS muon spectrometer in pp collisions at $\sqrt
  s$=13\,\,TeV}},  {\em Phys. Rev. D} {\bf 106} (2022), no.~3 032005,
  [\href{http://arxiv.org/abs/2203.00587}{{\tt arXiv:2203.00587}}].

\bibitem{Chacko:2005pe}
Z.~Chacko, H.-S. Goh, and R.~Harnik, {\it {The Twin Higgs: Natural electroweak
  breaking from mirror symmetry}},  {\em Phys. Rev. Lett.} {\bf 96} (2006)
  231802, [\href{http://arxiv.org/abs/hep-ph/0506256}{{\tt hep-ph/0506256}}].

\bibitem{Craig:2015pha}
N.~Craig, A.~Katz, M.~Strassler, and R.~Sundrum, {\it {Naturalness in the Dark
  at the LHC}},  {\em JHEP} {\bf 07} (2015) 105,
  [\href{http://arxiv.org/abs/1501.05310}{{\tt arXiv:1501.05310}}].

\bibitem{Curtin:2015fna}
D.~Curtin and C.~B. Verhaaren, {\it {Discovering Uncolored Naturalness in
  Exotic Higgs Decays}},  {\em JHEP} {\bf 12} (2015) 072,
  [\href{http://arxiv.org/abs/1506.06141}{{\tt arXiv:1506.06141}}].

\bibitem{Batz:2023zef}
A.~Batz, T.~Cohen, D.~Curtin, C.~Gemmell, and G.~D. Kribs, {\it {Dark Sector
  Glueballs at the LHC}},  \href{http://arxiv.org/abs/2310.13731}{{\tt
  arXiv:2310.13731}}.

\bibitem{Curtin:2013fra}
D.~Curtin et~al., {\it {Exotic decays of the 125 GeV Higgs boson}},  {\em Phys.
  Rev. D} {\bf 90} (2014), no.~7 075004,
  [\href{http://arxiv.org/abs/1312.4992}{{\tt arXiv:1312.4992}}].

\bibitem{Coccaro:2016lnz}
A.~Coccaro, D.~Curtin, H.~J. Lubatti, H.~Russell, and J.~Shelton, {\it
  {Data-driven Model-independent Searches for Long-lived Particles at the
  LHC}},  {\em Phys. Rev. D} {\bf 94} (2016), no.~11 113003,
  [\href{http://arxiv.org/abs/1605.02742}{{\tt arXiv:1605.02742}}].

\bibitem{CMS:2023arc}
{\bf CMS} Collaboration, {\it {Search for long-lived particles decaying in the
  CMS muon detectors in proton-proton collisions at
  $\sqrt{s}=13~\mathrm{TeV}$}}, .

\bibitem{GEANT4:2002zbu}
{\bf GEANT4} Collaboration, S.~Agostinelli et~al., {\it {GEANT4--a simulation
  toolkit}},  {\em Nucl. Instrum. Meth. A} {\bf 506} (2003) 250--303.

\bibitem{hllhclumi}
{\it {The High-Luminosity LHC Project. 298th Meeting of Scientific Policy
  Committee}},  tech. rep., 2016.

\bibitem{sato2008development}
T.~Sato, H.~Yasuda, K.~Niita, A.~Endo, and L.~Sihver, {\it Development of
  parma: Phits-based analytical radiation model in the atmosphere},  {\em
  Radiation research} {\bf 170} (2008), no.~2 244--259.

\bibitem{CRY}
C.~Hagmann, D.~Lange, and D.~Wright, {\it Cosmic-ray shower generator (cry) for
  monte carlo transport codes},  vol.~2, pp.~1143--1146, 01, 2007.

\bibitem{Tanabashi:2018oca}
{\bf Particle Data Group} Collaboration, M.~Tanabashi et~al., {\it {Review of
  Particle Physics}},  {\em Phys. Rev. D} {\bf 98} (2018), no.~3 030001.

\bibitem{Alwall:2011uj}
J.~Alwall, M.~Herquet, F.~Maltoni, O.~Mattelaer, and T.~Stelzer, {\it {MadGraph
  5 : Going Beyond}},  {\em JHEP} {\bf 06} (2011) 128,
  [\href{http://arxiv.org/abs/1106.0522}{{\tt arXiv:1106.0522}}].

\bibitem{Andreopoulos:2009rq}
C.~Andreopoulos et~al., {\it {The GENIE Neutrino Monte Carlo Generator}},  {\em
  Nucl. Instrum. Meth.} {\bf A614} (2010) 87--104,
  [\href{http://arxiv.org/abs/0905.2517}{{\tt arXiv:0905.2517}}].

\bibitem{battistoni2003fluka}
G.~Battistoni, A.~Ferrari, T.~Montaruli, and P.~Sala, {\it The fluka
  atmospheric neutrino flux calculation},  {\em Astroparticle Physics} {\bf 19}
  (2003), no.~2 269--290.

\bibitem{Curtin:2023skh}
D.~Curtin and J.~S. Grewal, {\it {Long Lived Particle Decays in MATHUSLA}},
  \href{http://arxiv.org/abs/2308.05860}{{\tt arXiv:2308.05860}}.

\bibitem{Alwall:2014hca}
J.~Alwall, R.~Frederix, S.~Frixione, V.~Hirschi, F.~Maltoni, O.~Mattelaer,
  H.~S. Shao, T.~Stelzer, P.~Torrielli, and M.~Zaro, {\it {The automated
  computation of tree-level and next-to-leading order differential cross
  sections, and their matching to parton shower simulations}},  {\em JHEP} {\bf
  07} (2014) 079, [\href{http://arxiv.org/abs/1405.0301}{{\tt
  arXiv:1405.0301}}].

\bibitem{Sjostrand:2006za}
T.~Sjostrand, S.~Mrenna, and P.~Z. Skands, {\it {PYTHIA 6.4 Physics and
  Manual}},  {\em JHEP} {\bf 05} (2006) 026,
  [\href{http://arxiv.org/abs/hep-ph/0603175}{{\tt hep-ph/0603175}}].

\bibitem{Sjostrand:2007gs}
T.~Sjostrand, S.~Mrenna, and P.~Z. Skands, {\it {A Brief Introduction to PYTHIA
  8.1}},  {\em Comput. Phys. Commun.} {\bf 178} (2008) 852--867,
  [\href{http://arxiv.org/abs/0710.3820}{{\tt arXiv:0710.3820}}].

\bibitem{Bozzi:2005wk}
G.~Bozzi, S.~Catani, D.~de~Florian, and M.~Grazzini, {\it {Transverse-momentum
  resummation and the spectrum of the Higgs boson at the LHC}},  {\em Nucl.
  Phys. B} {\bf 737} (2006) 73--120,
  [\href{http://arxiv.org/abs/hep-ph/0508068}{{\tt hep-ph/0508068}}].

\bibitem{deFlorian:2011xf}
D.~de~Florian, G.~Ferrera, M.~Grazzini, and D.~Tommasini, {\it
  {Transverse-momentum resummation: Higgs boson production at the Tevatron and
  the LHC}},  {\em JHEP} {\bf 11} (2011) 064,
  [\href{http://arxiv.org/abs/1109.2109}{{\tt arXiv:1109.2109}}].

\bibitem{LHCHiggsCrossSectionWorkingGroup:2016ypw}
{\bf LHC Higgs Cross Section Working Group} Collaboration, D.~de~Florian
  et~al., {\it {Handbook of LHC Higgs Cross Sections: 4. Deciphering the Nature
  of the Higgs Sector}},  \href{http://arxiv.org/abs/1610.07922}{{\tt
  arXiv:1610.07922}}.

\bibitem{ATLAS:2016fij}
{\bf ATLAS} Collaboration, G.~Aad et~al., {\it {Measurement of $W^{\pm}$ and
  $Z$-boson production cross sections in $pp$ collisions at $\sqrt{s}=13$ TeV
  with the ATLAS detector}},  {\em Phys. Lett. B} {\bf 759} (2016) 601--621,
  [\href{http://arxiv.org/abs/1603.09222}{{\tt arXiv:1603.09222}}].

\bibitem{Andreopoulos:2015wxa}
C.~Andreopoulos, C.~Barry, S.~Dytman, H.~Gallagher, T.~Golan, R.~Hatcher,
  G.~Perdue, and J.~Yarba, {\it {The GENIE Neutrino Monte Carlo Generator:
  Physics and User Manual}},  \href{http://arxiv.org/abs/1510.05494}{{\tt
  arXiv:1510.05494}}.

\bibitem{fruhwirth1987application}
R.~Fr{\"u}hwirth, {\it Application of kalman filtering to track and vertex
  fitting},  {\em Nuclear Instruments and Methods in Physics Research Section
  A: Accelerators, Spectrometers, Detectors and Associated Equipment} {\bf 262}
  (1987), no.~2-3 444--450.

\bibitem{lynch1991approximations}
G.~R. Lynch and O.~I. Dahl, {\it Approximations to multiple coulomb
  scattering},  {\em Nuclear Instruments and Methods in Physics Research
  Section B: Beam Interactions with Materials and Atoms} {\bf 58} (1991), no.~1
  6--10.

\bibitem{wolin1993covariance}
E.~Wolin and L.~Ho, {\it Covariance matrices for track fitting with the kalman
  filter},  {\em Nuclear Instruments and Methods in Physics Research Section A:
  Accelerators, Spectrometers, Detectors and Associated Equipment} {\bf 329}
  (1993), no.~3 493--500.

\bibitem{ATLAS:2023tkt}
{\bf ATLAS} Collaboration, {\it {Combination of searches for invisible decays
  of the Higgs boson using 139 fb\ensuremath{-}1 of proton-proton collision
  data at s=13 TeV collected with the ATLAS experiment}},  {\em Phys. Lett. B}
  {\bf 842} (2023) 137963, [\href{http://arxiv.org/abs/2301.10731}{{\tt
  arXiv:2301.10731}}].

\bibitem{Dainese:2019rgk}
A.~Dainese, M.~Mangano, A.~B. Meyer, A.~Nisati, G.~Salam, and M.~A. Vesterinen,
  eds., vol.~7/2019 of {\em CERN Yellow Reports: Monographs}.
\newblock CERN, Geneva, Switzerland, 2019.

\bibitem{CMS:2024bvl}
{\bf CMS} Collaboration, A.~Hayrapetyan et~al., {\it {Search for long-lived
  particles decaying in the CMS muon detectors in proton-proton collisions at
  s=13\,\,TeV}},  {\em Phys. Rev. D} {\bf 110} (2024), no.~3 032007,
  [\href{http://arxiv.org/abs/2402.01898}{{\tt arXiv:2402.01898}}].

\bibitem{Aielli:2019ivi}
G.~Aielli et~al., {\it {Expression of interest for the CODEX-b detector}},
  {\em Eur. Phys. J. C} {\bf 80} (2020), no.~12 1177,
  [\href{http://arxiv.org/abs/1911.00481}{{\tt arXiv:1911.00481}}].

\bibitem{Burdman:2006tz}
G.~Burdman, Z.~Chacko, H.-S. Goh, and R.~Harnik, {\it {Folded supersymmetry and
  the LEP paradox}},  {\em JHEP} {\bf 02} (2007) 009,
  [\href{http://arxiv.org/abs/hep-ph/0609152}{{\tt hep-ph/0609152}}].

\bibitem{MATHUSLAFastSim}
W.~Cui, D.~Curtin, J.~S. Grewal, and L.~Luo, {\it {\texttt{MATHUSLA FastSim}}},
   8, 2023.

\bibitem{Curtin:2022tou}
D.~Curtin, C.~Gemmell, and C.~B. Verhaaren, {\it {Simulating glueball
  production in Nf=0 QCD}},  {\em Phys. Rev. D} {\bf 106} (2022), no.~7 075015,
  [\href{http://arxiv.org/abs/2202.12899}{{\tt arXiv:2202.12899}}].

\bibitem{Mu2e:2014fns}
{\bf Mu2e} Collaboration, L.~Bartoszek et~al., {\it {Mu2e Technical Design
  Report}},  \href{http://arxiv.org/abs/1501.05241}{{\tt arXiv:1501.05241}}.

\bibitem{SIPM_active}
E.~Kuznetsov, {\it {Temperature-compensated silicon photomultiplier}},  {\em
  NIMA Volume 912, 21 December 2018, Pages 226-230} (2018).

\bibitem{SIPM_passive}
H.~Miyamoto and al., {\it {SiPM development and application for astroparticle
  physics experiments}},  {\em PROCEEDINGS OF THE 31st ICRC, \L\'{O}D\'{Z}}
  (2009).

\bibitem{Barron:2020kfo}
J.~Barron and D.~Curtin, {\it {On the Origin of Long-Lived Particles}},  {\em
  JHEP} {\bf 12} (7, 2020) 061, [\href{http://arxiv.org/abs/2007.05538}{{\tt
  arXiv:2007.05538}}].

\bibitem{CERN-LHCC-2020-004}
{\bf CMS} Collaboration, {\it {The Phase-2 Upgrade of the CMS Level-1
  Trigger}},  tech. rep., CERN, Geneva, 2020.
\newblock Final version.

\bibitem{Alkhatib:2019eyo}
I.~Alkhatib, {\it {Geometric Optimization of the MATHUSLA Detector}},
  \href{http://arxiv.org/abs/1909.05896}{{\tt arXiv:1909.05896}}.

\bibitem{MATHUSLA:2019qpy}
{\bf MATHUSLA} Collaboration, H.~Lubatti et~al., {\it {Explore the lifetime
  frontier with MATHUSLA}},  {\em JINST} {\bf 15} (2020), no.~06 C06026,
  [\href{http://arxiv.org/abs/1901.04040}{{\tt arXiv:1901.04040}}].

\bibitem{Alidra:2020thg}
M.~Alidra et~al., {\it {The MATHUSLA test stand}},  {\em Nucl. Instrum. Meth.
  A} {\bf 985} (2021) 164661, [\href{http://arxiv.org/abs/2005.02018}{{\tt
  arXiv:2005.02018}}].

\bibitem{Curtin:2017izq}
D.~Curtin and M.~E. Peskin, {\it {Analysis of Long Lived Particle Decays with
  the MATHUSLA Detector}},  {\em Phys. Rev. D} {\bf 97} (2018), no.~1 015006,
  [\href{http://arxiv.org/abs/1705.06327}{{\tt arXiv:1705.06327}}].

\bibitem{Alpigiani:2023qjb}
C.~Alpigiani et~al., {\it {Cosmic-ray searches with the MATHUSLA detector}},
  \href{http://arxiv.org/abs/2311.07704}{{\tt arXiv:2311.07704}}.

\bibitem{Alekhin:2015byh}
S.~Alekhin et~al., {\it {A facility to Search for Hidden Particles at the CERN
  SPS: the SHiP physics case}},  {\em Rept. Prog. Phys.} {\bf 79} (2016),
  no.~12 124201, [\href{http://arxiv.org/abs/1504.04855}{{\tt
  arXiv:1504.04855}}].

\bibitem{Baldini:2021hfw}
W.~Baldini et~al., {\it {SHADOWS (Search for Hidden And Dark Objects With the
  SPS)}},  \href{http://arxiv.org/abs/2110.08025}{{\tt arXiv:2110.08025}}.

\bibitem{Feng:2022inv}
J.~L. Feng et~al., {\it {The Forward Physics Facility at the High-Luminosity
  LHC}},  {\em J. Phys. G} {\bf 50} (2023), no.~3 030501,
  [\href{http://arxiv.org/abs/2203.05090}{{\tt arXiv:2203.05090}}].

\end{thebibliography}\endgroup
\bibliographystyle{JHEP}
%%%%%%%%%%%%%%%%%

\end{document}